\documentclass[11pt]{article}%

\usepackage[bookmarks, linkcolor=blue, urlcolor=blue]{hyperref}
\usepackage{amsmath}
\usepackage{amsthm}
\usepackage{mathtools}
\usepackage{amssymb}
\usepackage{amsfonts}
\usepackage{pgffor}
\usepackage{pgf}
\usepackage{tikz,pgfplots}
\usepackage{graphicx}
\usepackage{comment}
\usepackage{caption}
\usepackage{adjustbox}
\usepackage[authoryear]{natbib}
\usepackage{dcolumn}
\usepackage{rotating}
\usepackage[flushleft]{threeparttable}
\usepackage{pdflscape}
\usepackage{afterpage}
\usepackage[pass]{geometry}
\usepackage{setspace}
\usepackage{bigstrut}
\usepackage{footnote}%
\providecommand{\U}[1]{\protect\rule{.1in}{.1in}}
\newcolumntype{L}{D{.}{.}{2,2}}
\newcommand\mc[1]{\multicolumn{1}{c}{#1}}
\hypersetup{
	colorlinks   = true,
	citecolor    = blue
}
\specialcomment{mycomment} {\begingroup\color{gray}}{\endgroup}
\excludecomment{mycomment}
\usepgflibrary{plothandlers}
\newtheorem{theorem}{Theorem}
\newtheorem{proposition}{Proposition}[section]
\newtheorem{lemma}{Lemma}[section]

\newenvironment{pff}[1][Proof]{\vspace{1ex}{\noindent\textbf{#1.} }\hspace{.1em}}
{\hfill\qed\vspace{1ex}}

\newtheoremstyle{TheoremNum}
{\topsep}{\topsep}
{\upshape}
{}
{\bfseries}
{.}
{ }
{\thmname{#1}\thmnote{ \bfseries #3}}
\theoremstyle{definition}

\newtheorem{example2}{Example}[section]
\newtheorem{rem}{Remark}[section]
\newtheorem{assumption}{Assumption}

\numberwithin{equation}{section}
\newcommand*\diff{\mathop{}\!\mathrm{d}}
\DeclareMathOperator*{\argmax}{argmax}
\let\originalleft\left
\let\originalright\right
\renewcommand{\left}{\mathopen{}\mathclose\bgroup\originalleft}
\renewcommand{\right}{\aftergroup\egroup\originalright}
\begin{document}

\begin{center}
{\Large \textbf{Adjusted QMLE for the Spatial Autoregressive Parameter}}

\vspace{1cm}

\renewcommand{\thefootnote}{\fnsymbol{footnote}}

\begin{savenotes}
\begin{tabular}
[c]{ccc}%
Federico Martellosio\footnote{{\Large {\footnotesize Corresponding author. School of Economics, University of Surrey, Guildford, Surrey, GU2 7XH, UK. Tel: +44 (0) 1483
			683473}}} & \hspace{0.54cm} & Grant Hillier\\
University of Surrey, UK &  & CeMMAP and\\
{\Large {\footnotesize \texttt{f.martellosio@surrey.ac.uk}}} &  & University
of Southampton, UK\\
&  & {\Large {\footnotesize \texttt{ghh@soton.ac.uk}}}%
\end{tabular}
\end{savenotes}

\setcounter{footnote}{0}

\bigskip

February 28, 2019

\end{center}

\noindent\textbf{Abstract}

\noindent One simple, and often very effective, way to attenuate the impact of
nuisance parameters on maximum likelihood estimation of a parameter of
interest is to recenter the profile score for that parameter. We apply this
general principle to the quasi-maximum likelihood estimator (QMLE) of the
autoregressive parameter $\lambda$ in a spatial autoregression. The resulting
estimator for $\lambda$ has better finite sample properties compared to the
QMLE for $\lambda$, especially in the presence of a large number of
covariates. It can also solve the incidental parameter problem that arises,
for example, in social interaction models with network fixed effects, or in
spatial panel models with individual or time fixed effects. However, spatial
autoregressions present specific challenges for this type of adjustment,
because recentering the profile score may cause the adjusted estimate to be
outside the usual parameter space for $\lambda$. Conditions for this to happen
are given, and implications are discussed. For inference, we propose
confidence intervals based on a Lugannani--Rice approximation to the
distribution of the adjusted QMLE of $\lambda$. Based on our simulations, the
coverage properties of these intervals are excellent even in models with a
large number of covariates.

\section{Introduction}

Among the several difficulties posed by nuisance parameters, a fundamental
problem in a frequentist framework is that the profile likelihood function for
a parameter of interest is typically not a genuine likelihood, in the sense
that it does not correspond to the density of any observable random variable.
To tackle this issue, a number of \textit{modified profile likelihoods} have
been proposed \citep[see, e.g.,][]{LaskarKing2001, Pace2006}. Such modified
profile likelihoods are genuine likelihoods only in special cases, but they
can often be interpreted as approximations to a genuine likelihood. In
practice, they tend to perform better than the profile likelihood,
particularly when there is little information in the data about the nuisance
parameters, which is likely to happen, for example, if the number of nuisance
parameters is large relative to the sample size. The better performance of
modified profile likelihoods is not necessarily captured by first-order
asymptotic theory. Indeed, if the number of nuisance parameters does not
depend on the sample size, modified profile likelihoods usually produce
estimators that are first-order asymptotically equivalent to the maximum
likelihood estimator (MLE). On the other hand, if the number of nuisance
parameters increases with the sample size, a modified profile likelihood may
be preferable even in terms of first-order asymptotic properties \citep[see, e.g.,][]{Jiang96,Sartori2003, Arellano2007}.

One consequence of a profile likelihood not being a genuine likelihood is that
the expectation of the profile score is generally nonzero, which means that
setting the profile score to zero does not provide an unbiased estimating
equation. A simple modified profile likelihood is therefore obtained by
centering the profile score. We refer to the modified profile likelihood
obtained in this way as the \textit{adjusted profile likelihood}, and to the
associated MLE as the \textit{adjusted MLE}. The principle underlying this
type of adjustment has been motivated from various perspectives
\citep[e.g.,][]{Neyman1948, Conniffe1987, McCullagh1990}, and has been applied
to several statistical models \citep[e.g.,][]{Macaskill1993, DhaeneJoch2015}.

\begin{mycomment}
	or use the other dhaene jochmans
\end{mycomment}

\begin{mycomment}
	The adjusted MLE is defined by setting the profile score equal to
	its expectation rather than to zero, and hence it coincides with the expected
	MLE of \cite{Conniffe1987}.
\end{mycomment}

\begin{mycomment}
	delete:  see, e.g.,
	{Horrace2015} and {Lin2015} for just two examples.
\end{mycomment}

The present paper is concerned with the adjusted MLE of the autoregressive
parameter $\lambda$ in a spatial autoregression. Reliable estimation of
$\lambda$ is important in many applications in economics, as well as in other
fields. For example, in social interaction models $\lambda$ captures the
endogenous effect, the assessment of which may be crucial for policy purposes
\citep[see][]{Mansky93,Moffitt01,LeeLiuLin2010}. More precisely, we will
consider the quasi-MLE (QMLE), that is, the MLE obtained by maximizing the
Gaussian likelihood, but without assuming that the error distribution is truly
Gaussian. The literature on estimating (cross-sectional or panel) spatial
autoregressions is now very large, the QMLE being perhaps the most popular
estimator. An early reference for the QMLE in spatial autoregressions is
\cite{Ord1975}, whereas a rigorous first-order asymptotic analysis of the QMLE
was given only much later, in an influential paper by \cite{Lee2004}, under
conditions that have become standard in the literature. Recently, higher-order
approximations to the distribution of the QMLE of $\lambda$ have become
available \citep{Robinson2015}, and, motivated by the fact that the QMLE of
$\lambda$ can suffer from substantial bias, a number of studies have suggested
bias reduction procedures \citep[e.g.,][]{BaoUllah2007, Bao2013, Yang2015}. In
fact, part of the bias in the QMLE of a parameter of interest can be
attributed to the bias in the profile likelihood estimating equation, so
centering the profile score can itself be interpreted as a bias reduction
technique. \cite{LeeYu2010}, while discussing bias reduction in spatial panel
data models, state explicitly that correction methods based on modifying the
score \textquotedblleft might be possible and would be of interest in future
research\textquotedblright\ (see their footnote 24). \cite{Yu2015} show that,
in the absence of incidental parameters, the adjusted QMLE of $\lambda$ is
first-order asymptotically equivalent to the QMLE of $\lambda$. \cite{Yu2015}
also derive the second-order bias of the adjusted QMLE, and compare by
simulation the adjusted QMLE with other bias reduction techniques. Prior to
that article, \cite{Durban2000} had provided a preliminary investigation of
the adjusted QMLE in a class of models that include spatial autoregressions.
Notwithstanding the last two papers, the general distributional properties of
the adjusted QMLE of $\lambda$ remain unclear. Indeed, \cite{Yu2015} conclude
their analysis by suggesting that the adjusted QMLE of $\lambda$
\textquotedblleft deserves a deep study in future\textquotedblright.

\begin{mycomment}
concerning the sentence in the prev paragraph referring to \cite{LeeYu2010}, note that they're talking about correcting asy bias due to time fixed eff there
\end{mycomment}

Our main contributions are as follows. \textit{First}, on studying the
properties of the adjusted profile likelihood, we find that the distributions
of the QMLE of $\lambda$ and its adjusted version may be supported on
different intervals. This is due to the fact that, in spatial autoregressions,
the parameter $\lambda$ is usually restricted to a certain interval containing
the origin. Such a restriction is incorporated in the QMLE (which, strictly
speaking, should therefore be referred to as a restricted QMLE), but may be
violated once the profile score is recentered. We discuss the implications of
the supports being different, and give conditions for this to happen.
\textit{Second}, for inference about $\lambda$, we propose confidence
intervals based on the \cite{Lugannani80} saddlepoint approximation to the cdf
of the adjusted QMLE. Contrary to the commonly employed Wald confidence
intervals, these confidence intervals have excellent coverage properties even
when the dimension of the nuisance parameter is large. \textit{Third}, we
consider the social interaction model of \cite{LeeLiuLin2010}, and show that
the adjusted QMLE solves the incidental parameter problem due to the network
fixed effects without requiring the condition on $W$ (row normalization) that
is needed for the estimator in \cite{LeeLiuLin2010}. \textit{Fourth}, we
compare the QMLE and the adjusted QMLE by Monte Carlo simulation, and find
evidence that the latter is preferable in a variety of circumstances.

The rest of the paper is structured as follows. Section \ref{sec SLM}
introduces the spatial autoregressive (SAR) model and the QMLE of $\lambda$.
Section \ref{sec SLM adj prof lik} discusses the properties of the adjusted
profile likelihood for $\lambda$, and introduces the confidence intervals
based on the adjusted QMLE. Section \ref{sec SEM} briefly considers the
spatial error model, which is less popular than the SAR model in economic
applications, but provides important motivation for the adjusted ML procedure.
Section \ref{sec Monte Carlo} contains simulation evidence on the performance
of the adjusted QMLE and associated confidence intervals. Section
\ref{sec concl} discusses extensions. Appendix \ref{app add aux} contains some
auxiliary results needed for the proofs, which can be found in Appendix
\ref{app proofs}. Various technical materials related to the paper can be
found in the online Supplement.

Throughout the paper, all vectors and matrices are real valued unless
otherwise indicated. The null space of a matrix $A$ is denoted by
$\operatorname{null}(A)$, the column space by $\operatorname{col}(A),$ and the
orthogonal complement of $\operatorname{col}(A)$ by $\operatorname{col}%
^{\perp}(A)$. Also, $M_{A}$ denotes the orthogonal projector onto
$\operatorname{col}^{\perp}(A)$ ($M_{A}\coloneqq I_{n}-A(A^{\prime}A)^{-1}A^{\prime}$
if $A$ has full column rank). Finally, $\mu_{\mathbb{R}^{n}}$ denotes the
Lebesgue measure on $\mathbb{R}^{n}$, and \textquotedblleft
a.s.\textquotedblright\ stands for almost surely, with respect to
$\mu_{\mathbb{R}^{n}}$.

\begin{mycomment}
\begin{enumerate}
\item IMPORTANT: in the simulations, check how many times adjMLE%
$>$%
1 corresponds to global max of lik
$>$%
1. If most, then the diff between the two estimators is just a distortion due
to the fact that the lambdaMLE is a restricted MLE, and the correct comparison
would be between global MLE and adj MLE (or restricted adjMLE and MLE)
\item looks like in most cases $\hat{\lambda}_{\mathrm{aML}}\in\Lambda$ when
$\hat{\lambda}_{\mathrm{ML}}$ is not the unrestricted QMLE (but not always,
see plot\_lik\_score\_SLM\_with\_adj.m with adjustedMLEgreaterthan1\_QMLE=MLEunrestr.mat)
\item important: estimates outside $\Lambda$: use bayesian methods, collect
more data, model is wrong.\ Under specific distrib assumption the prob
estimator outside $\Lambda$ can be computed quickly using the cdf
representations ( a good disc of what to do when a negative estimate of a
variance is obtained is Searle (1971) p 407, which is the same as searle
casella p130
\item important: for the matter of support of adjusted MLE, cf Lee 2007
particularly p348. in his MC Lee reports both results restricting to -1,1 and
results without the restriction - would be nice to compare adj MLE to
conditional MLE in Lee 2007, and to understand exact relation to the estimator
in Lee Liu Lin 2010
\item durban currie: Bias and optimistic standard errors are two common
problems associated with the profile likelihood
\item maybe compute coverage conf intervals based on signed LR for adj lik and
compare them with those for prof lik (see SARTORI, R. BELLIO, A. SALVAN pace
1999 for this, and brazzale davison 2008) signed sq root of generaliz LR:
$R_{\mathrm{a}}(\lambda)=sgn(\hat{\lambda}_{\mathrm{aML}}-\lambda)\left[
2(l_{\mathrm{a}}(\hat{\lambda}_{\mathrm{aML}})-l_{\mathrm{a}}(\lambda
))\right]  ^{\frac{1}{2}}$ (this is a statistic once you fix $\lambda)$. For
background look at Severini Chap 7. model parametrized by scalar $\theta$:
$R(\theta)=sgn(\hat{\theta}-\theta)\left[  2(l(\hat{\theta})-l(\theta
))\right]  ^{\frac{1}{2}}$ Test of $\theta=\theta_{0}$ based on $R(\theta
_{0})$, conf region for $\theta$ is $\{\theta\in\Theta:R(\theta)\leq k\}$
(interval if $l(\lambda$) is unimodal) model parametrized by $\theta=\left(
\lambda,\eta\right)  $, $\lambda$ scalar. generalized LR is $R(\lambda
)=sgn(\hat{\lambda}_{\mathrm{ML}}-\lambda)\left[  2(l(\hat{\theta}%
)-l(\hat{\theta}_{\lambda}))\right]  ^{\frac{1}{2}}=sgn(\hat{\lambda
}_{\mathrm{ML}}-\lambda)\left[  2(l_{p}(\hat{\lambda}_{\mathrm{ML}}%
)-l_{p}(\lambda))\right]  ^{\frac{1}{2}}$ ($\hat{\theta}_{\lambda}%
=(\lambda,\hat{\eta}_{\lambda})$) - two approaches to improving $R(\lambda)$:
1) barndorff nielsen r$^{\ast}$ or use a modified prof lik
\item remember that when pivoting the cdf we need cdf to be monotonic in
$\theta$ see e.g. casella berger
\item $R_{\mathrm{a}}(\lambda)$ closer to being normal than $R(\lambda
)=sgn(\hat{\lambda}_{\mathrm{ML}}-\lambda)\left[  2(l_{\mathrm{a}}%
(\hat{\lambda}_{\mathrm{ML}})-l_{\mathrm{a}}(\lambda))\right]  ^{\frac{1}{2}}%
$, so more approximately pivotal. Conf interval $\{\lambda\in:R_{\mathrm{a}%
}(\lambda)\}$
\item one could also do Wald c.i. but they are not invariant - could plot
$l(\lambda)$ and $l_{\mathrm{a}}(\lambda)$ and horizontal lines for the
\item say results for beta are pretty much the same for MLe and adjMLE
\item Following \cite{McCullagh1990}, instead of correcting the QMLE, we
correct the profile likelihood function for $\lambda$.
\end{enumerate}
\end{mycomment}

\begin{mycomment}
CHECK\ Liu Yang Modified QML estimation of spatial autoregressive models with
unknown heteroskedasticity and nonnormality AND\ RELATED (a related correction
e.g. in Yang Shen082014 A Simple and Robust Method of Inference.pdf but for
the LM test not for score)
perhaps repeat design in lin and lee 2010 JoE - see also Jin and Lee RSUE
2012, that is also look at a case when QMLE is not consist. from preliminary
simul it looks like adj MLE is much better than MLE in the presence of
(unaccounted heterosk) in some cases (not with W = k\_ahead\_behind(n,5);
\bigskip(adj lik does strictly speaking solve the support problem, but in
those cases when supp of QMLE is restricted the adj QMLE has very large
variability) adj lik may be consist even when k$\rightarrow\infty$ as
n$\rightarrow\infty$.............. actually in this case OLS should do very
well...... looks like OLS does well when k is large (recall lee's result that
OLS can be consistent when there are regressors but not in a pure model) SAY
WHY THIS APPROACH iS APPEALING AND MAYBE ALSO EXPLAIN THAT BIAS IS NOT REALLY
MUCH OF A PROBLEM IN THESE MODELS IN GENERAL (BECAUSE IT IS EASY TO CORRECT
FOR BIAS without compromising the variance, GIVEN THE SHAPE OF THE BIAS
FUNCTION)
\bigskip....In short, when Assumption \ref{assum id} does not hold,
the QMLE of $\lambda$ either does not exist or is non-random. Also, the
adjusted likelihood function proposed later in the paper is flat if Assumption
\ref{assum id} fails, which makes the identifiability role of Assumption
\ref{assum id} very transparent.
\end{mycomment}

\section{\label{sec SLM}Preliminaries}

\subsection{The SAR model}

We consider the spatial autoregressive (SAR) model%
\begin{equation}
y=\lambda Wy+X\beta+\sigma\varepsilon, \label{SLM}%
\end{equation}
where $y$ is the $n\times1$ vector of observed random variables, $\lambda$ is
a scalar parameter, $W$ is a spatial weights matrix, $X$ is an $n\times k$
matrix of regressors with full column rank and with $k\leq n-2$, $\beta
\in\mathbb{R}^{k}$, $\sigma$ is a positive scale parameter, and $\varepsilon$
is a zero mean $n\times1$ random vector. For simplicity, we take $X$ and $W$
to be non-stochastic, and we assume that $W$ is completely known.
Alternatively, we could allow $X$ and $W$ to be random, and interpret the
analysis as conditional on them, provided that they are independent of
$\varepsilon$. Some of the columns of $X$ may be spatial lags of some other
columns, to allow for the estimation of, in the terminology of social network
analysis, contextual effects. When there is no $X$, the model is called a pure
SAR model. We allow for equation (\ref{SLM}) to also represent a spatial panel
data model, or a model in which individuals are divided in several networks.
In both cases, $W$ is a block diagonal matrix, with the number of blocks being
given by, respectively, the number of time points and the number of networks.
Additive fixed effects along one or both of the panel dimensions, or network
fixed effects, can be added. For the purpose of estimation, fixed effects are
treated as parameters, and hence can be included in $\beta$. Throughout the
paper we assume that $W$ has at least one (real) negative eigenvalue and at
least one (real) positive eigenvalue. This assumption is virtually always
satisfied in applications, especially because the diagonal entries of $W$ are
usually set to zero. The smallest real eigenvalue of $W$ is denoted by
$\omega_{\min}$, and the largest real eigenvalue is normalized, without loss
of generality, to 1.

On rewriting equation (\ref{SLM}) as $S(\lambda)y=X\beta+\sigma\varepsilon$,
where $S(\lambda)\coloneqq I_{n}-\lambda W$, it is clear that in order for $y$ to be
uniquely determined, given $X$\ and $\varepsilon$, it is necessary that
$S(\lambda)$ is nonsingular. This requires $\lambda\neq\omega^{-1}$, for any
nonzero real eigenvalue $\omega$ of $W$ (nonreal complex eigenvalues of $W$ do
not need to be considered here, because $\lambda$ is assumed to be real, and
$\omega^{-1}$ is real if and only $\omega$ is). In both applications and
theoretical studies, the parameter space for $\lambda$ is usually restricted
much further, namely to the largest interval containing the origin in which
$S(\lambda)$ is nonsingular, that is,
\[
\Lambda\coloneqq (\omega_{\min}^{-1},1),
\]
or a subset thereof (possibly independent of $n$) such as $(-1,1)$. Without
such restrictions the models are believed to be too erratic to be useful in
practice, and $\lambda$ is difficult to interpret.\footnote{From a large
sample perspective, the restriction to $(-1,1)$ is often imposed, along with a
uniform boundeness condition on $W$, to guarantee that the variances of the
$y_{i}$'s do not explode as $n$ grows. Also, note that $\Lambda$ and the
entries of $W$ are allowed to depend on $n$, although this is not emphasized
in our notation.}

The following assumption is required to rule out some pathological
combinations of $W$ and $X$.

\begin{assumption}
\label{assum id}There is no real eigenvalue $\omega$ of $W$ for which
$M_{X}(\omega I_{n}-W)=0.$
\end{assumption}

\begin{mycomment}
	PREVIOUS VERSION OF THE ASSUMPTION: For any real eigenvalue $\omega$ of $W$,
	$\operatorname{col}(\omega I_{n}-W)\nsubseteq\operatorname{col}(X)$.
\end{mycomment}

To provide some intuition for Assumption \ref{assum id}, we note that the
condition $M_{X}(\omega I_{n}-W)=0$ is equivalent to
$\mathrm{\operatorname{col}}(\omega I_{n}-W)\subseteq\operatorname{col}(X)$,
and we distinguish two cases. First, if Assumption \ref{assum id} is violated
for the eigenvalue $\omega=0$ (i.e., $\mathrm{\operatorname{col}}%
(W)\subseteq\operatorname{col}(X)$), it is evident from equation (\ref{SLM})
that estimating $\beta$ and $\lambda$ separately must be problematic. Second,
if Assumption \ref{assum id} is violated for some eigenvalue $\omega\neq0$,
then for any $y\in\mathbb{R}^{n}$ it is possible to find a $\beta\in
\mathbb{R}^{k}$ such that $S(\omega^{-1})y=X\beta$, meaning that a SAR model
with $\lambda=\omega^{-1}$ can provide perfect fit for any $y$. It is clear
that in this case any sensible inferential procedure should suggest
that\textit{ }$\lambda=\omega^{-1}$ and $\sigma=0$, for any $y$, and whatever
the true values of $\lambda$ and $\sigma$ are.\footnote{For the specific case
of the QMLE, see part (i) of Lemma \ref{lemma l(lambda) violation} in Section
\ref{sec suppl Assum1} of the Supplement. Section \ref{sec suppl Assum1} of
the Supplement also contains further technical remarks about Assumption
\ref{assum id}.}

The following example provides a simple illustration of Assumption
\ref{assum id}.

\begin{example2}
[Group interaction]\label{ex BGI}There are $R\geq1$ groups of individuals.
Individuals interact uniformly within their group, and do not interact across
different groups. If each group has the same size, say $m>1$, this type of
interaction can be represented by the block-diagonal weights matrix
\begin{equation}
W=I_{R}\otimes B_{m}, \label{W BGI}%
\end{equation}
where $B_{m}\coloneqq \frac{1}{m-1}\left(  \iota_{m}\iota_{m}^{\prime}-I_{m}\right)
$, with $\iota_{m}$ an $m\times1$ vector of all ones. See \cite{Lee2007b} and
\cite{HillierMartellosio2016} for theoretical studies of this model, and
\cite{Carrell2013} and \cite{boucher2014} for recent applications. For matrix
(\ref{W BGI}), one can easily verify that $\omega_{\min}=-\frac{1}{m-1}$ and
$\operatorname{col}(\omega_{\min}I_{n}-W)=\operatorname{col}(I_{R}\otimes
\iota_{m}).$ Noting that $I_{R}\otimes\iota_{m}$ is the design matrix of the
group fixed effects, it follows that, in a SAR model with weights matrix
(\ref{W BGI}), Assumption \ref{assum id} is violated (for $\omega=\omega
_{\min}$) whenever group fixed effects are included in the model (recall that
the fixed effects are treated as parameters to be estimated and hence are
included in $\beta$). It should be noted that the presence of group intercepts
does not cause a violation of Assumption \ref{assum id} when the model is
unbalanced, that is, not all groups have the same size.
\end{example2}

Indeed, it is well known that the model of Example \ref{ex BGI} suffers from
an identifiability problem. Specifically, \cite{Lee2007b} and
\cite{Bramoulle2009} show that the parameters of a SAR model with weights
matrix (\ref{W BGI}), group fixed effects, and contextual effects are not
identifiable after removal of the group fixed effects by a within
transformation. It is easily verified that, in this model, the identifiability
problem occurs even without contextual effects.

\begin{mycomment}
To see $\operatorname{col}(\omega_{\min}I_{n}-W)=\operatorname{col}%
(I_{R}\otimes\iota_{m})$:
EITHER: using $B=\left(  m-1\right)  ^{-1}\left(  \iota_{m}\iota_{m}^{\prime
}-I_{m}\right)  $, you get
\begin{align*}
\omega_{\min}I_{n}-W  & =-\left(  m-1\right)  ^{-1}(I_{R}\otimes I_{m}%
+I_{R}\otimes\left(  \iota_{m}\iota_{m}^{\prime}-I_{m}\right)  )\\
& =-\left(  m-1\right)  ^{-1}(I_{R}\otimes\iota_{m})(I_{R}\otimes\iota
_{m})^{\prime}%
\end{align*}
OR: first observe that matrix (\ref{W BGI}) has two eigenspaces:
$\operatorname{col}(I_{R}\otimes\iota_{m})$, associated to the eigenvalue $1$,
and $\operatorname{col}^{\perp}(I_{R}\otimes\iota_{m})$, associated to the
eigenvalue $\omega_{\min}=-1/(m-1)$. Since $\operatorname{col}(\omega
_{\min}I_{n}-W)=\mathrm{null}^{\perp}(W-\omega_{\min}I_{n})$, it follows that
$\operatorname{col}(\omega_{\min}I_{n}-W)=\mathrm{null}(W-I_{n}%
)=\operatorname{col}(I_{R}\otimes\iota_{m})$.
The violation of Assumption \ref{assum id} for the model in Example
\ref{ex BGI} is a reflection of the identifiability failure discussed in
\cite{Lee2007b} ($\lambda$ is not identifiable from the within equation of the
model if there is no variation in the group sizes). In fact, Assumption
\ref{assum id} rules out a (non-standard) identification issue that may affect
any SAR model, not only the model with weights matrix (\ref{W BGI}). We defer
a detailed discussion of this identification issue to Section
\ref{sec identif}. Finally, we note that $\mathrm{\operatorname{col}}(\omega
I_{n}-W)\subseteq\operatorname{col}(X)$ can be checked via the equivalent
condition $M_{X}(\omega I_{n}-W)=0$, where $M_{X}\coloneqq I_{n}-X(X^{\prime}%
X)^{-1}X^{\prime}.$
\end{mycomment}

\begin{mycomment}
	Other conditions that are equivalent to
	$\mathrm{\operatorname{col}}(\omega I_{n}-W)\subseteq\operatorname{col}(X)$
	are collected in Lemma \ref{lemma remarks} of the Supplement
\end{mycomment}

\subsection{\label{sec QMLE}The QMLE}

We now define the QMLE of the parameters in model (\ref{SLM}). By
quasi-likelihood we mean the likelihood that would prevail under the condition
$\varepsilon\sim\mathrm{N}(0,I_{n})$. Omitting additive constants, the
quasi-log-likelihood is
\begin{equation}
l(\beta,\sigma^{2},\lambda)\coloneqq -\frac{n}{2}\log(\sigma^{2})+\log\left\vert
\det\left(  S(\lambda)\right)  \right\vert -\frac{1}{2\sigma^{2}}%
(S(\lambda)y-X\beta)^{\prime}(S(\lambda)y-X\beta), \label{loglik}%
\end{equation}
for any $\lambda$ such that $S(\lambda)$ is nonsingular. To avoid tedious
repetitions, in the remainder of the paper we will often omit the
\textquotedblleft quasi-\textquotedblright\ in front of \textquotedblleft
log-likelihood\textquotedblright.

The QMLE in most common use is the maximizer of $l(\beta,\sigma^{2},\lambda)$
under the condition that $\lambda$ is in $\Lambda$ (or in a subset thereof).
That is, the QMLE is
\[
(\hat{\beta}_{\mathrm{ML}},\hat{\sigma}_{\mathrm{ML}}^{2},\hat{\lambda
}_{\mathrm{ML}})=\underset{\beta\in\mathbb{R}^{k},\hspace{0.1667em}\sigma
^{2}>0,\hspace{0.1667em}\lambda\in\Lambda}{\argmax}l(\beta,\sigma^{2}%
,\lambda).
\]
Maximization with respect to $\beta$ and $\sigma^{2}$ gives $\hat{\beta
}_{\mathrm{ML}}(\lambda)\coloneqq (X^{\prime}X)^{-1}X^{\prime}S(\lambda)y$ and
$\hat{\sigma}_{\mathrm{ML}}^{2}(\lambda)\coloneqq \frac{1}{n}y^{\prime}S^{\prime
}(\lambda)M_{X}S(\lambda)y$. The corresponding profile, or concentrated,
log-likelihood for $\lambda$ is, again omitting additive constants,
\begin{equation}
l(\lambda)\coloneqq l(\hat{\beta}_{\mathrm{ML}}(\lambda),\hat{\sigma}_{\mathrm{ML}%
}^{2}(\lambda),\lambda)=-\frac{n}{2}\log\left(  \hat{\sigma}_{\mathrm{ML}}%
^{2}(\lambda)\right)  +\log\left\vert \det\left(  S(\lambda)\right)
\right\vert . \label{prof lik}%
\end{equation}
The QMLE of $\lambda$ can be equivalently defined as%
\begin{equation}
\hat{\lambda}_{\mathrm{ML}}\coloneqq \argmax_{\lambda\in\Lambda}l(\lambda).
\label{MLE}%
\end{equation}
The function $l(\lambda)$ is a.s.\ well defined (see Section \ref{app smooth}
in the Supplement), and, clearly, is continuously differentiable whenever it
is well defined. Also, it is easy to see that, under Assumption \ref{assum id}%
, $l(\lambda)$ a.s.\ goes to $-\infty$ at each real zero of $\det(S(\lambda))$
\citep[cf.][]{Hillier2017}. Thus, $l(\lambda)$ has a.s.\ at least one critical
point corresponding to a maximum in any interval between two consecutive real
zeros of $\det\left(  S(\lambda)\right)  $. We also define the unrestricted
QMLE $\hat{\lambda}_{\mathrm{uML}}\coloneqq \argmax_{\lambda\in\Lambda_{u}}l(\lambda
)$, where $\Lambda_{u}\coloneqq \{\lambda\in\mathbb{R}:\det(S(\lambda))\neq0\}$. The
estimators $\hat{\lambda}_{\mathrm{ML}}$ and $\hat{\lambda}_{\mathrm{uML}}$
are different, because the global maximum of $l(\lambda)$ over $\Lambda_{u}$
is not necessarily in $\Lambda$, even when the true value of $\lambda$ is in
$\Lambda$.

\begin{mycomment}
	It is also worth noting that if
	maximization were restricted to a proper subset of $\Lambda$, then a censored
	version of $\hat{\lambda}_{\mathrm{ML}}$ would be obtained.
\end{mycomment}

\section{\label{sec SLM adj prof lik}The Adjustment}

\subsection{\label{subsec SLM adj prof lik}The adjusted profile likelihood}

A profile likelihood for a parameter of interest does not take into account
the sampling variability associated to the estimation of the nuisance
parameters, and hence, as mentioned in the introduction, is generally not a
genuine likelihood. As a consequence, a profile score estimating equation is
generally biased, which is likely to induce bias in the QMLE of the parameter
of interest. The adjusted QMLE solves the estimating equation obtained by
recentering the profile score. We now apply this general adjustment to the
estimation of $(\sigma^{2},\lambda)$ in the SAR model. The score for
$(\sigma^{2},\lambda)$ is centered assuming only that $\mathrm{E}%
(\varepsilon)=0$ and $\mathrm{var}(\varepsilon)=I_{n}$. Note that we treat
only $\beta$ as the nuisance parameter, not $(\beta,\sigma^{2})$, because an
adjusted estimator of $\sigma^{2}$ is required for estimation of $\lambda
$.\footnote{Treating $\sigma^{2}$ as a nuisance parameter too, and
consequently recentering the score for $\lambda$ only (i.e., the score
associated to the log-likelihood (\ref{prof lik})) would not produce an
adjusted estimator for $\sigma^{2}$. Also, the (exact) recentering of
$s(\lambda)$ would require stronger assumptions than what required for the
recentering of $s(\sigma^{2},\lambda)$; see Section \ref{sec suppl s(lambda)}
of the Supplement.}

On concentrating just $\beta$ out of the Gaussian log-likelihood
(\ref{loglik}), the profile log-likelihood for $(\sigma^{2},\lambda)$ is%
\begin{equation}
l(\sigma^{2},\lambda)\coloneqq l(\hat{\beta}_{\mathrm{ML}}(\lambda),\sigma^{2}%
,\lambda)=-\frac{n}{2}\log(\sigma^{2})+\log\left\vert \det\left(
S(\lambda)\right)  \right\vert -\frac{1}{2\sigma^{2}}y^{\prime}S^{\prime
}(\lambda)M_{X}S(\lambda)y, \label{lik sig lam SAR}%
\end{equation}
with profile score
\begin{equation}
s(\sigma^{2},\lambda)\coloneqq \left[
\begin{array}
[c]{c}%
-\frac{n}{2\sigma^{2}}+\frac{1}{2\sigma^{4}}y^{\prime}S^{\prime}(\lambda
)M_{X}S(\lambda)y\\
\frac{1}{\sigma^{2}}y^{\prime}W^{\prime}M_{X}S(\lambda)y-\mathrm{tr}%
(G(\lambda))
\end{array}
\right]  , \label{s sig lam}%
\end{equation}
where $G(\lambda)\coloneqq WS^{-1}(\lambda)$. We now compute the expectation of
$s(\sigma^{2},\lambda)$ under the SAR model $y=\lambda Wy+X\beta
+\sigma\varepsilon$. Assuming that $\mathrm{E}(\varepsilon)=0$ and
$\mathrm{var}(\varepsilon)=I_{n}$, we have\footnote{If the matrices $X$ or $W$
were stochastic, but independent of $\varepsilon$, we would be conditioning on
them here.}%
\begin{equation}
\mathrm{E}(s(\sigma^{2},\lambda))=\left[
\begin{array}
[c]{c}%
-\frac{n}{2\sigma^{2}}+\frac{n-k}{2\sigma^{2}}\\
\mathrm{tr}(M_{X}G(\lambda))-\mathrm{tr}(G(\lambda))
\end{array}
\right]  . \label{Es}%
\end{equation}

\begin{mycomment}
	$\mathrm{E(}\tilde{y}^{\prime}S^{-1}(\lambda)W^{\prime}M_{X}%
	\tilde{y})=\frac{1}{2}\mathrm{E(}\tilde{y}^{\prime}\left(  G(\lambda)%
	M_{X}+M_{X}G(\lambda)^{\prime}\right)  \tilde{y})=\sigma^{2}\mathrm{tr}%
	(M_{X}G(\lambda))$ and $\mathrm{E(}\tilde{y}^{\prime}M_{X}\tilde{y}%
	)=\sigma^{2}\mathrm{tr}(M_{X})$
	.............$\mathrm{E}$ denotes expectation
	w.r.t. any distribution of $y$ ........
\end{mycomment}

For a pure model, $\mathrm{E}(s(\sigma^{2},\lambda))=0$ for any $\lambda$ such
that $S(\lambda)$ is nonsingular, meaning that the estimating equation
$s(\sigma^{2},\lambda)=0$ is unbiased.\footnote{Depending on the sample size
and on the structure of $W$, $\hat{\lambda}_{\mathrm{ML}}$ can be considerably
biased even in the pure case, despite the fact that the estimating equation
$s(\sigma^{2},\lambda)=0$ is unbiased in that case. The adjustment studied in
this paper is not designed to reduce this type of bias in $\hat{\lambda
}_{\mathrm{ML}}$. Instead, it aims at reducing the bias due to the nuisance
parameter $\beta$.} When regressors are present, however, the unaccounted
variability in the estimation of $\beta$ causes the estimating equation
$s(\sigma^{2},\lambda)=0$ to be biased. Note that the expectation (\ref{Es})
does not depend on $\beta$, so recentering the score is straightforward. The
adjusted profile score for $(\sigma^{2},\lambda)$, defined as $s_{\mathrm{a}%
}(\sigma^{2},\lambda)\coloneqq s(\sigma^{2},\lambda)-\mathrm{E}(s(\sigma^{2}%
,\lambda))$, is
\begin{equation}
s_{\mathrm{a}}(\sigma^{2},\lambda)=\left[
\begin{array}
[c]{c}%
s_{\mathrm{a}1}(\sigma^{2},\lambda)\\
s_{\mathrm{a}2}(\sigma^{2},\lambda)
\end{array}
\right]  =\left[
\begin{array}
[c]{c}%
-\frac{n-k}{2\sigma^{2}}+\frac{1}{2\sigma^{4}}y^{\prime}S^{\prime}%
(\lambda)M_{X}S(\lambda)y\\
\frac{1}{\sigma^{2}}y^{\prime}W^{\prime}M_{X}S(\lambda)y-\mathrm{tr}%
(M_{X}G(\lambda))
\end{array}
\right]  . \label{sa sig lam}%
\end{equation}

Setting $s_{\mathrm{a}1}(\sigma^{2},\lambda)=0$ gives $\hat{\sigma
}_{\mathrm{aML}}^{2}(\lambda)\coloneqq \frac{n}{n-k}\hat{\sigma}_{\mathrm{ML}}%
^{2}(\lambda)$. That is, recentering the score automatically delivers the
usual degrees of freedom correction for $\hat{\sigma}_{\mathrm{ML}}%
^{2}(\lambda)$. The adjusted QMLE for $\lambda$ must then be a zero of%
\begin{equation}
s_{\mathrm{a}2}(\lambda)\coloneqq s_{\mathrm{a}2}(\hat{\sigma}_{\mathrm{aML}}%
^{2}(\lambda),\lambda)=(n-k)\frac{y^{\prime}W^{\prime}M_{X}S(\lambda
)y}{y^{\prime}S^{\prime}(\lambda)M_{X}S(\lambda)y}-\mathrm{tr}(M_{X}%
G(\lambda)), \label{sa2}%
\end{equation}
or, which is a.s.\ the same, must solve the estimating equation%
\begin{equation}
y^{\prime}S^{\prime}(\lambda)R(\lambda)S(\lambda)y=0, \label{unbesteq}%
\end{equation}
where
\[
R(\lambda)\coloneqq M_{X}\left(  G(\lambda)-\frac{\mathrm{tr}(M_{X}G(\lambda))}%
{n-k}I_{n}\right)  .
\]

We emphasize that, by construction, the estimating equation (\ref{unbesteq})
is exactly unbiased provided that $\mathrm{E}(\varepsilon)=0$ and
$\mathrm{var}(\varepsilon)=I_{n}$ (no further distributional assumptions are required).

Given the adjusted profile score $s_{\mathrm{a}}(\sigma^{2},\lambda)$, one can
define the function with gradient equal to $s_{\mathrm{a}}(\sigma^{2}%
,\lambda)$, which we refer to as the adjusted likelihood for $(\sigma
^{2},\lambda)$, denoted by $l_{\mathrm{a}}(\sigma^{2},\lambda)$. Letting
$\operatorname{Re}\left[  \cdot\right]  $ denote the real part of a complex
number, the following result gives a closed form expression for $l_{\mathrm{a}%
}(\sigma^{2},\lambda)$.

\begin{mycomment}
	Such an expression can be helpful for graphical or optimization purposes, but
	is not required for the results in this paper.
\end{mycomment}

\begin{proposition}
\label{prop l_a SLM}In a SAR model, the adjusted log-likelihood for
$(\sigma^{2},\lambda)$, up to an additive constant, is
\begin{equation}
l_{\mathrm{a}}(\sigma^{2},\lambda)=-\frac{n-k}{2}\log(\sigma^{2})-\frac
{1}{2\sigma^{2}}y^{\prime}S^{\prime}(\lambda)M_{X}S(\lambda
)y+\operatorname{Re}\left[  \mathrm{tr}\left(  M_{X}\log S(\lambda)\right)
\right]  , \label{l_a(sig,lambda)}%
\end{equation}
for any $\lambda$ such that $S(\lambda)$ is nonsingular, and for a suitable
choice (made in the proof) of the branch of the matrix logarithm $\log
S(\lambda)$.
\end{proposition}

\begin{mycomment}
	see comparison\_adjusted\_lik\_using\_matrixlog.m and Proposition
	\ref{prop adj lik} for the case W diagonaliz and all eigenval are real
\end{mycomment}

From expression (\ref{l_a(sig,lambda)}), we immediately obtain the adjusted
likelihood for $\lambda$ only,%
\begin{equation}
l_{\mathrm{a}}(\lambda)\coloneqq l_{\mathrm{a}}(\hat{\sigma}_{\mathrm{aML}}%
^{2}(\lambda),\lambda)=-\frac{n-k}{2}\log\left(  y^{\prime}S^{\prime}%
(\lambda)M_{X}S(\lambda)y\right)  +\operatorname{Re}\left[  \mathrm{tr}\left(
M_{X}\log S(\lambda)\right)  \right]  \label{l a lam}%
\end{equation}
(note that this is the likelihood with score $s_{\mathrm{a}2}(\lambda)$).
Equations (\ref{l_a(sig,lambda)}) and (\ref{l a lam}) may be useful for
graphical or optimization purposes, but are not used for the analytical
results that follow in the paper. Indeed, even the results that (for
simplicity and to aid intuition) are stated in terms of the adjusted
likelihood, are proved using only the expression for the adjusted score, and
could equally be formulated in terms of the adjusted score.

\begin{mycomment}
this is consistent with the fact that $s_{\mathrm{a}2}(\hat{\sigma
}_{\mathrm{aML}}^{2}(\lambda),\lambda)=\frac{n-k}{n}s_{\mathrm{a}}(\lambda)$,
(see above) and one of mycomment above. also see
check\_transformation\_approach\_panel\_indiv\_fixed\_effects.m,
spatial\_panel\_two\_way\_fixed\_effects.m, check\_transformation\_approach\_network\_fixed\_effects\_unbalanced.
\end{mycomment}

\begin{mycomment}
recall that in the multiparam case, the adj lik is not guaranteed to exist see mccullagh and tibshir
\end{mycomment}

\begin{mycomment}
we use $l_{\mathrm{a}}%
(\lambda)\coloneqq \int s_{\mathrm{a}}(\lambda)\diff\lambda$, where, remember, $\int s_{\mathrm{a}}(\lambda)\diff\lambda$ denotes the family of antiderivatives. Alternatively we could have defined  $l_{\mathrm{a}}%
(\lambda)\coloneqq \int_{\omega_{\min}^{-1}}^{\lambda} s_{\mathrm{a}}(t)dt$, which fixes the constant. BY THE WAY, the latter formulation is obviously more appropriate when the lik needs to be computed numerically
\end{mycomment}

\begin{rem}
Like its unadjusted version $l(\sigma^{2},\lambda)$, the adjusted profile
log-likelihood $l_{\mathrm{a}}(\sigma^{2},\lambda)$ is not, in general, a
genuine likelihood function. Further adjustments could be implemented to make
$l_{\mathrm{a}}(\sigma^{2},\lambda)$ closer to a genuine likelihood. In
particular, one could normalize $s_{\mathrm{a}}(\sigma^{2},\lambda)$ to make
it information unbiased (i.e., variance equal to minus the expectation of the
derivative of the score), as suggested by \cite{McCullagh1990}. Such
adjustments might improve the performance of first-order asymptotic
approximations to estimators or test statistics, but are not considered in
this paper because they do not affect the location of the zeros of
$s_{\mathrm{a}}(\sigma^{2},\lambda)$.
\end{rem}

So far, the adjusted QMLE of $\lambda$ has been introduced as a zero of
$s_{\mathrm{a}2}(\lambda)$, but of course $s_{\mathrm{a}2}(\lambda)$ may have
many (real) zeros. The question therefore arises as to how exactly the
adjusted QMLE should be defined. Recall from Section \ref{sec QMLE} that
$\hat{\lambda}_{\mathrm{ML}}$ is defined as the maximizer of $l(\lambda)$ over
$\Lambda$. One may therefore be tempted to define the adjusted QMLE of
$\lambda$ as the maximizer of $l_{\mathrm{a}}(\lambda)$ over $\Lambda$.
However, we shall see that $s_{\mathrm{a}2}(\lambda)$ may have no zeros in
$\Lambda$ (or, equivalently, the adjusted profile likelihood $l_{\mathrm{a}%
}(\lambda)$ may have no maximum on $\Lambda$), so it makes sense to define the
adjusted QMLE on an interval larger than $\Lambda$, $\Lambda_{\mathrm{a}}$
say. By analogy with the unadjusted case, we shall define $\Lambda
_{\mathrm{a}}$ to be the shortest open interval containing the origin with the
property that $l_{\mathrm{a}}(\lambda)\rightarrow-\infty$ a.s.\ at both
extremes of $\Lambda_{\mathrm{a}}.$ But first, in order to fully understand
the problem, we need to study the behavior of the profile score $s_{\mathrm{a}%
2}(\lambda)$, or, equivalently, of $l_{\mathrm{a}}(\lambda)$, near the zeros
of $\det(S(\lambda))$.

\begin{mycomment}
	 which is also
	stated in Jin Lee 2012 RSUE; also see
	expectation\_profile\_score\_SLM\_is\_bigO1.m
\end{mycomment}

\begin{mycomment}
	Such
	adjustments might improve the performance of first-order asymptotic
	approximations to estimators or test statistics, but this is not relevant for our purposes
	\citep[and it is found not to be very important in
	practice in many cases; see, e.g.,][]{Durban2000}. Ghosh1994
\end{mycomment}

\begin{mycomment}
	A necessary and
	sufficient condition for $l_{\mathrm{a}}(\lambda)$ to be a genuine likelihood is given
	in Remark \ref{rem genuine} below.
\end{mycomment}

\begin{mycomment}
	conjecture: $\mathrm{E}_{\lambda}(s(\lambda))=0$ for all $\lambda$ such that
	$S(\lambda)$ is nonsingular iff $X=0$
	attempts:%
	\begin{equation}
	\mathrm{E}_{\lambda}(s(\lambda))=\frac{n}{n-k}\mathrm{tr}(M_{X}G_{\lambda
	})-\mathrm{tr}(G(\lambda))=\frac{k}{n-k}\mathrm{tr}(G(\lambda)%
	)-\mathrm{tr}(P_{X}G(\lambda))
	\end{equation}
\end{mycomment}

\begin{mycomment}
	$s_{\mathrm{a}}(\sigma,\lambda)=0$ is unbiased under $\mathrm{E}(S(\lambda)y)=X\beta$
	and $\mathrm{var}(S(\lambda)y)=\sigma^{2}I_{n}$, but in order for
	$s_{\mathrm{a}}(\lambda)=0$ to be unbiased we also need normality (spherical symmetry
	actually); see adj\_lik\_adjusting\_the\_score.tex (the problem is that
	without normality the ratio $\frac{y^{\prime}W^{\prime}M_{X}S(\lambda
		)y}{y^{\prime}S^{\prime}(\lambda)M_{X}S(\lambda)y}$ and its denominator may be
	independent - see also griffith, The Moran coefficient for non-normal data). I
	guess expectation $s(\lambda)$ is quite robust to departures from normality
	(as long as $\mathrm{E}(S(\lambda)y)=X\beta$ and $\mathrm{var}(S(\lambda
	)y)=\sigma^{2}I_{n}$).... exact vs approximate centering
	when $\mathrm{var}(S(\lambda)y)\neq\sigma^{2}I_{n}$ $\hat{\lambda
	}_{\mathrm{ML}}$ is inconsist and $\mathrm{E}_{\lambda}(s(\lambda))$ does not
	go to $0$. can I\ compute $p\lim s(\lambda)$
	group fixed eff: based on http://www.cemmap.ac.uk/forms/ml\_jochmans.pdf,
	maybe I\ get $p\lim_{r\rightarrow\infty}s(\lambda)=\frac{n}{n-k}%
	\mathrm{tr}(M_{X}G(\lambda))-\mathrm{tr}(G(\lambda))$ (without normality)
\end{mycomment}

\subsection{\label{sec singular SLM}Behavior of $l_{\mathrm{a}}(\lambda)$ near
a zero of $\det(S(\lambda))$}

Perhaps unexpectedly, the profile score $s(\sigma^{2},\lambda)$ and its
adjusted version $s_{\mathrm{a}}(\sigma^{2},\lambda)$ may have very different
behavior as $\lambda$ approaches a real zero of $\det(S(\lambda))$. Obviously,
different behavior of $s(\sigma^{2},\lambda)$ and $s_{\mathrm{a}}(\sigma
^{2},\lambda)$ implies different behavior of the functions $l(\sigma
^{2},\lambda)$ and $l_{\mathrm{a}}(\sigma^{2},\lambda)$, and hence of their
profile versions $l(\lambda)$ and $l_{\mathrm{a}}(\lambda)$. For simplicity,
we state the results in this section in terms of the univariate functions
$l(\lambda)$ and $l_{\mathrm{a}}(\lambda)$. We shall see that the different
behavior of $l(\lambda)$ and $l_{\mathrm{a}}(\lambda)$ accounts for important
differences in the properties of the associated estimators for $\lambda$.

\begin{mycomment}
	In Appendix \ref{app l(lambda) SLM} we
	show that both functions are a.s. continuous on any open interval between two
	consecutive real zeros of $\det(S(\lambda))$, and therefore a.s. bounded on any
	closed subset of that interval.
\end{mycomment}

To start with, note that the analysis is straightforward for the pure model.
In that case, we trivially have $l_{\mathrm{a}}(\lambda)=l(\lambda)$, and
plugging $\hat{\sigma}_{\mathrm{ML}}^{2}(\lambda)=\frac{1}{n}y^{\prime
}S^{\prime}(\lambda)S(\lambda)y$ in equation (\ref{prof lik}) reveals that
$l(\lambda)$ a.s.\ approaches $-\infty$ near any real zero of $\det
(S(\lambda))$. The presence of regressors complicates the analysis. We shall
confine attention to semisimple eigenvalues of $W$. An eigenvalue is said to
be \textit{semisimple} if its algebraic and geometric multiplicities are equal
\citep[for the matrix theoretic definitions and results used in this section see, for instance,][]{Meyer2000}.
While it simplifies the analysis considerably, the restriction to semisimple
eigenvalues does not imply a great loss of generality. For example, all
eigenvalues of a diagonalizable matrix are semisimple, and any simple
eigenvalue is semisimple (an eigenvalue being \textit{simple} if it has
algebraic multiplicity equal to one). The behavior of $l_{\mathrm{a}}%
(\lambda)$ close to the eigenvalue $\omega=1$ is often particularly important,
and that eigenvalue is in most cases semisimple in applications. For example,
it is semisimple when $W$ is \textit{row stochastic} (even if $W$ is not
diagonalizable, and the algebraic multiplicity of $\omega=1$ is larger than
one), or when $W$ is \textit{irreducible}%
.\footnote{\label{foot row stoch irred}A row stochastic matrix is a square
nonnegative matrix whose row sums are all equal to 1. Irreducibility can be
defined in terms of the graph of a matrix as follows. Let the graph of an
$n\times n$ matrix $A$ be the directed graph on $n$ vertices in which there is
an an edge from vertex $i$ to vertex $j$ if and only $A(i,j)\neq0$. Also, call
a graph \textit{strongly connected} if there is a sequence of directed edges
from any vertex $i$ to any vertex $j$. Then, $A$ is \textit{irreducible} if
and only if the graph of $A$ is strongly connected
\citep[e.g.,][p. 671]{Meyer2000}.}

The simplification afforded by the restriction to semisimple eigenvalues is
that we can express the conditions for $l_{\mathrm{a}}(\lambda)$ to diverge or
be bounded near a real zero of $\det(S(\lambda))$ in terms of a projector onto
an eigenspace of $W$. Without the restriction to semisimple eigenvalues, the
conditions would need to be stated in terms of projections onto generalized
eigenspaces. If the eigenvalue $\omega$ of $W$ is semisimple, then
$\mathrm{null}\left(  W-\omega I_{n}\right)  $ (the eigenspace of $W$
associated to the eigenvalue $\omega$) and $\operatorname{col}\left(  W-\omega
I_{n}\right)  $ are complementary subspaces of $\mathbb{C}^{n}$, and therefore
we can define a (unique) projector, denoted by $Q_{\omega}$, onto
$\mathrm{null}\left(  W-\omega I_{n}\right)  $ along $\operatorname{col}%
\left(  W-\omega I_{n}\right)  $.\footnote{The matrix $Q_{\omega}$ is a
\textquotedblleft spectral projector\textquotedblright\ of $W$; see, for
instance, \cite{Meyer2000}, where explicit general representations for such
projectors can also be found. Particular cases are given in the proof of Lemma
\ref{lemma null} and in the proof of Lemma \ref{lemma left right} in the
Supplement.} Recalling that the zeros of $\det(S(\lambda))$ are the
reciprocals of the nonzero eigenvalues of $W$, we can prove the following result.

\begin{theorem}
\label{lemma lim gen W}Suppose Assumption \ref{assum id} holds. In a SAR
model, for any semisimple nonzero real eigenvalue $\omega$ of $W$,
$\lim_{\lambda\rightarrow\omega^{-1}}l_{\mathrm{a}}(\lambda)$ is a.s.

\begin{enumerate}
\item[(i)] $-\infty$ if $\mathrm{tr}(M_{X}Q_{\omega})>0$;

\item[(ii)] bounded if $\mathrm{tr}(M_{X}Q_{\omega})=0;$

\item[(iii)] $+\infty$ if $\mathrm{tr}(M_{X}Q_{\omega})<0$.
\end{enumerate}
\end{theorem}

We can now compare the behavior of the adjusted profile log-likelihood
$l_{\mathrm{a}}(\lambda)$ near the real zeros of of $\det(S(\lambda))$, as
established by Theorem \ref{lemma lim gen W}, with the behavior of the
unadjusted profile log-likelihood $l(\lambda)$ near those points. Recall from
Section \ref{sec QMLE} that $\lim_{\lambda\rightarrow\omega^{-1}}%
l(\lambda)=-\infty$ a.s., for any nonzero real eigenvalue $\omega$ of $W$
(under Assumption \ref{assum id}). Thus, $l(\lambda)$ and its adjusted version
$l_{\mathrm{a}}(\lambda)$ have the same behavior near a point $\omega^{-1}$,
for a semisimple nonzero real eigenvalue $\omega$, \textit{only} in case (i)
of Theorem \ref{lemma lim gen W}. In case (ii), $l_{\mathrm{a}}(\lambda)$ can
be extended to a function that is a.s.\ continuous at $\lambda=\omega^{-1}$.
This can be achieved by extending the domain of $l_{\mathrm{a}}(\lambda)$ to
include $\omega^{-1}$, and setting $l_{\mathrm{a}}(\omega^{-1})\coloneqq \lim
_{\lambda\rightarrow\omega^{-1}}l_{\mathrm{a}}(\lambda)$. From now on, when we
say that $l_{\mathrm{a}}(\lambda)$ is continuous at $\lambda=\omega^{-1}$ we
implicitly assume that this extension has been performed. In case (iii) of
Theorem \ref{lemma lim gen W}, $l_{\mathrm{a}}(\lambda)$ is unbounded from
above near $\omega^{-1}$. From unreported numerical experiments, it appears
that $\mathrm{tr}(M_{X}Q_{\omega})<0$ is an extremely rare occurrence for
pairs $(W,X)$ which are likely to be encountered in
applications.\footnote{\label{foot neg}For example, for all simulation designs
in Section \ref{sec Monte Carlo}, we find that $l_{\mathrm{a}}(\lambda)$ is
always bounded from above on $\Lambda_{\mathrm{a}}$. A complete understanding
of when $l_{\mathrm{a}}(\lambda)$ may be unbounded from above would be of
interest, but is left for future research. Note that the fact that
$l_{\mathrm{a}}(\lambda)$ can be unbounded from above is not surprising, since
$l(\lambda)$ itself can be unbounded from above (see Lemma
\ref{lemma l(lambda) violation} in the Supplement).} We shall also see shortly
that $\mathrm{tr}(M_{X}Q_{\omega})<0$ cannot occur if $W$ is symmetric.

\begin{mycomment}
	In the same way as $\lambda$ is unidentifiable from $l(\lambda)$ when
	Assumption \ref{assum id} fails, $\lambda$ is unidentifiable from
	$l_{\mathrm{a}}(\lambda)$ in case (iii) of Theorem \ref{lemma lim gen W}.
\end{mycomment}

\begin{mycomment}
	remark for myself\emph{: }$tr(M_{X}Q_{\omega})<0$ for
	$l_{\mathrm{a}}(\lambda)$ is not the same as \emph{ASSUMP C for }$l_{\mathrm{a}}(\lambda)$ in
	the sense that it does not imply that $s_{\mathrm{a}}(\lambda)$ indep of $y$ (see
	plot\_lik\_score\_SLM\_with\_adj\_trMQneg.m). any other connection between
	$tr(M_{X}Q_{\omega})<0$ and ASSUMP C?
	remember that what happens is that when $l$ is unb from above $l_a$ is flat (this requires (ii) not (iii))
\end{mycomment}

Ruling out the pathological case (iii), it is useful to try and understand
which of cases (i) and (ii) in Theorem \ref{lemma lim gen W} is likely to
occur in applications. At first sight, the condition $\mathrm{tr}%
(M_{X}Q_{\omega})=0$ in case (ii) may look very restrictive. Indeed, Lemma
\ref{lemma null set} in Appendix \ref{app add aux} establishes that
$\mathrm{tr}(M_{X}Q_{\omega})\neq0$ for \textit{generic, }in the measure
theoretic sense, full column rank $X$ (and for any fixed $W$). However, $X$
typically contains an intercept (or group intercepts), and this implies that
$\mathrm{tr}(M_{X}Q_{\omega})=0$ occurs, at $\omega=1$, for a very large class
of matrices $W$ used in practice. To see why this is the case, the next lemma
provides a condition for $\mathrm{tr}(M_{X}Q_{\omega})=0$ in terms of the
eigenspace $\mathrm{null}(W-\omega I_{n})$ (the eigenspace of $W$ associated
to the eigenvalue $\omega$).

\begin{mycomment}
	See also Lemma \ref{lemma left right} of Appendix
	\ref{app add aux} which gives a condition for $\mathrm{tr}(M_{X}Q_{\omega})=0$
	in terms of right and left eigenvectors of $W$ associated to $\omega$.
\end{mycomment}

\begin{samepage}
\begin{lemma}
\label{lemma null}\leavevmode
\begin{enumerate}
\item[(i)] For any semisimple eigenvalue $\omega$ of $W$, $\mathrm{tr}
(M_{X}Q_{\omega})=0$ if $\mathrm{null}(W-\omega I_{n})\subseteq
\operatorname{col}(X)$;
\item[(ii)] For any eigenvalue $\omega$ of a symmetric $W$, $\mathrm{tr}
(M_{X}Q_{\omega})=0$ if $\mathrm{null}(W-\omega I_{n})\subseteq
\operatorname{col}(X)$, $\mathrm{tr}(M_{X}Q_{\omega})>0$ otherwise.
\end{enumerate}
\end{lemma}
\end{samepage}

Part (i) of Lemma \ref{lemma null} establishes that $\mathrm{null}(W-\omega
I_{n})\subseteq\operatorname{col}(X)$ is sufficient for case (ii) of Theorem
\ref{lemma lim gen W} to apply, and hence for $l(\lambda)$ and $l_{\mathrm{a}%
}(\lambda)$ to have different behaviour near $\lambda=\omega^{-1}$ (for any
semisimple nonzero real eigenvalue $\omega$ of $W$, and provided that
Assumption \ref{assum id} holds). It turns out that this sufficient condition
is very often satisfied in applications. Two examples are given next, the
second one being a generalization of the first.

\begin{example2}
[Row stochastic and irreducible weights matrix]\label{exa row-stoch irred}In
applications of spatial autoregressions, $W$ is often row stochastic and
irreducible (cf. footnote \ref{foot row stoch irred}), and an intercept is
included in the regression. By the Perron--Frobenius Theorem
\citep[e.g.,][Theorem 8.4.4]{Horn1985}, $\omega=1$ is a simple (and hence
semisimple) eigenvalue of $W$, and the associated eigenspace $\mathrm{null}
(W-I_{n})$ is spanned by a vector of identical entries, and therefore is in
$\operatorname{col}(X)$. It follows, under Assumption \ref{assum id}, that
$l(\lambda)$ a.s.\ approaches $-\infty$ as $\lambda\rightarrow1$, while
$l_{\mathrm{a}}(\lambda)$ is a.s.\ continuous at $\lambda=1$.
\end{example2}

\begin{example2}
[Block diagonal weights matrix]\label{exa block diag}Example
\ref{exa row-stoch irred} generalizes immediately to the case when $W$ is a
block diagonal matrix whose blocks are row stochastic and irreducible
matrices, of, say, size $m_{r}\times m_{r}$. That is, using direct sum
notation, $W=\bigoplus_{r=1}^{R}W_{r}$. This situation arises, for instance,
in a social interaction model on $R$ networks \citep[e.g.,][]{LeeLiuLin2010},
or in a spatial panel model where individuals are followed over time
\citep[in which case $r$ is the time dimension; see, e.g.,][]{LeeYu2010}. When
$W$ has this structure, the eigenspace $\mathrm{null}(W-I_{n})$ is spanned by
the columns of the (network or time) fixed effects matrix $\bigoplus_{r=1}%
^{R}\iota_{m_{r}}$, and therefore is in $\operatorname{col}(X)$ as long as the
regressions contains those fixed effects. In that case, and provided that
Assumption \ref{assum id} holds, $l(\lambda)$ a.s.\ approaches $-\infty$ as
$\lambda\rightarrow1$, while $l_{\mathrm{a}}(\lambda)$ is a.s.\ continuous at
$\lambda=1$.
\end{example2}

According to part (ii) of Lemma \ref{lemma null}, when $W$ is symmetric, the
condition $\mathrm{null}(W-\omega I_{n})\subseteq\operatorname{col}(X)$ is
also necessary for $l_{\mathrm{a}}(\lambda)$ to be a.s.\ continuous at
$\lambda=\omega^{-1}$. Thus, for symmetric $W$, Theorem \ref{lemma lim gen W}
reduces to the simple statement that $\lim_{\lambda\rightarrow\omega^{-1}%
}l_{\mathrm{a}}(\lambda)$ is a.s.\ bounded if $\mathrm{null}(W-\omega
I_{n})\subseteq\operatorname{col}(X)$, $-\infty$ otherwise, for any semisimple
nonzero real eigenvalue $\omega$.

\begin{figure}[ptbh]
\centering
\makebox[\columnwidth]{	\includegraphics[scale=1]{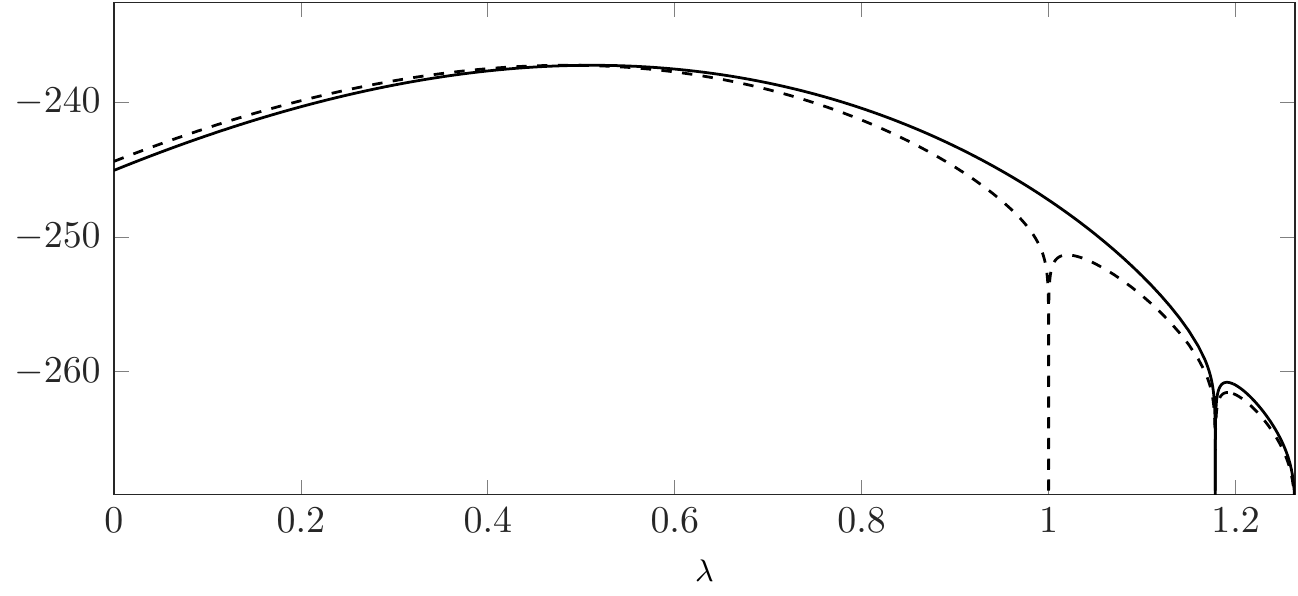} }\caption{The
profile likelihood $l(\lambda)$ (dashed) and its adjusted version
$l_{\mathrm{a}}(\lambda)$ (solid) for a SAR model on a connected $G(100,0.05)$
graph.}%
\label{fig Erdos}%
\end{figure}

We end this section by providing a graphical comparison of $l(\lambda)$ and
$l_{\mathrm{a}}(\lambda)$. Consider a SAR model with weights matrix $W$ equal
to the row normalized adjacency matrix of an Erd{\H{o}}s-R{\'{e}}nyi $G(n,p)$
graph \citep{Erdos1959}. The $G(n,p)$ graph is a random graph on $n$ vertices,
with an edge between any two vertices being present with probability $p$,
independently of every other edge. Suppose that the regression contains an
intercept, and that, for simplicity, the graph is connected. Then, since $W$
is row-stochastic and irreducible, $l_{\mathrm{a}}(\lambda)$ is
a.s.\ continuous at $\lambda=1$ and a.s.\ approaches $-\infty$ as $\lambda$
approaches any other singularity of $S(\lambda)$. Figure \ref{fig Erdos}
displays $l(\lambda)$ and $l_{\mathrm{a}}(\lambda)$, for one random draw of
$G(n,p)$ and for one random draw from the intercept-only model $y=.5Wy+\iota
_{n}+\varepsilon$, with $\varepsilon\sim\mathrm{N}(0,I_{n})$. The
log-likelihood functions are plotted for $\lambda\in(0,\omega_{3}^{-1})$,
where $\omega_{3}$ is the third largest eigenvalue of $W$, and $l_{\mathrm{a}%
}(\lambda)$ has been lowered so that the two likelihoods have the same maximum
value.\footnote{For the particular random draw of $\varepsilon$ underlying
Figure \ref{fig Erdos}, $\hat{\lambda}_{\mathrm{ML}}$ and its adjusted version
$\hat{\lambda}_{\mathrm{a}\mathrm{ML}}$ (defined as in Section
\ref{sec adjMLE} below) are about 0.478 and 0.506, respectively. Over $10^{6}$
draws from $\varepsilon\sim\mathrm{N}(0,I_{n})$ (and for the same draw of
$G(n,p)$ used for Figure \ref{fig Erdos}), empirical bias and RMSE are
$-0.039$ and $0.128$ for $\hat{\lambda}_{\mathrm{ML}}$ and $-0.007$ and
$0.126$ for $\hat{\lambda}_{\mathrm{a}\mathrm{ML}}$.} Note that, contrary to
$l(\lambda)$, $l_{\mathrm{a}}(\lambda)$ does not go to $-\infty$ as
$\lambda\rightarrow1$.

\begin{mycomment}
lik_v_adj_lik2_v2.pdf is the version of lik_v_adj_lik2.pdf with raised adj lik as requested by the referee	
lik_v_adj_lik3.tikz is the same with $l_{\mathrm{a}}(\hat{\sigma}_{\mathrm{aML}}^{2}(\lambda),\lambda)=\frac
{n-k}{n}l_{\mathrm{a}}(\lambda)$, resulting plot is virtually undistingishable
	\end{mycomment}

\begin{mycomment}
	remark (this is established in Lemma \ref{lemma left right}): if $W$ is nonnegative and irreducible, then $\mathrm{tr}(M_{X}%
	Q_{1})=0$ if $\mathrm{null}(W-I_{n})\subseteq\operatorname{col}(X)$ or
	$\mathrm{null}(W^{\prime}-I_{n})\subseteq\operatorname{col}(X)$
	in that case $Q_{1}=pq^{\prime}/q^{\prime}p$ with $p$ and $q$ (the
	perroneigenv for W and W') both positive (Meyer p 677) - positivity actually
	doesn't help, I\ can still get $\mathrm{tr}(M_{X}Q_{1})<0$!\ (see
	diag\_entries\_HMH.m and then plotlikaftercheck\_unimodality.m)
	$\mathrm{tr}(M_{X}Q_{1})=q^{\prime}M_{X}p/q^{\prime}p$
	more generally, if $\omega$ simple (algebraic multiplicity equal to one) then
	$Q_{\omega}=q_{\omega}^{\prime}p_{\omega}/q_{\omega}^{\prime}p_{\omega}$
	(meyer p518) so $\mathrm{tr}(M_{X}Q_{\omega})=\mathrm{tr}(q_{\omega}^{\prime
	}M_{X}p_{\omega})/q_{\omega}^{\prime}p_{\omega}=q_{\omega}^{\prime}%
	M_{X}p_{\omega}/q_{\omega}^{\prime}p_{\omega}$, so $\mathrm{tr}(M_{X}%
	Q_{\omega})$ if $\mathrm{null}(W-\omega I_{n})\subseteq\operatorname{col}(X)$
	or $\mathrm{null}(W^{\prime}-\omega I_{n})\subseteq\operatorname{col}(X)$ (but
	there can also be other cases)
\end{mycomment}

\begin{mycomment}
	looks like when $\mathrm{tr}(M_{X}Q_{\omega})\ $is small but not zero
	$l_{\mathrm{a}}(\lambda)$ is "almost continuous" at $\lambda=\omega^{-1}$ in the sense
	that $l_{\mathrm{a}}(\lambda)$ goes down to -$\infty$ very quickly. apart from a small
	neighborhood of $\omega^{-1}$ $l_{\mathrm{a}}(\lambda)$ looks continuous. this will
	cause an accumulation of prob mass for $\hat{\lambda}_{\mathrm{aML}}$ around
	$\omega^{-1}$ (on the side of $\omega^{-1}$ that is in $\Lambda_{\mathrm{a}}$)
\end{mycomment}

\begin{mycomment}
	$\omega$ is simple, $Q_{\omega}=h_{\omega}l_{\omega}^{\prime
	}/l_{\omega}^{\prime}h_{\omega}$ so $\mathrm{tr}(M_{X}Q_{\omega})=\frac
	{1}{l_{\omega}^{\prime}h_{\omega}}\mathrm{tr}(l_{\omega}^{\prime}%
	M_{X}h_{\omega})=\frac{1}{l_{\omega}^{\prime}h_{\omega}}l_{\omega}^{\prime
	}M_{X}h_{\omega}...........$so $\mathrm{tr}(M_{X}Q_{\omega})<0$ when angle
	between $M_{X}l_{\omega}$ and $M_{X}h_{\omega}$ is obtuse ($\mathrm{tr}%
	(M_{X}Q_{\omega})=0$ iff $M_{X}l_{\omega}$ are orthogonal $M_{X}h_{\omega}$,
	which includes the case one of the two is zero)
	RECALL\ $l_{\omega}^{\prime}h_{\omega}$ can never be zerol $l_{\omega}%
	^{\prime}h_{\omega}$ close to zero means $\omega$ is ill conditioned.
	$\mathrm{tr}(M_{X}Q_{\omega})=0$ if $l_{\omega}$ $h_{\omega}$ become
	orthogonal after projection onto col$^{\perp}$(X),
\end{mycomment}

\subsection{\label{sec adjMLE}The adjusted QMLE}

Theorem \ref{lemma lim gen W} establishes that the adjusted profile
log-likelihood $l_{\mathrm{a}}(\lambda)$ may, in contrast to $l(\lambda)$, be
a.s.\ continuous at the extremes of the parameter space $\Lambda$. As a
consequence, there is no guarantee that $l_{\mathrm{a}}(\lambda)$ has a
maximum over $\Lambda$, which suggests that $l_{\mathrm{a}}(\lambda)$ should
be maximized over a larger set, $\Lambda_{\mathrm{a}}$ say. As anticipated in
Section \ref{subsec SLM adj prof lik}, it is natural to define $\Lambda
_{\mathrm{a}}$ as the shortest open interval containing the origin with the
property that $l_{\mathrm{a}}(\lambda)\rightarrow-\infty$ a.s.\ at both
extremes of $\Lambda_{\mathrm{a}}$. For example, in the case of Figure
\ref{fig Erdos}, $\Lambda=(-1.195,1)$ and $\Lambda_{\mathrm{a}}%
=(-1.195,1.178).$ Note that the extremes of $\Lambda_{\mathrm{a}}$ must always
be zeros of $\det(S(\lambda))$, because $l_{\mathrm{a}}(\lambda)$ is
a.s.\ continuous between consecutive real zeros of $\det(S(\lambda
))$.\footnote{The fact that $l_{\mathrm{a}}(\lambda)$ is a.s.\ continuous
between consecutive real zeros of $\det(S(\lambda))$ also means that, ignoring
the pathological cases in which $l_{\mathrm{a}}(\lambda)$ is a.s.\ unbounded
from above, $\Lambda_{\mathrm{a}}$ is the smallest open set containing the
origin on which $l_{\mathrm{a}}(\lambda)$ is guaranteed to have a maximum
a.s.} Hence, our definition of $\Lambda_{\mathrm{a}}$ requires the following assumption.

\begin{mycomment}
	because in that case the maximum of $l_{\mathrm{a}}(\lambda)$ over $\Lambda$ can be on the boundary of $\Lambda$
\end{mycomment}

\begin{mycomment}
	we are led to introduce a modification of $\Lambda$: the set...
\end{mycomment}

\begin{mycomment}
	Note that the same as in previous footnote holds for $\Lambda$, if
	the maintained assumption that $W$ has at least one negative and at least one
	positive eigenvalue did not hold.
\end{mycomment}

\begin{assumption}
\label{assump Lambda_a}There is at least one negative zero and at least one
positive zero of $\det(S(\lambda))$ such that $l_{\mathrm{a}}(\lambda
)\rightarrow-\infty$ a.s.\ as $\lambda$ approaches those zeros.
\end{assumption}

It is clear from Section \ref{sec singular SLM} that Assumption
\ref{assump Lambda_a} can be violated only in very special cases. In those
cases, one could take the left (resp., right) endpoint of $\Lambda
_{\mathrm{a}}$ to be $-\infty$ (resp., $+\infty$), but we refrain from doing
this, for simplicity.

\begin{mycomment}
	don't need this: In fact, we shall see
	below that a simple condition is sufficient for $l_{\mathrm{a}}(\lambda)$ to have a
	unique maximum on $\Lambda_{\mathrm{a}}$.
\end{mycomment}

\begin{mycomment}
	Note that the same as in previous footnote holds for $\Lambda$, if
	the maintained assumption that $W$ has at least one negative and at least one
	positive eigenvalue did not hold.
\end{mycomment}

It is natural to define the adjusted QMLE of $\lambda$ as the maximizer of
$l_{\mathrm{a}}(\lambda)$ over $\Lambda_{\mathrm{a}}$, that is,
\[
\hat{\lambda}_{\mathrm{aML}}\coloneqq \argmax_{\lambda\in\Lambda_{\mathrm{a}}%
}l_{\mathrm{a}}(\lambda).
\]

Maximization of $l_{\mathrm{a}}(\lambda)$ over a subset of $\Lambda
_{\mathrm{a}}$ would yield a censored version of $\hat{\lambda}_{\mathrm{aML}%
}$. In particular, let $\bar{\Lambda}$ be $\Lambda$ augmented with one of its
endpoints if $l_{\mathrm{a}}(\lambda)$ is bounded near that endpoint
(augmented with both endpoints if $l_{\mathrm{a}}(\lambda)$ is bounded near
both endpoints), and let $\bar{\lambda}_{\mathrm{aML}}\coloneqq \argmax_{\lambda
\in\bar{\Lambda}}l_{\mathrm{a}}(\lambda)$ be the estimator that is obtained by
maximizing $l_{\mathrm{a}}(\lambda)$ over $\bar{\Lambda}$. Since
$\Lambda\subseteq\Lambda_{\mathrm{a}}$, $\bar{\lambda}_{\mathrm{aML}}$ is a
censored version of $\hat{\lambda}_{\mathrm{aML}}$.

\begin{mycomment}
	($\bar{\lambda
	}_{\mathrm{aML}}$ has nonzero probability of being equal to an extreme of
	$\Lambda$ if $\Lambda\subset\Lambda_{\mathrm{a}}$)
\end{mycomment}

Note that the set $\Lambda_{\mathrm{a}}$, contrary to $\Lambda$, may depend on
$X$ (because, by Theorem \ref{lemma lim gen W}, whether or not $l_{\mathrm{a}%
}(\lambda)\rightarrow-\infty$ a.s.\ at some zero of $\det(S(\lambda))$ depends
on $X$). For a \textit{fixed} $X$, $\Lambda_{\mathrm{a}}=\Lambda$ if
$l_{\mathrm{a}}(\lambda)\rightarrow-\infty$ a.s.\ as $\lambda$ approaches the
extremes of $\Lambda$, and $\Lambda_{\mathrm{a}}\supset\Lambda$ otherwise. But
it is also possible to compare $\Lambda$ and $\Lambda_{\mathrm{a}}$ for
\textit{generic} $X$ (in the measure theoretic sense), or better, since the
model typically contains an intercept, for a generic matrix $X$ containing an
intercept. The following example is important.

\begin{example2}
\label{exa generic}Suppose $W$ is row-stochastic and irreducible, and that an
intercept is included in the model. Let $X=(\iota_{n},\widetilde{X})$ and
$\widetilde{\mathcal{X}}\coloneqq \{\widetilde{X}\in\mathbb{R}^{n\times(k-1)}
:\mathrm{rank}(X)=k\}$. We know from Example \ref{exa row-stoch irred} that,
in this case, $l_{\mathrm{a}}(\lambda)$ is a.s.\ continuous at $\lambda=1$,
for all $\widetilde{X}\in\widetilde{\mathcal{X}}$. Consider now an arbitrary
semisimple nonzero real eigenvalue $\omega\neq1$ of $W$. By Lemma
\ref{lemma null set} in Appendix \ref{app add aux} and Theorem
\ref{lemma lim gen W}, $l_{\mathrm{a}}(\lambda)$ is a.s.\ unbounded from below
or from above near $\omega^{-1}$ for $\mu_{\mathbb{R}^{n\times(k-1)}}$-almost
every $\widetilde{X}\in\widetilde{\mathcal{X}}$. Assuming that $\omega_{\min}$
and the second largest (positive) eigenvalue of $W$, denoted by $\omega_{2}$,
are semisimple, it follows that $\Lambda_{\mathrm{a}}=(\omega_{\min}%
^{-1},\omega_{2}^{-1})$, for $\mu_{\mathbb{R}^{n\times(k-1)}}$-almost every
$\widetilde{X}\in\widetilde{\mathcal{X}}\setminus\left\{  \mathcal{P}%
_{\omega_{\min}} \cup\mathcal{P}_{\omega_{2}}\right\}  ,$ where $\mathcal{P}%
_{\omega}$ is the set of pathological $\widetilde{X}$ such that $l_{\mathrm{a}%
}(\lambda)$ is a.s.\ unbounded from above near $\omega^{-1}$ (i.e.,
$\mathcal{P}_{\omega}\coloneqq \{\widetilde{X}\in\mathbb{R}^{n\times(k-1)}%
:\mathrm{rank}(X)=k$ and $\mathrm{tr}(M_{X}Q_{\omega})<0\}$). As discussed
earlier, $\mathcal{P} _{\omega}$ is expected to be very small or even
$\mu_{\mathbb{R} ^{n\times(k-1)}}$-null in cases of interest in applications.
By Lemma \ref{lemma null}, $\mathcal{P}_{\omega}$ is empty if $W$ is
symmetric, so in that case $\Lambda_{\mathrm{a}}=(\omega_{\min}^{-1}%
,\omega_{2}^{-1})$ for $\mu_{\mathbb{R}^{n\times(k-1)}}$-almost every
$\widetilde{X}\in\widetilde{\mathcal{X}}$.
\end{example2}

Three remarks about the result in Example \ref{exa generic} that
$\Lambda_{\mathrm{a}}=(\omega_{\min}^{-1},\omega_{2}^{-1})$, for generic
non-pathological $\widetilde{X}$ are in order. First, note that $\omega
_{2}^{-1}>1$ and recall that $\Lambda=(\omega_{\min}^{-1},1)$, independently
of $X$. Thus, how much larger $\Lambda_{\mathrm{a}}$ is compared to $\Lambda$
depends only on the \textit{eigenvalue gap} $1-\omega_{2}$. There is
considerable evidence in the graph theory literature that the eigenvalue gap
$1-\omega_{2}$ tends to be large when the graph underlying $W$ has good
connectivity and randomness properties
\citep[especially when $W$ is the normalized adjacency matrix of a regular, undirected, loopless graph; see, e.g.][Chapter 4]{Brouwer11},
and this is indeed something we will come upon in our Monte Carlo experiments
later. Second, on replacing the intercept $\iota_{n}$ with group intercepts
$\bigoplus_{r=1}^{R}\iota_{m_{r}}$, as in Example \ref{exa block diag}, the
result in Example \ref{exa generic} generalizes immediately to the case when
$W$ is block diagonal with row stochastic and irreducible blocks. Third, the
argument used in Example \ref{exa generic} can also be applied to compare
$\Lambda_{\mathrm{a}}$ and $\Lambda$ for weights matrices that are not
row-stochastic and irreducible (or are not block diagonal with row-stochastic
and irreducible blocks). To do this, note that the result $\Lambda
_{\mathrm{a}}=(\omega_{\min}^{-1},\omega_{2}^{-1})$ for generic
non-pathological $\widetilde{X}$ in Example \ref{exa generic} arises because,
in that case, $\iota_{n}\in\operatorname{col}(X)$ and $\iota_{n}$ spans the
eigenspace $\mathrm{null}(W-I_{n})$, so that the condition in Part (i) of
Lemma \ref{lemma null} is satisfied. When $W$ is not row-stochastic and
irreducible, such a special interaction between $\operatorname{col}(X)$ and
$W$ will typically not occur, and as a result $\Lambda_{\mathrm{a}}%
=\Lambda=(\omega_{\min}^{-1},1)$ for generic non-pathological $\widetilde{X}$.
This is most easily seen when $W$ is symmetric. In that case, by part (ii) of
Lemma \ref{lemma null} and Theorem \ref{lemma lim gen W}, $\Lambda
_{\mathrm{a}}$ is different from $\Lambda$ only if $\mathrm{null}(W-I_{n})$ or
$\mathrm{null}(W-\omega_{\min}I_{n})$ are in $\operatorname{col}(X)$. In
general there is no reason why $\operatorname{col}(X)$ should contain those eigenspaces.

\begin{mycomment}
	See Appendix \ref{app l(lambda) SLM}.
\end{mycomment}

\subsection{Relationship between the QMLE and the adjusted QMLE}

The original motivation for seeking an unbiased estimating equation is, of
course, bias reduction, or more generally, improved inference on $\lambda$
(and possibly $\sigma^{2}).$ The simulation evidence reported in Section
\ref{sec Monte Carlo} below is unequivocal that the hoped-for bias reduction
is certainly achieved, and without detriment to the mean squared error.
However, there is one caveat that must be mentioned here, which we discuss next.

We have seen that the intervals $\Lambda\ $and $\Lambda_{\mathrm{a}}$ on which
the distributions of $\hat{\lambda}_{\mathrm{ML}}$ and $\hat{\lambda
}_{\mathrm{aML}}$ are supported are different in many cases of interest. That
is, there are circumstances in which $\hat{\lambda}_{\mathrm{aML}}$ lies
outside $\Lambda.$ This raises the question of whether, from the point of view
of interpreting the parameter $\lambda$, that outcome is
acceptable.\footnote{The situation is somewhat similar to, but more subtle
than, that in which the MLE for a variance (or covariance matrix) satisfies
the expected non-negativity (or positive definiteness) requirement, but a
bias-corrected version of it does not. For example, subtraction of the
estimated first (bias) term in an asymptotic expansion for the expectation of
the MLE can have this undesirable outcome. We also note that several other
estimators of $\lambda$, for example IV, GMM, or indirect inference
estimators, may have support larger than $\Lambda$
\citep[see e.g.,][]{Kelejian98,Lee2007a,Kyriacou2017}.} Ultimately, this will
depend on the context, but censoring the estimator to ensure that it lies in
$\Lambda$ will\ clearly entail sacrifice in terms of bias. The extent of the
censoring would usually be greater the larger is $\lambda$ in absolute value,
and would also depend on characteristics of $W$ such as its sparseness (see
the Monte Carlo simulations in Section \ref{sec Monte Carlo}). But,
presumably, the greater the censoring, the more sacrifice there will be in
terms of bias-reduction. In fact, simulations reported in Section
\ref{sec sup unr} of the Supplement suggest that, when $\hat{\lambda
}_{\mathrm{aML}}$ is outside $\Lambda$, often the unrestricted maximizer of
$l(\lambda)$ is also outside $\Lambda$ (i.e., $\hat{\lambda}_{\mathrm{ML}}%
\neq\hat{\lambda}_{\mathrm{uML}}$). This suggests that, whenever
$\Lambda_{\mathrm{a}}\neq\Lambda$, $\hat{\lambda}_{\mathrm{aML}}$ should be
regarded as a modification to the QMLE that maximizes $l(\lambda)$ over
$\Lambda_{\mathrm{a}}$, not over $\Lambda$.

\begin{mycomment}
looks like in most cases $\hat{\lambda}_{\mathrm{aML}}\notin\Lambda$ when
$\hat{\lambda}_{\mathrm{ML}}$ is not the unrestricted QMLE (but not always,
see plot\_lik\_score\_SLM\_with\_adj.m with adjustedMLEgreaterthan1\_QMLE=MLEunrestr.mat)
\end{mycomment}

\subsection{\label{sec ci}Confidence intervals}

Confidence intervals for a parameter of interest based on the QMLE may perform
poorly if the data contains little information about the nuisance parameters.
This may be the case, for instance, when the number of nuisance parameters is
large relative to the sample size. The theory presented in the paper so far
allows the construction of confidence intervals for $\lambda$ with accurate
coverage even in the case of little information about $\beta$.

We say that a differentiable function is \textit{single-peaked} on an open
interval if, on that interval, it has a maximum and no other stationary points
corresponding to minima or maxima (so the function is non-decreasing to the
left of the peak, and non-increasing to the right of the peak). We shall see
later in this subsection that $l_{\mathrm{a}}(\lambda)$ is single-peaked on
$\Lambda_{\mathrm{a}}$ quite generally (Proposition \ref{theo cdf MLE}), but
before doing that we show how single-peakedness can be used to construct
confidence intervals for $\lambda.$ If $l_{\mathrm{a}}(\lambda)$ is
single-peaked on $\Lambda_{\mathrm{a}}$, then the cdf of $\hat{\lambda
}_{\mathrm{aML}}$ admits the representation
\begin{equation}
\Pr(\hat{\lambda}_{\mathrm{aML}}\leq z;\beta,\sigma^{2},\lambda)=\Pr
(y^{\prime}S^{\prime}(z)R(z)S(z)y\leq0), \label{cdf}%
\end{equation}
for any $z\in\Lambda_{\mathrm{a}}$.\footnote{The notation $\Pr(\hat{\lambda
}_{\mathrm{aML}}\leq z;\beta,\sigma^{2},\lambda)$ emphasizes that, of course,
we are assuming that $y$ is generated by the SAR model (\ref{SLM}). But it is
worth remarking that, for a general random vector $y$, we would still have
$\Pr(\hat{\lambda}_{\mathrm{aML}}\leq z)=\Pr(y^{\prime}S^{\prime
}(z)R(z)S(z)y\leq0)$; all that is required is that the likelihood
(\ref{loglik}) is used for estimation.} That is, the cdf of $\hat{\lambda
}_{\mathrm{aML}}$ at $z$ equals the cdf of the quadratic form $y^{\prime
}S^{\prime}(z)R(z)S(z)y$ at zero. Thus, one can use some approximation to the
cdf (at zero) of $y^{\prime}S^{\prime}(z)R(z)S(z)y$ to obtain an approximation
to the cdf (at $z$) of $\hat{\lambda}_{\mathrm{aML}}$. Since the cumulant
generating function of a quadratic form is available in closed form, at least
under normality, one natural candidate is the Lugannani--Rice saddlepoint
approximation \citep{Lugannani80}. In fact, using a saddlepoint approximation
to the cdf of a score to recover the cdf of the corresponding estimator had
already been investigated in \cite{Daniels1983}
\citep[see also][Chapter 12]{Butler2007}. The approximate cdf of $\hat
{\lambda}_{\mathrm{aML}}$ can then be inverted to obtain confidence intervals
for $\lambda$. This approach to constructing confidence intervals has been
applied by \cite{Hillier2017} to the (unadjusted) QMLE of $\lambda$. Whilst
revising the present paper we discovered that essentially the same approach
had previously been suggested by \cite{Paige2009} for general quadratic
estimating equations, and then by \cite{Jeganathan2015} specifically for some
spatial models.

\begin{rem}
Let $\hat{\theta}$ an estimator obtained from the estimating equation
$q(\theta)=0$. \cite{Paige2009} use the device $\Pr(\hat{\theta}\leq
z)=\Pr(q(z)\leq0)$ under the assumption that $q(\theta)$ is monotonically
decreasing in $\theta$, whereas we have used it under the more general
assumption that the function with derivative $q(\theta)$ is single-peaked. It
is worth pointing out that, for the case of $\hat{\lambda}_{\mathrm{ML}}$ in a
SAR model, the difference between the two assumptions is immaterial when all
the eigenvalues of $W$ are real. Indeed, the score associated to the profile
log-likelihood (\ref{prof lik}) is $s(\lambda)\coloneqq n(y^{\prime}W^{\prime}%
M_{X}S_{\lambda}y)/(y^{\prime}S_{\lambda}^{\prime}M_{X}S_{\lambda
}y)-\mathrm{tr}\left(  G_{\lambda}\right)  $, and hence$\ $the equation
$s(\lambda)=0$ is a.s. equivalent to $f(\lambda)=0$, where the function
$f(\lambda)\coloneqq ny^{\prime}W^{\prime}M_{X}S_{\lambda}y-\mathrm{tr}\left(
G_{\lambda}\right)  y^{\prime}S_{\lambda}^{\prime}M_{X}S_{\lambda}y$ is
monotonically decreasing in $\lambda$ if all eigenvalues of $W$ are real \citep[see][]{LiYuBai2013}.
\end{rem}

\begin{mycomment}
	Two
	remarks concerning the comparison between the approaches to the construction
	of confidence intervals in \cite{Paige2009} and in \cite{Hillier2017} are in
	order. First, \cite{Paige2009} discuss a formal justification for replacing
	the nuisance parameters with their QMLEs given $\lambda$, which was missing in
	\cite{Hillier2017}. Second, \cite{Paige2009} assume monotonicity of the
	estimating equation, which is stronger than single-peakedness of the profile
	likelihood. In fact, monotonicity of the profile score is not needed for
	equation (\ref{cdf}), and is generally not satisfied for $s(\lambda)$
	\citep[see Remark S.2.2 in the Supplementary Material to][]{Hillier2017} or
	$s_{\mathrm{a}2}(\lambda)$.....yeah but satisfied for the est eq see li yu
	bai........see monotonicity\_score.tex
\end{mycomment}

\begin{mycomment}
	The
	Lugannani--Rice approximation had also been previously applied in a context
	similar to the present one by cite{Tiefelsdorf2002}
\end{mycomment}

Let $\widetilde{\Pr}(\hat{\lambda}_{\mathrm{aML}}\leq z;\beta,\sigma
^{2},\lambda)$ denote the approximation to the cdf of $\hat{\lambda
}_{\mathrm{aML}}$, obtained by the Lugannani--Rice formula (see Section
\ref{sec suppl ci} of the Supplement for details). The parameters $\beta$ and
$\sigma^{2}$ in the approximation can be replaced with their QMLEs given
$\lambda$, $\hat{\beta}_{\mathrm{ML}}(\lambda)$ and $\hat{\sigma
}_{\mathrm{aML}}^{2}(\lambda)$, to give the approximation $\widetilde{\Pr
}(\hat{\lambda}_{\mathrm{aML}}\leq z;\lambda)\coloneqq \widetilde{\Pr}(\hat{\lambda
}_{\mathrm{aML}}\leq z;\hat{\beta}_{\mathrm{ML}}(\lambda),\hat{\sigma
}_{\mathrm{aML}}^{2}(\lambda),\lambda)$. Confidence intervals for $\lambda$
based on $\hat{\lambda}_{\mathrm{aML}}$ can then be constructed by inverting
$\widetilde{\Pr}(\hat{\lambda}_{\mathrm{aML}}\leq z;\lambda)$. More
specifically, replace $z$ with the observed value of $\hat{\lambda
}_{\mathrm{aML}}$, say $\hat{\lambda}_{\mathrm{aML}}^{\mathrm{obs}}$, and let
$\lambda_{1}\coloneqq \inf\{\lambda:\widetilde{\Pr}(\hat{\lambda}_{\mathrm{aML}}%
\leq\hat{\lambda}_{\mathrm{aML}}^{\mathrm{obs}};\lambda)=1-\alpha_{1}\}$ and
$\lambda_{2}=\sup\{\lambda:$ $\widetilde{\Pr}(\hat{\lambda}_{\mathrm{aML}}%
\leq\hat{\lambda}_{\mathrm{aML}}^{\mathrm{obs}};\lambda)=\alpha_{2}\}$. Then
$(\lambda_{1},\lambda_{2})$ is an approximate $\left(  1-\alpha_{1}-\alpha
_{2}\right)  \%$ two-sided confidence interval for $\lambda$.\footnote{Note
that the set $\{\lambda:\alpha_{1}\leq\widetilde{\Pr}(\hat{\lambda
}_{\mathrm{aML}}\leq z;\lambda)\leq\alpha_{2}\}$ is in general a union of
intervals. It is an interval if $\widetilde{\Pr}(\hat{\lambda}_{\mathrm{aML}%
}\leq z;\lambda)$ is monotonic in $\lambda$. It seems reasonable to expect
this monotonicity to hold in many cases, at least approximately.}

\begin{mycomment}
	this was in the prev footnote:This is the reason why we have defined
	$\lambda_{1}$ as the \textit{smallest} value of $\lambda$ such that
	$\widetilde{\Pr}(\hat{\lambda}_{\mathrm{aML}}\leq\hat{\lambda}_{\mathrm{aML}%
	}^{\mathrm{obs}};\lambda)=1-\alpha_{1}$, and $\lambda_{2}$ the
	\textit{largest} value of $\lambda$ such that $\widetilde{\Pr}(\hat{\lambda
	}_{\mathrm{aML}}\leq\hat{\lambda}_{\mathrm{aML}}^{\mathrm{obs}};\lambda
	)=\alpha_{2}$.
\end{mycomment}

\begin{mycomment}
	it is reasonable to expect
	this to be the case quite generally, it is likely that monotonicity in $\lambda$ can be
	established only in special cases.
\end{mycomment}

The Lugannani--Rice approximation we employ is, as is frequently the case, a
normal based one. That is, it is constructed on the basis of the cumulant
generating function of $y^{\prime}S^{\prime}(z)R(z)S(z)y$ that obtains if
$\varepsilon\sim\mathrm{N}(0,I_{n})$. It is important to emphasize, however,
that we intend the approximation to be used generally. Indeed, there is
considerable evidence in the literature that a normal-based Lugannani--Rice
approximation is typically very accurate for the distribution of a statistic
that has a limiting normal distribution. There is also evidence that a
normal-based Lugannani--Rice approximation works generally very well for the
distribution of a quadratic form in nonnormal variables as long as we are far
from the lower tail of the distribution \citep{Wood1993}. This seems to be
relevant in our case as we are only interested in the cdf of $y^{\prime
}S^{\prime}(z)R(z)S(z)y$ at $0$, and $R(z)$ is indefinite.

The key condition for representation (\ref{cdf}) to hold is that
$l_{\mathrm{a}}(\lambda)$ is single-peaked on $\Lambda_{\mathrm{a}}$. We now
discuss this condition. The following very mild assumption allows us to make
use of the results derived earlier for semisimple eigenvalues.\footnote{When
$\Lambda_{\mathrm{a}}=\Lambda$, there is nothing to assume, because in that
case $\Lambda$ does not contain any zero of $\det(S(\lambda))$. When
$\Lambda_{\mathrm{a}}\supset\Lambda$, in most cases of practical interest the
only zero of $\det(S(\lambda))$ in $\Lambda_{\mathrm{a}}$ is the one
corresponding to the eigenvalue $1$, which, as pointed out earlier, is
virtually always semisimple in applications.}

\begin{mycomment}
	numerator of adj score not necessarily monotonic when all eigenvalues of W are real; see monotonicity_score.tex
\end{mycomment}

\begin{assumption}
\label{assump semis}Every eigenvalue of $W$ corresponding to a zero of
$\det(S(\lambda))$ in $\Lambda_{\mathrm{a}}$ is semisimple.
\end{assumption}

\begin{mycomment}
	Assumption \ref{assump semis} requires all nonzero eigenvalues $\omega$ of $W$
	such that $\omega^{-1}\in\Lambda_{\mathrm{a}}$ to be semisimple.
\end{mycomment}

\begin{proposition}
\label{theo cdf MLE}Suppose Assumptions \ref{assum id}, \ref{assump Lambda_a},
and \ref{assump semis} hold. If
\begin{equation}
(n-k)\mathrm{tr}(M_{X}G^{2}(\lambda))>\left[  \mathrm{tr}\left(
M_{X}G(\lambda)\right)  \right]  ^{2}\text{ for all }\lambda\in\Lambda
_{\mathrm{a}}\text{ such that }\det(S(\lambda))\neq0\text{,} \label{C1}%
\end{equation}
the adjusted profile log-likelihood function $l_{\mathrm{a}}(\lambda)$ is a.s.
single-peaked on $\Lambda_{\mathrm{a}}$.
\end{proposition}

It turns out that if $W$ is symmetric, condition (\ref{C1}) is satisfied for
any $X$ (see Lemma \ref{lemma delta_a<0} in Appendix \ref{app add aux}), and
hence $\Lambda_{\mathrm{a}}$ is a.s. single-peaked in that case. If $W$ is not
symmetric, the condition depends on both $W$ and $X$.\footnote{It is
interesting to note that, for the unadjusted profile likelihood $l(\lambda)$,
single-peakedness over $\Lambda$ holds if $n\mathrm{tr}(G^{2}(\lambda
))>\left[  \mathrm{tr}(G(\lambda))\right]  ^{2}$ for all $\lambda\in\Lambda$,
a condition that does \textit{not} depend on $X$ \citep[see][]{Hillier2017}.}
Extensive numerical experimentation with nonsymmetric $W$ suggests that for
the vast majority of pairs $(W,X)$ likely to be met in applications,
$l_{\mathrm{a}}(\lambda)$ is a.s.\ single-peaked on $\Lambda_{\mathrm{a}}$.
And, for pairs $(W,X)$ such that condition (\ref{C1}) is not satisfied, the
probability (for some distribution of $y$ that is absolutely continuous w.r.t.
$\mu_{\mathbb{R}^{n}}$) that $l_{\mathrm{a}}(\lambda)$ is multi-peaked is
generally very small. Note that the right hand side of representation
(\ref{cdf}) is likely to provide a good approximation to the cdf of
$\hat{\lambda}_{\mathrm{aML}}$ whenever the probability that $l_{\mathrm{a}%
}(\lambda)$ is multi-peaked is nonzero but small. The performance of the
saddlepoint confidence intervals is assessed by numerical simulation in
Section \ref{sec Monte Carlo}. Naturally, the confidence intervals can be
inverted to obtain hypothesis tests on $\lambda$, but we do not investigate
the power properties of such tests here.

\begin{mycomment}
I had this footnote: For example, condition (\ref{C1}) is
	satisfied in all repetitions of all Monte Carlo experiments in Section
	\ref{sec Monte Carlo}. but this does not apply anylonger as X is now fixed across reps
\end{mycomment}

\begin{mycomment}
	to check condition (\ref{C1}) USE check\_unimodality\_adj\_lik\_delta\_a\_negative.m
\end{mycomment}

\begin{mycomment}
	for any $z\in\Lambda_{\mathrm{a}}$. This cdf representation is exact if
	$l_{\mathrm{a}}(\lambda)$ is a.s.\ single peaked, and is likely to be highly accurate if
	the probability of multi-peakedeness is low. One important application of the
	representation is to the construction of higher-order confidence intervals, by
	means of the Lugannani--Rice saddlepoint approximation (\cite{Lugannani80}).
	expected to be more accurate than the first-order asymptotic ones quite generally.
\end{mycomment}

\begin{mycomment}
	\[
	y^{\prime}R_{\mathrm{a}}(\lambda)y=0
	\]
	where $R_{\mathrm{a}}(\lambda)\coloneqq S^{\prime}(\lambda)M_{X}\left(  G(\lambda
	)-\frac{\mathrm{tr}(M_{X}G(\lambda))}{n-k}I_{n}\right)  S(\lambda)$.
	for the MLE replace $\frac{\mathrm{tr}(M_{X}G(\lambda))}{n-k}$ replaced by
	$\frac{\mathrm{tr}(G(\lambda))}{n}$
	for the MLE $R(\lambda)\coloneqq S^{\prime}(\lambda)M_{X}\left(  G(\lambda
	)-\frac{\mathrm{tr}(G(\lambda))}{n}I_{n}\right)  S(\lambda)$%
	\begin{equation}
	\Pr(\hat{\lambda}_{\mathrm{aML}}\leq z)=\Pr(\tilde{y}^{\prime}R(z)\tilde
	{y}_{z}\leq0).
	\end{equation}
	for any $z\in\Lambda$, \bigskip\bigskip
\end{mycomment}

\begin{mycomment}
	FOR MYSELF:\ in fact I suspect whether
	we have single-peakedness or not depends only on $W$ for $l(\lambda)$ and
	depends on both $W$ and $X$ for $l_{\mathrm{a}}(\lambda)$, but to establish this I'd
	need to prove that $\delta(\lambda)<0$ for any $\lambda\in\Lambda$ and
	$\delta_{\mathrm{a}}(\lambda)<0$ for any $\lambda\in\Lambda_{\mathrm{a}}$ are nec and suff for
	single peakdness of $l(\lambda)$ and $l_{\mathrm{a}}(\lambda)$, respectively (I think
	this is true but not simple to prove)
\end{mycomment}

\begin{mycomment}
	\[
	C(\lambda)\coloneqq G(\lambda)-\frac{\mathrm{tr}(G(\lambda))}{n}I_{n},
	\]
	\begin{align*}
	tr\left(  C^{2}(\lambda)\right)    & =tr\left(  G^{2}(\lambda)+\frac{\left[
		\mathrm{tr}(G(\lambda))\right]  ^{2}}{n^{2}}I_{n}-2\frac{\mathrm{tr}%
		(G(\lambda))}{n}G(\lambda)\right)  \\
	& =tr(G^{2}(\lambda))+\frac{\left[  \mathrm{tr}(G(\lambda))\right]  ^{2}}%
	{n}-2\frac{\left[  \mathrm{tr}(G(\lambda))\right]  ^{2}}{n}\\
	& =tr(G^{2}(\lambda))-\frac{\left[  \mathrm{tr}(G(\lambda))\right]  ^{2}}%
	{n}=-n\delta(\lambda)
	\end{align*}
	\begin{align*}
	tr\left(  C_{a\lambda}^{2}\right)    & =tr\left(  G^{2}(\lambda)%
	+\frac{\left[  \mathrm{tr}(M_{X}G(\lambda))\right]  ^{2}}{\left(  n-k\right)
		^{2}}I_{n}-2\frac{\mathrm{tr}(M_{X}G(\lambda))}{n-k}G(\lambda)\right)  \\
	& =tr(G^{2}(\lambda))+\frac{\left[  \mathrm{tr}(M_{X}G(\lambda))\right]
		^{2}}{\left(  n-k\right)  ^{2}}n-2\frac{\mathrm{tr}(M_{X}G(\lambda))}%
	{n-k}tr(G(\lambda))
	\end{align*}
	\bigskip%
\end{mycomment}

\begin{mycomment}
	FOR\ MYSELF:\ Lemma \ref{lemma delta_a<0} does not extend
	to the case of similar to symm W
	\par
	single-peakedness of $l_{\mathrm{a}}$ on $\Lambda_{\mathrm{a}}$ HOLDS WHEN W IS SYMM not
	necessarily when W is sim symm matrix ............
	\par
	Under Assumption \ref{assum sim symm} $G(\lambda)-tI_{n}=T(A(I_{n}-\lambda
	A)^{-1}-tI_{n})^{2}T^{-1}$ so $C(G(\lambda)-tI_{n})^{2}C^{\prime}%
	=CT(A(I_{n}-\lambda A)^{-1}-tI_{n})^{2}T^{-1}C^{\prime}$. Since $(A(I_{n}%
	-\lambda A)^{-1}-tI_{n})^{2}$ is positive semidefinite, $t^{\prime}%
	(A(I_{n}-\lambda A)^{-1}-tI_{n})^{2}t\geq0$ for any $t$. so $y^{\prime
	}C(G(\lambda)-tI_{n})^{2}C^{\prime}y\geq0$ for any $y$ $\mathrm{tr}\left(
	C(G(\lambda)-tI_{n})^{2}C^{\prime}\right)  =\mathrm{tr}\left(  CG_{\lambda
	}^{2}C^{\prime}+t^{2}M_{X}-2tCG(\lambda)C^{\prime}\right)  =\mathrm{tr(}%
	M_{X}G^{2}(\lambda))+t^{2}(n-k)-2t\mathrm{tr}(M_{X}G(\lambda))$ $t=\frac
	{1}{n-k}\mathrm{tr}(M_{X}G(\lambda))$ $\mathrm{tr}\left(  C(G_{\lambda
	}-tI_{n})^{2}C^{\prime}\right)  =\mathrm{tr(}M_{X}G^{2}(\lambda))+\frac
	{1}{n-k}\left[  \mathrm{tr}(M_{X}G(\lambda))\right]  ^{2}-2\frac{1}%
	{n-k}\left[  \mathrm{tr}(M_{X}G(\lambda))\right]  ^{2}$ ntp$\frac{1}%
	{n-k}\left[  \mathrm{tr}\left(  M_{X}G(\lambda)\right)  \right]
	^{2}-\mathrm{tr}(M_{X}G^{2}(\lambda))\leq0$ $\mathrm{tr}(M_{X}G^{2}(\lambda)
	)-\frac{1}{n-k}\left[  \mathrm{tr}\left(  M_{X}G(\lambda)\right)
	\right]  ^{2}\geq0$
\end{mycomment}

\begin{mycomment}
	FOR\ MYSELF: Lemma \ref{lemma delta_a<0}%
	corresponds to the result in \cite{Hillier2017} that if all
	eigenvalues of $W$ are real then $\delta(\lambda)<0$ for all $\lambda
	\in\Lambda$.
\end{mycomment}

\begin{mycomment}
	In general, both $l(\lambda)$ and $l_{\mathrm{a}}(\lambda)$ may have multiple peaks on,
	respectively, $\Lambda$ and $\Lambda_{\mathrm{a}}$. Lemma \ref{lemma unimod} is the
	analog of a result in \cite{Hillier2017} for $l(\lambda)$. Proving
	single-peakedness for $l_{\mathrm{a}}(\lambda)$ is complicated by the fact that
	$\Lambda_{\mathrm{a}}$, contrary to $\Lambda$, may contain zeros of $\det(S(\lambda))$,
	and, as discussed earlier, $l_{\mathrm{a}}(\lambda)$ may be unbounded from above at
	these points.
\end{mycomment}

\begin{mycomment}
	.......HM $l(\lambda)$ single-peaked if $tr(C^{2})>0$ iff $tr(G^{2})-\frac
	{1}{n}\left[  tr(G)\right]  ^{2}>0$ iff $\left[  tr(G)\right]  ^{2}%
	-ntr(G^{2})<0$%
	\begin{equation}
	\delta_{\mathrm{a}}(\lambda)\coloneqq \left[  \mathrm{tr}\left(  M_{X}G(\lambda)\right)
	\right]  ^{2}-(n-k)\mathrm{tr}(M_{X}G^{2}(\lambda)),
	\end{equation}
\end{mycomment}

\subsection{\label{sec network f.e. model}Network fixed effects}

In this section we apply the score adjustment to a social interaction model
with network fixed effects and contextual effects
\citep[e.g.,][]{Lee2007b, Bramoulle2009, LeeLiuLin2010, Lin2015, Fortin2015}.
There are $R$ networks, with network $r$ having $m_{r}$ individuals. The model
is%
\begin{equation}
y_{r}=\lambda W_{r}y_{r}+\widetilde{X}_{r}\gamma+W_{r}\widetilde{X}_{r}%
\delta+\alpha_{r}\iota_{m_{r}}+\sigma\varepsilon_{r},\text{ }r=1,\ldots,R,
\label{network model}%
\end{equation}
where $W_{r}$ is the weights matrix of network $r$, $\alpha_{r}$ is an
unobserved network fixed effect, $\widetilde{X}_{r}$ is an $m_{r}\times
\tilde{k}$ matrix of regressors, $\gamma$ and $\delta$ are $\tilde{k}\times1$
parameters. In terms of equation (\ref{SLM}), $y=(y_{1}^{\prime}%
,\ldots,y_{R}^{\prime})^{\prime}$, $W=\bigoplus_{r=1}^{R}W_{r}$, $\beta
=(\gamma^{\prime},\delta^{\prime},\alpha_{1},\ldots,\alpha_{R})^{\prime}$,
$\varepsilon=(\varepsilon_{1}^{\prime},\ldots,\varepsilon_{R}^{\prime})^{\prime}%
$, and $X=(\widetilde{X},W\widetilde{X},\bigoplus_{r=1}^{R}\iota_{m_{r}})$,
with $\widetilde{X}\coloneqq (\widetilde{X}_{1}^{\prime},\ldots,\widetilde{X}_{R}%
^{\prime})^{\prime}$. We assume that $X$ has full column rank.

\begin{mycomment}
	this was not true (to see this you can use spatial panel two way fixed effects.m): Here, we are assuming that
		$\mathrm{rank}(\widetilde{X}_{r},W_{r}\widetilde{X}_{r})=2k$. If
		$\mathrm{rank}(\widetilde{X}_{r},W_{r}\widetilde{X}_{r})$ were smaller than
		$2k,$ say equal to $s$, then of course only $s$ linearly independent columns
		of $(\widetilde{X}_{r},W_{r}\widetilde{X}_{r})$ should be included in the
		regression.
\end{mycomment}

To avoid the incidental parameter problem that arises in the case of many
networks of fixed dimensions, inference in this model usually proceeds by
first eliminating the network fixed effects. Define the orthogonal projector
$M_{r}\coloneqq I_{m_{r}}-\frac{1}{m_{r}}\iota_{m_{r}}\iota_{m_{r}}^{\prime}$, and let
$F_{r}$ be the $m_{r}\times\left(  m_{r}-1\right)  $ matrix of orthonormal
eigenvectors of $M_{r}$ corresponding to the eigenvalue $1$, so that
$F_{r}^{\prime}F_{r}=I_{m_{r}-1}$, $F_{r}F_{r}^{\prime}=M_{r}$, and
$F_{r}^{\prime}\iota_{m_{r}}=0$. In a likelihood framework, the standard
procedure is that proposed by \cite{LeeLiuLin2010}, which eliminates the fixed
effects by premultiplying each equation in (\ref{network model}) by
$F_{r}^{\prime}$. Under the condition that each weights matrix $W_{r}$ has all
row sums equal to one (i.e., $W_{r}\iota_{m_{r}}=\iota_{m_{r}}$ for all
$r=1,\ldots,R$), we have $F_{r}^{\prime}W_{r}=F_{r}^{\prime}W_{r}F_{r}%
F_{r}^{\prime}$, and therefore the transformed model is%
\begin{equation}
y_{r}^{\ast}=\lambda W_{r}^{\ast}y_{r}^{\ast}+\widetilde{X}_{r}^{\ast}%
\gamma+W_{r}^{\ast}\widetilde{X}_{r}^{\ast}\delta+\sigma\varepsilon_{r}^{\ast
},\text{ }r=1,\ldots,R, \label{transf LLL}%
\end{equation}
where $y_{r}^{\ast}\coloneqq F_{r}^{\prime}y_{r}$, $W_{r}^{\ast}\coloneqq F_{r}^{\prime}%
W_{r}F_{r},$ $\widetilde{X}_{r}^{\ast}\coloneqq F_{r}^{\prime}\widetilde{X}_{r}$, and
$\varepsilon_{r}^{\ast}\coloneqq F_{r}^{\prime}\varepsilon_{r}$. Note that
$\mathrm{var}(\varepsilon_{r}^{\ast})=I_{m-1}$ if $\mathrm{var}(\varepsilon
_{r})=I_{m}$. We denote the profile (quasi) log-likelihood for $(\sigma
^{2},\lambda)$ based on the transformed model (\ref{transf LLL}) by
$l_{\mathrm{LLL}}(\sigma^{2},\lambda)$. If the condition $W_{r}\iota_{m_{r}%
}=\iota_{m_{r}}$, for each $r=1,\ldots,R$, is not satisfied, the transformation
by $F_{r}^{\prime}$ does not yield a reduced form and hence a likelihood. In
that case, only non-likelihood procedures, such as the GMM of
\cite{LiuLee2010}, are available.

The score adjustment discussed in the present paper provides an alternative
likelihood solution to the large-$R$ incidental parameter problem. It provides
a large-$R$ consistent estimator of all model parameters (see Remark
\ref{rem consist}), like the \cite{LeeLiuLin2010} procedure, but, contrary to
\cite{LeeLiuLin2010}, it does not require the constant row sums condition. The
score adjustment is performed exactly as for the general SAR model, treating
the $\alpha_{r}$'s as parameters, profiling out the parameter $\beta
=(\gamma^{\prime},\delta^{\prime},\alpha_{1},\ldots,\alpha_{R})^{\prime}$ as in
equation (\ref{lik sig lam SAR}), and recalling that the expectation
(\ref{Es}) does not depend on $\beta$.

The next proposition shows that the score adjustment method is equivalent to
the \cite{LeeLiuLin2010} method, if the latter applies (i.e., $W_{r}%
\iota_{m_{r}}=\iota_{m_{r}}$ for all $r$) and there are no covariates (i.e.,
the model is $y_{r}=\lambda W_{r}y_{r}+\alpha_{r}\iota_{m_{r}}+\sigma
\varepsilon_{r}$, $r=1,\ldots,R$).

\begin{proposition}
\label{prop aML=LLL}Assume that, in model (\ref{network model}), there are no
regressors and $W_{r}\iota_{m_{r}}=\iota_{m_{r}}$, for each $r=1,\ldots,R$. Then,
$l_{\mathrm{a}}(\sigma^{2},\lambda)=l_{\mathrm{LLL}}(\sigma^{2},\lambda)$, for
any $y\in\mathbb{R}^{n}$.
\end{proposition}

\begin{mycomment}
\item recall $l_{\mathrm{a}}(\sigma^{2},\lambda)$ is a profile lik (fixed eff are profiled out) while $l_{\mathrm{LLL}}(\sigma^{2},\lambda)$ is the whole lik (no profiling)
\end{mycomment}

When covariates are present, the estimators of $\sigma^{2}$ and $\lambda$ (and
hence of $\beta$) obtained by the score adjustment method are different from
the ones obtained by the \cite{LeeLiuLin2010} method. Both methods solve the
large-$R$ incidental parameter problem, but, in addition to not requiring the
constant row sum condition, the score adjustment approach also implements a
correction that deals with the nuisance parameters $\gamma$ and $\delta$. The
two methods are compared by simulation in Section \ref{secd network f.e.}%
.\footnote{\label{foot asy equiv}Our focus is on finite sample results. From
the point of view of first-order asymptotics, we expect the two estimators to
be equivalent under the conditions in \cite{LeeLiuLin2010}. Such conditions
include, in particular, the fact that $\tilde{k}$ does not vary with $n$; if
$\tilde{k}$ increased with $n$, the adjusted QMLE might have an advantage even
from the point of view of first-order asymptotics.}

\begin{rem}
\label{rem consist}A formal consistency proof for the adjusted QMLE in model
(\ref{network model}) would require stating several regularity conditions,
which is not our aim here. Heuristically, however, consistency follows from a
standard argument that we can sketch here
\citep[see, e.g.,][]{Neyman1948, DhaeneJoch2015}. To start with, note that in
SAR\ models without incidental parameters, the expectation $\mathrm{E}%
(s(\sigma^{2},\lambda))$ in equation (\ref{Es}) is $O(1)$ as $n\rightarrow
\infty$ under standard conditions \citep[see, e.g.,][Lemma A.9]{Lee2004b}, and
therefore the necessary condition $\mathrm{plim}_{n\rightarrow\infty}\frac
{1}{n}s(\sigma^{2},\lambda)=0$ for $(\hat{\sigma}_{\mathrm{ML}}^{2}%
,\hat{\lambda}_{\mathrm{ML}})$ to be consistent is satisfied. On the other
hand, when there is an incidental parameter of dimension $R$, as in model
(\ref{network model}), and $n$ increases only because $R$ increases,
$\mathrm{plim}_{n\rightarrow\infty} \frac{1}{n}s(\sigma^{2},\lambda)\neq0$
because the bias of the profile score $s(\sigma^{2},\lambda)$ is typically
$O(R)$; however, the fact that the expectation (\ref{Es}) is independent of
the nuisance parameter $\beta$ immediately implies that $\mathrm{plim}%
_{n\rightarrow\infty}\frac{1} {n}\left\{  s(\sigma^{2},\lambda)-\mathrm{E}%
(s(\sigma^{2},\lambda))\right\}  =0$, which is essentially what delivers
consistency of the adjusted QMLE.
\end{rem}

\begin{mycomment}
..........(which we refer to with the acronym $\mathrm{LLL}$)\bigskip
\end{mycomment}

\begin{mycomment}
OLD (slightly less general as MLEs could be the same even when the likelihoods
are diff, as in the case of direct and transf approach for individ fixed
effects spatial panel model)\ Assume that $W_{i}\iota_{m_{i}}=\iota_{m_{i}}$,
for all $i=1,\ldots,R$, and that model (\ref{network model}) does not contain any
regressors. Then, $(\hat{\lambda}_{\mathrm{LLL}},\hat{\sigma}_{\mathrm{LLL}%
}^{2})$ coincide with $(\hat{\lambda}_{\mathrm{aML}},\hat{\sigma
}_{\mathrm{aML}}^{2})$.
\end{mycomment}

\begin{mycomment}
	\begin{itemize}
		\item network fixed effects (see ECOD022 slides), premultiply by $I_{n}\otimes
		F^{\prime}$ (see \emph{sp\_lag\_sim\_fedeLLL\_Z\_X\_in\_each\_rep.m }and
		\emph{check\_transformation\_approach\_network\_fixed\_effects.m} and
		\emph{saddlepoint\_ci\_simulation\_network\_fix\_eff\_adj.m}) ($%
		{\textstyle\bigoplus_{i=1}^{r}}
		F_{i}^{\prime}$ in the unbalanced case)
		reml lik premultiply by $C$ s.t. $CC^{\prime}=I_{n-k}$ and $C^{\prime}C=M_{X}$
		yes verified using bias\_MLE\_panel\_table\_X\_in\_each\_rep\_k\_varies.m.
		Note that when the blocks W are row-stoch and irred adj lik is cont at
		lambda=1 as long as X contains group fixed eff (so in this case both
		lambda\_MLE\_adj and lambda\_LLL may be larger than 1)
		\item When covariates are present, adj lik (for SEM or SLM with fixed eff)
		goes one step further, as it takes care also of the regressors.
	\end{itemize}
\end{mycomment}

\section{\label{sec SEM}Spatial error model}

The spatial error model
\begin{equation}
y=X\beta+u,\text{ }u=\lambda Wu+\sigma\varepsilon\label{SEM}%
\end{equation}
is considerably less popular than the SAR model (\ref{SLM}) in economic
applications, but, as we shall see, provides important motivation for the
score adjustment considered in this paper. We now briefly consider this model,
leaving all details to Section \ref{suppl SEM} of the Supplement. It is
convenient to generalize the error structure in model (\ref{SEM}) to
$A(\theta)u=\sigma\varepsilon$, where $A(\theta)$ is a square matrix that is
invertible for any value of the parameter $\theta$ in a subset $\Theta$ of
$\mathbb{R}^{p}$. In the case of the spatial error model, we may take
$A(\theta)=S(\lambda).$

\begin{mycomment}
	a (not necessarily symmetric)\ square root of $\Sigma^{-1}(\theta)$ (i.e.,
	$A(\theta)$ is such that $A^{\prime}(\theta)A(\theta)=\Sigma^{-1}(\theta)$).
	For the spatial error model, we may take $A(\theta)=S(\lambda).$
	$-\frac{1}{2}\log\left(  \det\left(  \Sigma(\theta)\right)  \right)  =\frac
	{1}{2}\log\left(  \det\left(  A^{\prime}(\theta)A(\theta)\right)  \right)
	=\log\left(  \left\vert \det\left(  A(\theta)\right)  \right\vert \right)  $
\end{mycomment}

\begin{mycomment}
	Let $\Sigma(\theta)$ be a matrix that is symmetric, positive definite, and
	differentiable in $\theta$, for any value of the parameter $\theta$ in a
	subset $\Theta$ of $\mathbb{R}^{p}$. It is convenient to generalize model
	(\ref{SEM}) to $y=X\beta+\sigma V(\theta)\varepsilon,$ with $\mathrm{E}%
	(\varepsilon)=0$, where $V(\theta)$ is a square root of the matrix
	$\Sigma(\theta)$ (i.e., $V(\theta)$ is such that $V(\theta)V^{\prime}%
	(\theta)=\Sigma(\theta)$). For the spatial error model, we may take
	$V(\theta)=S^{-1}(\lambda).$
	$\Sigma(\theta)=S(\lambda)^{-1}S^{\prime}(\lambda)^{-1}=\left(  S^{\prime
	}(\lambda)S(\lambda)\right)  ^{-1}$
	$\Sigma^{-1}(\theta)=S^{\prime}(\lambda)S(\lambda)$
\end{mycomment}

\begin{mycomment}
	OLDER. It is convenient to generalize model (\ref{SEM}) to a general
	regression model with exogenous regressors and correlated errors, which we
	write as $y=X\beta+u,$ with $\mathrm{E}(u)=0$ and $\mathrm{var}(u)=\sigma
	^{2}\Sigma(\theta)$, where the parameter $\theta$ belongs to some subset
	$\Theta\subseteq\mathbb{R}^{p}$. It is assumed that $\Sigma(\theta)$ is
	positive definite for any $\theta\in\Theta$, and differentiable in $\theta$.
	Note that the model is invariant under the group $\mathcal{G}_{X}$ of
	transformations $y\rightarrow\kappa y+X\delta$ in the sample space, for any
	$\kappa>0$, any $\delta\in\mathbb{R}^{k}$
	\citep[for the relevant definition of invariance, see][Chapter 6]{Lehmann2005}.
	For the spatial error model, $\Sigma(\lambda)=\left(  S^{\prime}%
	(\lambda)S(\lambda)\right)  ^{-1}.$
\end{mycomment}

The profile (quasi) log-likelihood for $(\sigma^{2},\theta)$ based on the
assumption $\varepsilon\sim\mathrm{N}(0,I_{n})$ is (up to an additive
constant)%
\begin{equation}
l(\sigma^{2},\theta)\coloneqq l(\hat{\beta}_{\mathrm{ML}}(\theta),\sigma^{2}%
,\theta)=-\frac{n}{2}\log(\sigma^{2})+\log\left\vert \det\left(
A(\theta)\right)  \right\vert -\frac{1}{2\sigma^{2}}y^{\prime}U(\theta)y,
\label{lik sig theta}%
\end{equation}
where $U(\theta)\coloneqq A^{\prime}(\theta)M_{A(\theta)X}A(\theta)$. As for the case
of the SAR model, the score associated to (\ref{lik sig theta}) can be
(exactly) recentered assuming only that $\mathrm{E}(\varepsilon)=0$ and
$\mathrm{var}(\varepsilon)=I_{n}$. The corresponding adjusted likelihood is
\begin{equation}
l_{\mathrm{a}}(\sigma^{2},\theta)=-\frac{n-k}{2}\log(\sigma^{2})-\frac
{1}{2\sigma^{2}}y^{\prime}U(\theta)y+\log\left\vert \det\left(  A(\theta
)\right)  \right\vert -\frac{1}{2}\log\left(  \det\left(  X^{\prime}A^{\prime
}(\theta)A(\theta)X\right)  \right)  . \label{l_a_regr_sigma_theta}%
\end{equation}

Remarkably, and contrary to the case of a SAR model, $l_{\mathrm{a}}%
(\sigma^{2},\theta)$ is a genuine likelihood. Hence, the corresponding score
provides an unbiased and information unbiased estimating equation (under the
assumptions that $\mathrm{E}(\varepsilon)=0$ and $\mathrm{var}(\varepsilon
)=I_{n}$). In fact, $l_{\mathrm{a}}(\sigma^{2},\theta)$ is equivalent to the
likelihood used for restricted ML (REML) estimation, which has been shown to
be quite generally preferable to ML estimation in a regression model with
correlated errors \citep[see, e.g.,][]{Thompson1962, Patterson1971, RahmanKing1997}.

Maximization of (\ref{l_a_regr_sigma_theta}) for fixed $\theta$ gives
$\hat{\sigma}_{\mathrm{a}\mathrm{ML}}^{2}(\theta)=\frac{1}{n-k}y^{\prime
}U(\theta)y$, and thus the profile adjusted likelihood for $\theta$ only is%
\begin{equation}
l_{\mathrm{a}}(\theta)\coloneqq l_{\mathrm{a}}(\hat{\sigma}_{\mathrm{a}\mathrm{ML}%
}^{2}(\theta),\theta)=-\frac{n-k}{2}\log(y^{\prime}U(\theta)y)+\log\left\vert
\det\left(  A(\theta)\right)  \right\vert -\frac{1}{2}\log\left(  \det\left(
X^{\prime}A^{\prime}(\theta)A(\theta)X\right)  \right)  .
\label{l_a_regr_theta}%
\end{equation}

To fully understand the effect of the score adjustment, it is useful to relate
the likelihood $l_{\mathrm{a}}(\theta)$ to invariance properties of the model.
Assuming that the distribution of $\varepsilon$ does not depend on the
parameters $\beta$ and $\sigma$ (and that the parameters are identifiable), it
is straightforward to check that the model is invariant under the group
$\mathcal{G}_{X}$ of transformations $y\rightarrow\kappa y+X\delta$ in the
sample space, for any $\kappa>0$, any $\delta\in\mathbb{R}^{k}$, and for a
fixed $X$
\citep[for the relevant definition of invariance, see][Chapter 6]{Lehmann2005}.
The likelihood $l_{\mathrm{a}}(\theta)$ corresponds to the density of a
maximal invariant under the group $\mathcal{G}_{X}$ (and $l_{\mathrm{a}%
}(\sigma^{2},\theta)$ corresponds to the density of a maximal invariant under
the group of transformations $y\rightarrow y+X\delta$, $\delta\in
\mathbb{R}^{k}$). That is, using the adjusted likelihood $l_{\mathrm{a}%
}(\theta)$ corresponds to imposing that inference should be invariant with
respect to the group under which the model itself is invariant, as advocated
by the \textquotedblleft principle of invariance\textquotedblright.

\begin{mycomment}
	is considerably less popular than the SAR model (\ref{SLM}) in economic
	applications, but provides important motivation for the adjusted ML procedure
	considered in this paper. In this section we briefly consider this model,
	leaving all details to Section \ref{suppl SEM} of the Supplement. Let
	$\Sigma(\theta)$ be a matrix that is symmetric, positive definite, and
	differentiable in $\theta$, for any value of the parameter $\theta$ in a
	subset $\Theta$ of $\mathbb{R}^{p}$. It is convenient to generalize model
	(\ref{SEM}) to $y=X\beta+\sigma A^{-1}(\theta)\varepsilon,$ where $A(\theta)$
	is a (not necessarily symmetric)\ square root of $\Sigma^{-1}(\theta)$ (i.e.,
	$A(\theta)$ is such that $A^{\prime}(\theta)A(\theta)=\Sigma^{-1}(\theta)$).
	For the spatial error model, we may take $A(\theta)=S(\lambda).$
	The (quasi) profile log-likelihood for $\theta$ that obtains under the
	assumption $\varepsilon\sim\mathrm{N}(0,I_{n})$ is
	\begin{equation}
	l(\theta)\coloneqq -\frac{n}{2}\log\left(  \hat{\sigma}_{\mathrm{ML}}^{2}%
	(\theta)\right)  -\frac{1}{2}\log\left(  \det\left(  \Sigma(\theta)\right)
	\right)  ,
	\end{equation}
	with $\hat{\sigma}_{\mathrm{ML}}^{2}(\theta)\coloneqq (y^{\prime}\Sigma^{-1}%
	(\theta)\left(  I_{n}-X(X^{\prime}\Sigma^{-1}(\theta)X)^{-1}X^{\prime}%
	\Sigma^{-1}(\theta)\right)  y)/n$. Recentering the profile score and
	integrating produces the adjusted log-likelihood%
	\begin{equation}
	l_{\mathrm{a}}(\theta)\coloneqq -\frac{n-k}{2}\log\left(  \hat{\sigma}_{\mathrm{ML}%
	}^{2}(\theta)\right)  -\frac{1}{2}\log\left(  \det\left(  \Sigma
	(\theta)\right)  \right)  -\frac{1}{2}\log\left(  \det\left(  X^{\prime}%
	\Sigma^{-1}(\theta)X\right)  \right)  .
	\end{equation}
\end{mycomment}

\begin{mycomment}
	give $l(\lambda)$ \bigskip
\end{mycomment}

We do not give details here, but it can be shown that, as $\lambda$ approaches
any real zero of $\det\left(  S(\lambda)\right)  $, $l(\lambda)\rightarrow
-\infty$ a.s., whereas $l_{\mathrm{a}}(\lambda)$ may either a.s. approach
$-\infty$ or be a.s.\ bounded in a spatial error model.\footnote{Given the
correspondence between $l_{\mathrm{a}}(\lambda)$ and the density of a maximal
invariant under the group $\mathcal{G}_{X}$, the fact that $l_{\mathrm{a}%
}(\lambda)$ may be bounded as $\lambda\rightarrow\omega^{-1}$, for some
nonzero real eigenvalue $\omega$ of $W$, provides an explanation of why the
limiting power of an invariant test for $\lambda=0$ may be strictly in $(0,1)$
as $\lambda\rightarrow\omega^{-1}$ \citep[see][]{Martellosio2010}. On the other hand,
if $l_{\mathrm{a}}(\lambda)\rightarrow-\infty$, then the power of an invariant
test for $\lambda=0$ may approach 0 or 1 as $\lambda\rightarrow\omega^{-1}$
depending on the location of the critical region.} Thus, as might have been
expected, in the spatial error model the profile log-likelihood and its
adjusted version behave very much as in the SAR model.\footnote{There is one
important difference though. Contrary to the case of the SAR model, in the
spatial error model $l_{\mathrm{a}}(\lambda)$ can never approach $+\infty$
a.s.\ as $\lambda$ approaches a real zero of $\det\left(  S(\lambda)\right)
$. This is because $l_{\mathrm{a}}(\lambda)$ is a genuine likelihood, so
a.s.\ unboundedness from above of $l_{\mathrm{a}}(\lambda)$ as $\lambda$
approaches a zero of $\det\left(  S(\lambda)\right)  $ would imply the
existence of a density that diverges to +$\infty$ almost everywhere on
$\mathbb{R}^{n}$ as $\lambda$ approaches that zero.} So, in particular, it
makes sense to define the adjusted QMLE of $\lambda$ on a support different
from $\Lambda$, as in the SAR model. More generally, this implies that when
the covariance parameter $\theta$ is restricted to a certain set, the REML
estimator of $\theta$ does not necessarily respect that restriction, something
that, to the best of our knowledge, has not been noted before in the
literature.

\begin{mycomment}
OLD:\ For the spatial error model, we need the following assumption, which
plays the same role as Assumption \ref{assum id} for the SAR model, and is
discussed in detail in the Supplement.
\begin{assumption}
\label{assum id SEM}
(i) $M_{X}W\neq0$; (ii) there is no nonzero real eigenvalue $\omega$ of $W$
for which $M_{S(\omega^{-1})X}S(\omega^{-1})=0$.
\end{assumption}
[[[[[[[[[[REVISE]]]]]]]]]]Under Assumption \ref{assum id SEM}, it can be shown
that, as $\lambda$ approaches any real zero of $\det\left(  S(\lambda)\right)
$, $l(\lambda)\rightarrow-\infty$ a.s., but $l_{\mathrm{a}}(\lambda)$ may
either a.s. approach $-\infty$ or be a.s.\ bounded. Thus, as might have been
expected, in the spatial error model the profile log-likelihood and its
adjusted version behave very much as in the SAR model.\footnote{There is one
important difference though. Contrary to the case of the SAR model, in the
spatial error model $l_{\mathrm{a}}(\lambda)$ can never approach $+\infty$
a.s.\ as $\lambda$ approaches a real zero of $\det\left(  S(\lambda)\right)
$. This is because $l_{\mathrm{a}}(\lambda)$ is a genuine likelihood, so
a.s.\ unboundedness from above of $l_{\mathrm{a}}(\lambda)$ as $\lambda$
approaches a zero of $\det\left(  S(\lambda)\right)  $ would imply the
existence of a density that diverges to +$\infty$ almost everywhere on
$\mathbb{R}^{n}$ as $\lambda$ approaches that zero.} So, in particular, it
makes sense to define the adjusted QMLE of $\lambda$ on a support different
from $\Lambda$, as in the SAR model. More generally, this implies that when
the covariance parameter $\theta$ is restricted to a certain set, the REML
estimator of $\theta$ does not necessarily respect that restriction, something
that, to the best of our knowledge, has not been noted before in the literature.
\end{mycomment}

\begin{mycomment}
	PREVIOUS\ FORMULATION\ of Assumption \ref{assum id SEM}:\ (i)
	$\operatorname{col}(W)\nsubseteq\operatorname{col}(X)$; (ii) for any nonzero
	real eigenvalue $\omega$ of $W$, $\operatorname{col}(S(\omega^{-1}%
	))\neq\operatorname{col}(S(\omega^{-1})X).$\
	recall Lemma \ref{lemma SSX}:\ $M_{S(\omega^{-1})X}S(\omega^{-1})=0$ iff
	$\operatorname{col}(S(\omega^{-1}))=\operatorname{col}(S(\omega^{-1})X)$
\end{mycomment}

\begin{mycomment}
	INTERESTING: in a spatial error model with with weights matrix $W=I_{R}\otimes
	B_{m}$, $k\geq R$ is sufficient for $l(\lambda)$ to be unbounded from above
	near $\omega=-(m-1)$ for generic $X$ (this is not the case for the SAR model).
	At the moment this appears only in the suppl.
\end{mycomment}

\begin{mycomment}
	\emph{OLD1} It is convenient to generalize model (\ref{SLM}) to a general
	regression model with exogenous regressors and correlated errors, which we
	write as $y=X\beta+u,$ with $\mathrm{E}(u)=0$ and $\mathrm{var}(u)=\sigma
	^{2}\Sigma(\theta)$, where the parameter $\theta$ belongs to some subset
	$\Theta\subseteq\mathbb{R}^{p}$. It is assumed that $\Sigma(\theta)$ is
	positive definite for any $\theta\in\Theta$ and differentiable in $\theta$. We
	decompose $\Sigma^{-1}(\theta)$ as $A^{\prime}(\theta)A(\theta)$ where
	$A(\theta)$ is a (not necessarily symmetric) square root of $\Sigma
	^{-1}(\theta)$. In the spatial error model case, $\Sigma(\lambda)=\left(
	S^{\prime}(\lambda)S(\lambda)\right)  ^{-1}$ and we may take $A(\theta
	)=S(\lambda).$
	
	The (quasi) profile log-likelihood for $\theta$ is
	\begin{equation}
	l(\theta)\coloneqq -\frac{n}{2}\log\left(  \hat{\sigma}_{\mathrm{ML}}^{2}%
	(\theta)\right)  +\log\left\vert \det\left(  A(\theta)\right)  \right\vert ,
	\end{equation}
	with $\hat{\sigma}_{\mathrm{ML}}^{2}(\theta)$ now being given by $(y^{\prime
	}A^{\prime}(\theta)M_{A(\theta)X}A(\theta)y)/n$, with $M_{A(\theta)X}$
	denoting the orthogonal projector onto $\operatorname{col}^{\perp}%
	(A(\theta)X)$. Recentering the profile score and integrating produces the
	adjusted log-likelihood%
	\[
	l_{\mathrm{a}}(\theta)\coloneqq -\frac{n-k}{2}\log\left(  \hat{\sigma}_{\mathrm{ML}%
	}^{2}(\theta)\right)  +\log\left\vert \det\left(  A(\theta)\right)
	\right\vert -\frac{1}{2}\log\left(  \det\left(  X^{\prime}A^{\prime}%
	(\theta)A(\theta)X\right)  \right)  .
	\]
	
	Remarkably, and contrary to the case of a SAR model, $l_{\mathrm{a}}(\theta)$
	is a genuine likelihood, and hence the adjusted profile score provides an
	unbiased and information unbiased estimating equation. In fact, $l_{\mathrm{a}%
	}(\theta)$ corresponds to the density of the maximal invariant under the group
\end{mycomment}

\section{\label{sec Monte Carlo}Simulation evidence}

We conduct Monte Carlo experiments to investigate the performance of the
adjusted MLE of $\lambda$ in the SAR model. First, we consider the case of a
single network, and then the case of the multiple networks model of Section
\ref{sec network f.e. model}. The number of replications is $10^{6}$ in all experiments.

\subsection{\label{sec cross}Single network}

A Watts-Strogatz random graph is formed by rewiring with probability $p$ each
link in a $h$-ahead $h$-behind circular matrix \citep{Watts1998}.\footnote{A
$h$-ahead $h$-behind circular matrix $A$ is the adjacency matrix of a graph
made of $n$ vertices on a circle such that each vertex is linked to its $h$
nearest neighbors on each side. That is, $A(i,j)=1$ if $0<\left\vert
i-j\right\vert \operatorname{mod}(n-1-h)\leq h$), and $A(i,j)=0$ otherwise.}
The extreme cases $p=0$ and $p=1$ correspond, respectively, to a $h$-ahead
$h$-behind circular matrix and to a Erd{\H{o}}s-R{\'{e}}nyi random graph.
Watts-Strogatz graphs are popular in social network analysis because, for
relatively small values of $p$, they can reproduce important characteristics
of many real world networks, including high clustering and small distances
between most nodes. In our first numerical experiment, we take $W$ in model
(\ref{SLM}) to be the normalized adjacency matrix of (one realization of) a
Watts-Strogatz graph. Normalization of $W$ is either a row normalization or a
spectral normalization (by spectral normalization we mean that the matrix is
rescaled by its spectral radius). Note that the two normalizations are
equivalent when $p=0$, due to the fact that a $h$-ahead $h$-behind circular
matrix has constant row sums. The matrix $X$ contains an intercept, $\tilde
{k}$ regressors, and, to allow for contextual effects, the spatially lagged
version of those regressors. That is, $X=(\iota_{n},\widetilde{X}%
,W\widetilde{X})$, where $\widetilde{X}$ is $n\times\tilde{k}$. The matrix
$\widetilde{X}$ is drawn in each repetition; half of its columns are drawn
from independent $\mathrm{N}(0,I_{n})$ distributions, half from independent
uniform distributions on $[0,1]$.\footnote{Given our theoretical framework, it
could be argued that a design with $\widetilde{X}$ fixed across repetitions
would be more appropriate. However, we prefer to vary $\widetilde{X}$ across
repetitions, in an attempt to investigate the performance of the adjusted QMLE
in an environment where $\widetilde{X}$ is random.} For $p>0$, $W$ is drawn
once and then kept constant across repetitions. We set $\beta=\iota
_{2\tilde{k}+1}$, $\sigma=1$, and $n=200$. The errors $\varepsilon_{i}$ are
generated from independent standard normal distributions.

In the present context, $\hat{\lambda}_{\mathrm{ML}}$ is, subject to
regularity conditions, consistent and asymptotically normal as $n\rightarrow
\infty$ \citep{Lee2004}, and $\hat{\lambda}_{\mathrm{ML}}$ and $\hat{\lambda
}_{\mathrm{aML}}$ are first-order asymptotically equivalent \citep{Yu2015}.
Table \ref{Table cross landscape} compares the finite sample performance of
$\hat{\lambda}_{\mathrm{ML}}$ and $\hat{\lambda}_{\mathrm{aML}}$ for a range
of values of $p$, $h$, and $\lambda$, when $\tilde{k}=2$. The columns headed
by $\hat{\lambda}_{\mathrm{ML}}$ and $\hat{\lambda}_{\mathrm{aML}}$ give the
bias(s.d.) of the estimators. $\Delta\%$ stands for percentage change (of the
absolute bias or RMSE) from $\hat{\lambda}_{\mathrm{ML}}$ to $\hat{\lambda
}_{\mathrm{aML}}$. For the case of row normalization, we also report
$\omega_{2}^{-1}$, the percentage of times when $\hat{\lambda}_{\mathrm{aML}%
}>1$, denoted by $\%(\hat{\lambda}_{\mathrm{aML}}>1)$, and bias and s.d. of
$\bar{\lambda}_{\mathrm{aML}}$ (which, recall, is the estimator that is set to
1 when $\hat{\lambda}_{\mathrm{aML}}>1$). According to the results in Section
\ref{sec singular SLM}, in the present setting $\Lambda_{\mathrm{a}}$ must be
the same in each repetition, with its right endpoint being equal to
$\omega_{2}^{-1}$ when $W$ is row normalized, equal to 1 when $W$ is
spectrally normalized (and $p>0$). Thus, $\hat{\lambda}_{\mathrm{aML}}$ may be
greater than $1$ when $W$ is row normalized, while $\hat{\lambda
}_{\mathrm{aML}}<1$ in all repetitions when $W$ is spectrally normalized. The
results in Table \ref{Table cross landscape} show that, when $p=0$,
$\hat{\lambda}_{\mathrm{aML}}$ provides a significant improvement compared to
$\hat{\lambda}_{\mathrm{ML}}$ both in terms of bias and RMSE. As $p$
increases, the reduction in bias afforded by $\hat{\lambda}_{\mathrm{aML}}$
increases but at the cost of greater variability, for both normalizations. The
bias of $\hat{\lambda}_{\mathrm{aML}}$ is small, unless $h$ is large and $p$
is small. Note that $\omega_{2}^{-1}$ increases with $p$ and $h$ (cf. the
paragraph after Example \ref{exa generic}), and $\%(\hat{\lambda
}_{\mathrm{aML}}>1)$ is nondecreasing in $p$, $h,$ and $\lambda$, and can be
very large when $p$, $h,$ and $\lambda$ are large. As $\%(\hat{\lambda
}_{\mathrm{aML}}>1)$ increases, $\bar{\lambda}_{\mathrm{aML}}$ becomes less
biased, and more variable, compared to $\hat{\lambda}_{\mathrm{aML}}$.
\enlargethispage{\baselineskip}
\enlargethispage{\baselineskip}

Table \ref{Table ktilde} analyzes the impact of the number of regressors. We
take $\lambda=.5$ and $h=5$. For these values of $\lambda$ and $h$,
$\mathrm{Pr}(\hat{\lambda}_{\mathrm{aML}}>1)$ is either zero or negligible in
the row normalized case, so, contrary to the previous table, we do not report
$\omega_{2}^{-1}$, $\%(\hat{\lambda}_{\mathrm{aML}}>1)$, and $\bar{\lambda
}_{\mathrm{aML}}$. As expected, the bias reduction afforded by $\hat{\lambda
}_{\mathrm{aML}}$ increases as $\tilde{k}$ increases. As $\tilde{k}$
increases, the relative performance of $\hat{\lambda}_{\mathrm{aML}}$,
compared to $\hat{\lambda}_{\mathrm{ML}}$, improves also in terms of RMSE if
$p$ is not large.

\thispagestyle{empty}\newgeometry{left=3cm, right=1.5cm, top=2cm, bottom=1cm}
\begin{landscape}
	\begin{table}[p]
		
		\captionsetup{width=22.5cm,margin={.9cm,2.6cm}}
		\caption{Comparison of $\hat{\lambda}_{\mathrm{ML}}$ and $\hat{\lambda}_{\mathrm{aML}}$ on a
			Watts-Strogatz network of size $n=200$, with $\tilde{k}=2$ regressors. $\Delta\%$ refers to a percentage change from $\hat{\lambda}_{\mathrm{ML}}$ to $\hat{\lambda
			}_{\mathrm{aML}}$.}		\label{Table cross landscape}	
			\begin{threeparttable}	
			\begin{adjustbox}{width=.95\linewidth}		
				$
				\begin{array}[c]{llllcrcLcLrccrcL}\hline
				&  & && \multicolumn{7}{l}{\text{Row normalization}} &  & \multicolumn{4}{l}{\text{Spectral
						normalization}}\\
				\cline{5-11}\cline{13-16}
				 &  & && \hat{\lambda}_{\mathrm{ML}} & \multicolumn{1}{c}{\hat{\lambda
					}_{\mathrm{aML}}} &  &  &  & &
				\mc{\bar{\lambda}_{\mathrm{aML}}} & &		
				\hat{\lambda}_{\mathrm{ML}} & \multicolumn{1}{c}{\hat{\lambda
					}_{\mathrm{aML}}} &  & \bigstrut[t]\\
				p & h & \lambda && \mc{\text{bias(s.d.)}} &\mc{\text{bias(s.d.)}} & \Delta\%\left\vert \text{bias}\right\vert  & \mc{\Delta\%\mathrm{RMSE}} & \omega_{2}^{-1} & \mc{\%(\hat{\lambda}_{\mathrm{aML}}>1)} &
				\mc{\text{bias(s.d.)}} & &		
				\mc{\text{bias(s.d.)}} & \mc{\text{bias(s.d.)}} & \Delta\%\left\vert \text{bias}\right\vert  & \mc{\Delta\%\mathrm{RMSE}}\\
				\hline
				
				\multicolumn{1}{l}{0} & \multicolumn{1}{l}{5} & \multicolumn{1}{l}{0} &&-0.072(0.161)	&-0.020(0.160)	&-72.27	&-8.80	&1.01	&0.00	&-0.020(0.160)&&-0.072(0.161)	&-0.020(0.160)	&-72.27	&-8.80	\bigstrut[t] \\
				\multicolumn{1}{l}{} & \multicolumn{1}{l}{} & \multicolumn{1}{l}{0.5} &&-0.046(0.095)	&-0.015(0.094)	&-68.09	&-10.58	&1.01	&0.00	&-0.015(0.094)&& -0.046(0.095)	&-0.015(0.094)	&-68.09	&-10.58	\\
				\multicolumn{1}{l}{} & \multicolumn{1}{l}{} & \multicolumn{1}{l}{0.9} &&-0.013(0.025)	&-0.005(0.024)	&-61.17	&-12.09	&1.01	&0.00	&-0.005(0.024)&& -0.013(0.025)	&-0.005(0.024)	&-61.17	&-12.09 \\
				\multicolumn{1}{l}{} & \multicolumn{1}{l}{10} & \multicolumn{1}{l}{0} &&-0.168(0.253)	&-0.048(0.246)	&-71.24	&-17.50	&1.02	&0.00	&-0.048(0.246)&& -0.168(0.253)	&-0.048(0.246)	&-71.24	&-17.50\\
				\multicolumn{1}{l}{} & \multicolumn{1}{l}{} & \multicolumn{1}{l}{0.5} &&-0.105(0.153)	&
				0.034(0.145)	&-68.07	&-19.51	&1.02	&0.00	&-0.034(0.145)&& -0.105(0.153)	&-0.034(0.145)	&-68.07	&-19.51	\\
				\multicolumn{1}{l}{} & \multicolumn{1}{l}{} & \multicolumn{1}{l}{0.9} &&-0.032(0.045)	&-0.012(0.042)	&-62.49	&-21.12	&1.02	&0.00	&-0.012(0.042)&& -0.032(0.045)	&-0.012(0.042)	&-62.49	&-21.12	\\
				\multicolumn{1}{l}{} & \multicolumn{1}{l}{50} & \multicolumn{1}{l}{0} &&-1.150(0.734)	&-0.215(0.821)	&-81.33	&-37.78	&1.60	&3.58	&-0.220(0.813)&& -1.150(0.734)	&-0.215(0.821)	&-81.33	&-37.78	\\
				\multicolumn{1}{l}{} & \multicolumn{1}{l}{} & \multicolumn{1}{l}{0.5} &&-1.068(0.718)	&-0.254(0.696)	&-76.24	&-42.43	&1.60	&10.90	&-0.270(0.676)&& -1.068(0.718)	&-0.254(0.696)	&-76.24	&-42.43	\\
				\multicolumn{1}{l}{} & \multicolumn{1}{l}{} & \multicolumn{1}{l}{0.9} &&-0.849(0.609)	&-0.222(0.540)	&-73.81	&-44.11	&1.60	&31.36	&-0.277(0.490)&& -0.849(0.609)	&-0.222(0.540)	&-73.81	&-44.11\\
				\\[-2.5mm]
				\multicolumn{1}{l}{0.2} & \multicolumn{1}{l}{5} & \multicolumn{1}{l}{0}&&-0.045(0.151)	&-0.010(0.153)	&-78.84	&-2.91	&1.15	&0.00	&-0.010(0.153)&&-0.044(0.149)	&-0.009(0.151)	&-79.46	&-2.68	\\
				\multicolumn{1}{l}{} & \multicolumn{1}{l}{} & \multicolumn{1}{l}{0.5} &&-0.044(0.107)	&-0.013(0.108)	&-71.27	&-6.30	&1.15	&0.00	&-0.013(0.108)&& -0.039(0.101)	&-0.011(0.102)	&-72.62	&-5.42	\\
				\multicolumn{1}{l}{} & \multicolumn{1}{l}{} & \multicolumn{1}{l}{0.9}&&-0.035(0.050)	&-0.009(0.053)	&-74.54	&-12.14	&1.15	&0.59	&-0.009(0.053)&& -0.017(0.034)	&-0.004(0.036)	&-78.08	&-5.68	\\
				\multicolumn{1}{l}{} & \multicolumn{1}{l}{10} & \multicolumn{1}{l}{0}&&-0.099(0.221)	&-0.022(0.226)	&-78.03	&-6.29	&1.22	&0.00	&-0.022(0.226)&& -0.092(0.214)		&-0.019(0.219)	&-79.31	&-5.73	\\
				\multicolumn{1}{l}{} & \multicolumn{1}{l}{} & \multicolumn{1}{l}{0.5}&&-0.096(0.164)	&-0.027(0.166)	&-71.55	&-11.47	&1.22	&0.00	&-0.027(0.166)&&-0.079(0.148)	&-0.019(0.151)	&-75.36	&-9.28	\\
				\multicolumn{1}{l}{} & \multicolumn{1}{l}{} & \multicolumn{1}{l}{0.9}&&-0.082(0.087)	&-0.021(0.093)	&-74.06	&-20.30	&1.22	&6.49	&-0.023(0.090)&& -0.033(0.048)	&-0.005(0.053)	&-84.01	&-8.67	\\
				\multicolumn{1}{l}{} & \multicolumn{1}{l}{50} & \multicolumn{1}{l}{0}&&-0.563(0.582)	&-0.071(0.702)	&-87.40	&-12.77	&2.34	&5.22	&-0.081(0.684)&& -0.491(0.542)	&-0.061(0.640)	&-87.60	&-12.16	\\
				\multicolumn{1}{l}{} & \multicolumn{1}{l}{} & \multicolumn{1}{l}{0.5}&&-0.661(0.514)	&-0.108(0.637)	&-83.65	&-22.85	&2.34	&17.49	&-0.150(0.581)&& -0.467(0.406)	&-0.083(0.474)	&-82.31	&-22.23	\\
				\multicolumn{1}{l}{} & \multicolumn{1}{l}{} & \multicolumn{1}{l}{0.9}&&-0.737(0.445)	&-0.137(0.566)	&-81.47	&-32.44	&2.34	&38.25	&-0.251(0.451)&&  -0.164(0.126)	&-0.036(0.141)	&-78.17	&-29.41	\\
				\\[-2.5mm]
				\multicolumn{1}{l}{0.5} & \multicolumn{1}{l}{5} & \multicolumn{1}{l}{0}&&-0.028(0.151)	&-0.002(0.154)	&-92.77	& 0.59	&1.49	&0.00	&-0.002(0.154)&& -0.026(0.146)	&-0.002(0.149)	&-92.81	& 0.61	\\
				\multicolumn{1}{l}{} & \multicolumn{1}{l}{} & \multicolumn{1}{l}{0.5}&&-0.044(0.127)	&-0.008(0.132)	&-82.44	&-1.43	&1.49	&0.00	&-0.008(0.132)&& -0.031(0.107)	&-0.005(0.111)	&-82.94	&-0.90	\\
				\multicolumn{1}{l}{} & \multicolumn{1}{l}{} & \multicolumn{1}{l}{0.9}&&-0.077(0.078)	&-0.010(0.098)	&-86.53	&-9.95	&1.49	&12.46	&-0.015(0.090)&& -0.013(0.032)	&-0.002(0.034)	&-85.61	&-0.57	\\
				\multicolumn{1}{l}{} & \multicolumn{1}{l}{10} & \multicolumn{1}{l}{0}&&-0.064(0.218)	&-0.005(0.229)	&-91.37	& 0.66	&1.80	&0.00	&-0.005(0.229)&&  -0.056(0.207)	&-0.005(0.216)	&-91.90	& 0.73	\\
				\multicolumn{1}{l}{} & \multicolumn{1}{l}{} & \multicolumn{1}{l}{0.5}&&-0.098(0.186)	&-0.014(0.205)	&-85.97	&-2.21	&1.80	&0.27	&-0.014(0.205)&&  -0.066(0.155)	&-0.008(0.168)	&-87.27	&-0.66	\\
				\multicolumn{1}{l}{} & \multicolumn{1}{l}{} & \multicolumn{1}{l}{0.9}&&-0.161(0.125)	&-0.020(0.171)	&-87.87	&-15.89	&1.80	&25.19	&-0.041(0.143)&& -0.028(0.047)	&-0.003(0.053)	&-89.61	&-2.84	\\
				\multicolumn{1}{l}{} & \multicolumn{1}{l}{50} & \multicolumn{1}{l}{0}&&-0.435(0.540)	&-0.010(0.708)	&-97.64	& 2.15	&4.37	&7.69	&-0.033(0.665)&&  -0.352(0.492)	&-0.019(0.600)	&-94.71	&-0.80	\\
				\multicolumn{1}{l}{} & \multicolumn{1}{l}{} & \multicolumn{1}{l}{0.5}&&-0.585(0.469)	&-0.021(0.698)	&-96.38	&-6.75	&4.37	&23.23	&-0.111(0.573)&&  -0.349(0.351)	&-0.045(0.430)	&-87.09	&-12.84	\\
				\multicolumn{1}{l}{} & \multicolumn{1}{l}{} & \multicolumn{1}{l}{0.9}&&-0.742(0.399)	&-0.031(0.684)	&-95.86	&-18.71	&4.37	&43.44	&-0.243(0.454)&&  -0.112(0.099)	&-0.021(0.114)	&-81.04	&-22.26	\\
				\\[-2.5mm]
				\multicolumn{1}{l}{1} & \multicolumn{1}{l}{5} & \multicolumn{1}{l}{0}&&-0.024(0.151)	&-0.000(0.155)	&-99.25	& 1.32	&1.76	&0.00	&-0.000(0.155)&& -0.021(0.146)	&-0.000(0.149)	&-99.45	& 1.26	\\
				\multicolumn{1}{l}{} & \multicolumn{1}{l}{} & \multicolumn{1}{l}{0.5}&&-0.044(0.133)	&-0.005(0.140)	&-87.73	& 0.48	&1.76	&0.00	&-0.005(0.140)&&  -0.029(0.110)	&-0.004(0.114)	&-85.40	& 0.39	\\
				\multicolumn{1}{l}{} & \multicolumn{1}{l}{} & \multicolumn{1}{l}{0.9}&&-0.092(0.085)	&-0.008(0.115)	&-91.07	&-8.02	&1.76	&17.42	&-0.018(0.101)&& -0.012(0.031)	&-0.002(0.033)	&-84.70	&-0.18	\\
				\multicolumn{1}{l}{} & \multicolumn{1}{l}{10} & \multicolumn{1}{l}{0}&&-0.052(0.219)	& 0.000(0.231)	&-99.87	& 2.74	&2.45	&0.00	& 0.000(0.231)&& -0.043(0.204)	& 0.001(0.214)	&-97.97	& 2.59	\\
				\multicolumn{1}{l}{} & \multicolumn{1}{l}{} & \multicolumn{1}{l}{0.5}&&-0.097(0.191)	&-0.006(0.220)	&-93.57	& 2.60	&2.45	&0.84	&-0.007(0.219)&& -0.057(0.153)	&-0.003(0.168)	&-94.75	& 2.73	\\
				\multicolumn{1}{l}{} & \multicolumn{1}{l}{} & \multicolumn{1}{l}{0.9}&&-0.186(0.133)	&-0.010(0.201)	&-94.38	&-11.71	&2.45	&29.94	&-0.046(0.156)&& -0.023(0.044)	&-0.002(0.049)	&-91.20	&-0.78	\\
				\multicolumn{1}{l}{} & \multicolumn{1}{l}{50} & \multicolumn{1}{l}{0}&&-0.415(0.531)	& 0.002(0.710)	&-99.59	& 5.34	&7.27	&8.04	&-0.024(0.661)&&  -0.303(0.465)	&-0.009(0.565)	&-97.19	& 1.92	\\
				\multicolumn{1}{l}{} & \multicolumn{1}{l}{} & \multicolumn{1}{l}{0.5}&&-0.572(0.459)	&-0.005(0.709)	&-99.08	&-3.38	&7.27	&23.99	&-0.104(0.570)&&  -0.276(0.314)	&-0.028(0.383)	&-89.81	&-8.22	\\
				\multicolumn{1}{l}{} & \multicolumn{1}{l}{} & \multicolumn{1}{l}{0.9}&&-0.739(0.390)	&-0.011(0.704)	&-98.53	&-15.67	&7.27	&44.14	&-0.240(0.452)&&  -0.079(0.081)	&-0.012(0.094)	&-84.39	&-16.43\bigstrut[b]\\
				\hline
				
				\end{array}
				$
				
			\end{adjustbox}
		
		\end{threeparttable}
		
	\end{table}
\end{landscape}
\restoregeometry

\begin{mycomment}
	\emph{table 1. done with }bias\_MLE\_SLM\_cross\_sect\_watts\_strogatz\_X\_in\_each\_rep.m
to fix X use bias_MLE_SLM_cross_sect_watts_strogatz_fixed_X.m
	\emph{table 2 OK. done with }bias\_MLE\_SLM\_cross\_sect\_watts\_strogatz\_X\_in\_each\_rep\_k\_varies.m
\end{mycomment}

\begin{mycomment}
	- different n, $\lambda$, error distrib in suppl material (recall results
	depend on $\beta$ and $\sigma$ through $\beta/\sigma$)
\end{mycomment}

\begin{mycomment}
	remember $l_{\mathrm{a}}(\lambda)$ single peaked over $\Lambda_{\mathrm{a}}$ whenever $W$ is
	symm. Also check condition $\delta_{\mathrm{a}}(\lambda)<0$ for row-normalized cases
\end{mycomment}

\begin{mycomment}
	...............say results for beta are pretty much the same for MLe and adjMLE
\end{mycomment}

\begin{mycomment}
	For ci use
	\emph{saddlepoint\_ci\_simulation\_cross\_section\_Watts\_strogatz.m}
	In this model it doesn't look like adj lik is useful in general for CI as the
	saddlepoint CI with corrected (using n/(n-k)) sigma\symbol{94}2MLE work very
	well quite generally. Cases when adj lik CI are better than saddlepoint CI
	with corrected (using n/(n-k)) sigma\symbol{94}2MLE are when there are context
	eff and k is large and h is large - see
	output\_cross\_section\_large\_k\_h\_table.txt produced using saddlepoint\_ci\_simulation\_cross\_section\_varying\_h\_fixed\_n.m
\end{mycomment}
\begin{mycomment}
RECALL bias depend on nuis param
\end{mycomment}

\subsection{\label{secd network f.e.}Multiple networks}

We now generate data according to model (\ref{network model}). For simplicity,
we consider a balanced case: each of the $R$ networks has the same number,
$m$, of individuals. The matrices $W_{r}$ and $\widetilde{X}$ are generated as
in Section \ref{sec cross}. That is, each $W_{r}$ is the normalized adjacency
matrix of (a realization of) a Watts-Strogatz graph, and the regressors are
drawn in each repetition, $\tilde{k}/2$ of them from a standard normal
distribution, and the other $\tilde{k}/2$ from a $\mathrm{Unif}(0,1)$
distribution. Errors and the fixed effects $\alpha_{r}$ are drawn (in each
repetition) from $\mathrm{N}(0,1)$ distributions, and we set $\gamma
=\delta=\iota_{\tilde{k}}$ and $\sigma=1.$\footnote{The simulated model is one
in which fixed effects and regressors are random and independent of each
other, and $W$ is non-stochastic. Note that we could allow for some
correlation between fixed effects and regressors, but this is not needed for
our purposes.}

Table \ref{Table panel} compares $\hat{\lambda}_{\mathrm{aML}}$ to the
estimator $\hat{\lambda}_{\mathrm{LLL}}$ obtained by maximizing the
\cite{LeeLiuLin2010} likelihood $l_{\mathrm{LLL}}(\sigma^{2},\lambda)$, for a
range of values of $R$, $m$, and $\tilde{k}$.\footnote{Some results for larger
$R$ are given in Table \ref{Table panel R=100} in the Supplement.} We choose
$\lambda=0.5$, and set the parameters of the Watts-Strogatz random graph to
$h=5$ and $p=0.2$. When $W$ is row normalized, $\hat{\lambda}_{\mathrm{aML}}$
performs similarly to $\hat{\lambda}_{\mathrm{LLL}}$.\footnote{Note that, as
for the design in Section \ref{sec cross}, in the present setting the set
$\Lambda_{\mathrm{a}}$ is the same in all repetitions.} In fact, it has lower
bias (but note that the bias of $\hat{\lambda}_{\mathrm{LLL}}$ is already very
small in most of the cases considered in the table) and slightly slower RMSE.
As expected, the relative performance of $\hat{\lambda}_{\mathrm{LLL}}$
improves as $\tilde{k}$ increases.\footnote{Similarly to the cross-sectional
case, the reduction in RMSE is higher for lower values of $p$ (for example, in
the extreme case $p=0$, $\Delta\%\text{RMSE}$ is -4.417, -14.707, -24.278 when
$\tilde{k}$ is $2,6,10$, respectively, and $R=10$, $m=20$).} When $W$ is
spectrally normalized, $\hat{\lambda}_{\mathrm{LLL}}$ cannot be obtained, so
only results for $\hat{\lambda}_{\mathrm{aML}}$ are reported. The simulation
results show that $\hat{\lambda}_{\mathrm{aML}}$ performs very satisfactorily
even in that case.

\begin{table}[pth]
\caption{Comparison of $\hat{\lambda}_{\mathrm{ML}}$ and $\hat{\lambda
}_{\mathrm{aML}}$ on a Watts-Strogatz network of size $n=200$, with
$\lambda=.5$ and $h=5$. $\Delta\%$ refers to a percentage change from
$\hat{\lambda}_{\mathrm{ML}}$ to $\hat{\lambda}_{\mathrm{aML}}$.}%
\label{Table ktilde}
\captionsetup{width=23.5cm} \begin{threeparttable}	
		\begin{adjustbox}{width=1\textwidth,center=\textwidth}		
			$
			\begin{array}[c]{lllcccLccccL}
			\hline
			&                        &&                                           \multicolumn{4}{l}{\text{Row normalization}}                                            &  &                                             \multicolumn{4}{l}{\text{Spectral normalization}}             \\
			\cline{4-7}\cline{9-12}
				 &            && \hat{\lambda}_{\mathrm{ML}} & \hat{\lambda}_{\mathrm{aML}} &  &  &  & \hat{\lambda}_{\mathrm{ML}} & \hat{\lambda
			}_{\mathrm{aML}} & & \bigstrut[t]\\
			p & \tilde{k}              && \mc{\text{bias(s.d.)}} & \mc{\text{bias(s.d.)}} & \Delta\%\left\vert \text{bias}\right\vert & \mc{\Delta\%\text{RMSE}} &  & \mc{\text{bias(s.d.)}} & \mc{\text{bias(s.d.)}} & \Delta\%\left\vert \text{bias}\right\vert & \mc{\Delta\%\text{RMSE}} \\ \hline
			\multicolumn{1}{l}{0}    & \multicolumn{1}{l}{2}  && -0.053(0.104)&	-0.017(0.101)&	-67.29	&		-12.32	&&	-0.053(0.104)&	-0.017(0.101)&	-67.29	&		-12.32	\bigstrut[t] \\
			\multicolumn{1}{l}{}    & \multicolumn{1}{l}{6}  && -0.072(0.095)&	-0.013(0.092)&	-81.54	&		-22.13	&&	-0.072(0.095)&	-0.013(0.092)&	-81.54	&		-22.13	\\
			\multicolumn{1}{l}{}    & \multicolumn{1}{l}{10} && -0.086(0.091)&	-0.011(0.088)&	-87.22	&		-29.41	&&	-0.086(0.091)&	-0.011(0.088)&	-87.22	&		-29.41	 \\
			\\[-3.5mm]
			\multicolumn{1}{l}{0.2}   & \multicolumn{1}{l}{2}  && -0.050(0.116)&	-0.015(0.117)&	-71.09	&		-7.09	&&	-0.038(0.102)&	-0.010(0.103)&	-72.47	&		-4.96	 \\
			\multicolumn{1}{l}{}    & \multicolumn{1}{l}{6}  && -0.055(0.104)&	-0.011(0.104)&	-80.26	&		-10.97	&&	-0.039(0.087)&	-0.007(0.088)&	-82.03	&		-7.71	\\
			\multicolumn{1}{l}{}    & \multicolumn{1}{l}{10} && -0.059(0.096)&	-0.009(0.097)&	-84.55	&		-13.82	&&	-0.039(0.078)&	-0.005(0.078)&	-86.65	&		-9.80 \\
			\\[-3.5mm]
			\multicolumn{1}{l}{0.5}   & \multicolumn{1}{l}{2}  &&-0.049(0.132)&	-0.009(0.138)&	-82.39&			-1.72	&&-0.028(0.103)&	-0.005(0.106)&	-81.88&			-0.48	\\
			\multicolumn{1}{l}{}    & \multicolumn{1}{l}{6}  && -0.048(0.123)&	-0.008(0.128)&	-84.19	&		-2.83	&&-0.022(0.086)&	-0.003(0.088)&	-84.18	&		-0.85\\
			\multicolumn{1}{l}{}    & \multicolumn{1}{l}{10} && -0.048(0.116)&	-0.006(0.121)&	-86.38	&		-3.43	&&-0.018(0.074)&	-0.002(0.075)&	-86.07	&		-1.19	\\
			\\[-3.5mm]
			\multicolumn{1}{l}{1}    & \multicolumn{1}{l}{2}  &&  -0.047(0.137)&	-0.006(0.146)&	-88.13	&		 0.56	&&-0.025(0.105)&	-0.003(0.109)&	-87.54	&		 0.61	\\
			\multicolumn{1}{l}{}    & \multicolumn{1}{l}{6}  && -0.043(0.129)& -0.005(0.137)&	-88.03	&		 0.64	&&-0.018(0.088)&	-0.002(0.090)&	-88.64	&		 0.56	\\
			\multicolumn{1}{l}{}    & \multicolumn{1}{l}{10} && -0.041(0.124)&	-0.005(0.131)&	-88.62	&		 0.47	&&-0.014(0.075)&	-0.002(0.077)&	-88.69	&		 0.47	 \bigstrut[b]\\ \hline
			\end{array}
			$
			
		\end{adjustbox}
	
	\end{threeparttable}
\end{table}

\begin{table}[tbh]
\caption{Model (\ref{network model}) with $\lambda=0.5$, $h=5$ and $p=0.2$.
$\Delta\%$ refers to a percentage change from $\hat{\lambda}_{\mathrm{LLL}}$
to $\hat{\lambda}_{\mathrm{aML}}$.}%
\label{Table panel}
\captionsetup{width=23.5cm} \begin{threeparttable}	
		\begin{adjustbox}{width=.85\textwidth,center=\textwidth}
			$		
			\begin{array}[c]{llllcccLcc}\hline
			&  &  & &\multicolumn{4}{l}{\text{Row normalization}} &  & \text{Spectral normalization}\\\cline{5-8}%
			\cline{10-10}%
			 &  &  && \hat{\lambda}_{\mathrm{LLL}} &
			\hat{\lambda}_{\mathrm{aML}} &
			& &  & \hat{\lambda}_{\mathrm{aML}}\bigstrut[t]\\
			
			\tilde{k} & R & m && 	\mc{\text{bias(s.d.)}} &
				\mc{\text{bias(s.d.)}} & \Delta\%\left\vert \text{bias}\right\vert
			& \mc{\Delta\%\text{RMSE}} &  & 	\mc{\text{bias(s.d.)}}\\\hline

			\multicolumn{1}{l}{2}& \multicolumn{1}{l}{10} & \multicolumn{1}{l}{20} && -0.032(0.209)&	-0.016(0.210)& -49.93	&	-0.58 &&-0.007(0.170) \bigstrut[t]\\
			\multicolumn{1}{l}{} & \multicolumn{1}{l}{}   & \multicolumn{1}{l}{30} && -0.021(0.130)&	-0.011(0.130)& -49.68	&	-0.94&&-0.007(0.111)\\
			\multicolumn{1}{l}{} & \multicolumn{1}{l}{20} & \multicolumn{1}{l}{20} && -0.015(0.149)&	-0.008(0.149)&-48.86	&	-0.08&&-0.004(0.119)\\
			\multicolumn{1}{l}{} & \multicolumn{1}{l}{}   & \multicolumn{1}{l}{30} &&  -0.011(0.091)&-0.006(0.090)&	-46.48	&	-0.60&&-0.003(0.078)\\
			\multicolumn{1}{l}{} & \multicolumn{1}{l}{30} & \multicolumn{1}{l}{20} &&-0.010(0.120)&	-0.005(0.120)&	-48.23	&	-0.25&&-0.002(0.098)\\
			\multicolumn{1}{l}{} & \multicolumn{1}{l}{}   & \multicolumn{1}{l}{30} &&  -0.007(0.073)&-0.004(0.073)&	-44.79	&	-0.40&&-0.003(0.065)\\
			\\[-2.5mm]
			\multicolumn{1}{l}{6}& \multicolumn{1}{l}{10} & \multicolumn{1}{l}{20} && -0.053(0.200)&	-0.014(0.202)&	-73.94	&	-2.31 &&-0.005(0.141)\\
			\multicolumn{1}{l}{} & \multicolumn{1}{l}{}   & \multicolumn{1}{l}{30}&& -0.033(0.117)&	-0.010(0.117)& -70.68	&	-2.98 &&-0.004(0.093)\\
			\multicolumn{1}{l}{} & \multicolumn{1}{l}{20} & \multicolumn{1}{l}{20} &&-0.025(0.140)&	-0.006(0.141)&	-74.59	&	-0.87 &&-0.002(0.098)\\
			\multicolumn{1}{l}{} & \multicolumn{1}{l}{}   & \multicolumn{1}{l}{30} &&-0.016(0.081)&	-0.005(0.081)&	-70.50	&	-1.29 &&-0.002(0.065)\\
			\multicolumn{1}{l}{} & \multicolumn{1}{l}{30} & \multicolumn{1}{l}{20} &&-0.017(0.113)&	-0.004(0.114)&	-76.75	&	-0.55&&-0.002(0.080)\\
			\multicolumn{1}{l}{} & \multicolumn{1}{l}{}   & \multicolumn{1}{l}{30} && -0.011(0.066)&	-0.003(0.066)&	-69.45	&	-0.80 &&-0.002(0.053)\\
			\\[-2.5mm]
			\multicolumn{1}{l}{10}& \multicolumn{1}{l}{10}& \multicolumn{1}{l}{20} && -0.070(0.196)&	-0.013(0.199)&	-81.28	&	-4.12 &&-0.004(0.128)\\
			\multicolumn{1}{l}{} & \multicolumn{1}{l}{}   & \multicolumn{1}{l}{30} &&-0.041(0.110)&	-0.008(0.111)&	-80.24	&	-5.92 &&-0.005(0.084)\\
			\multicolumn{1}{l}{} & \multicolumn{1}{l}{20} & \multicolumn{1}{l}{20} &&-0.033(0.133)&	-0.005(0.134)&	-83.58	&	-2.30 &&-0.002(0.086)\\
			\multicolumn{1}{l}{} & \multicolumn{1}{l}{}   & \multicolumn{1}{l}{30} &&-0.020(0.074)&	-0.005(0.074)&	-77.00	&	-3.12&& -0.002(0.057)\\
			\multicolumn{1}{l}{} & \multicolumn{1}{l}{30} & \multicolumn{1}{l}{20} && -0.023(0.106)&	-0.004(0.106)&	-81.81	&	-2.00 &&-0.001(0.069)\\
			\multicolumn{1}{l}{} & \multicolumn{1}{l}{}   & \multicolumn{1}{l}{30} && -0.013(0.059)&	-0.003(0.059)&	-79.55	&	-2.36 &&-0.001(0.046) \bigstrut[b]\\
			\hline
			\end{array}
			$
		\end{adjustbox}
	
	\end{threeparttable}
\end{table}

\begin{mycomment}
	-included context effects as this gives more pronounced diff MLE / adj MLE
	- genereting regressors as all normal or half normal half unif makes little
	difference
	-ALSO GIVE GAMMA as in the simul for indiv fixed eff in MLE paper
\end{mycomment}

\begin{mycomment}
	\emph{table 3 done with }bias\_MLE\_panel\_table\_X\_in\_each\_rep\_k\_varies.m
\end{mycomment}

Table \ref{Table network fe CI} reports empirical coverages of Wald confidence
intervals based on first-order asymptotic normality of $\hat{\lambda
}_{\mathrm{LLL}}$ (in the columns headed by $\mathrm{W}_{\mathrm{LLL}}$),
empirical coverages of Wald confidence intervals based on first-order
asymptotic normality of $\hat{\lambda}_{\mathrm{aML}}$ (in the columns headed
by $\mathrm{W}_{\mathrm{aML}}$), and empirical coverages of saddlepoint
confidence intervals based on $\hat{\lambda}_{\mathrm{aML}}$ (in the columns
headed by $\mathrm{s}_{\mathrm{aML}}$). Since $\hat{\lambda}_{\mathrm{LLL}}$
is not available when $W$ is spectrally normalized, we only report the case of
row normalized $W$.\footnote{When $W$ is spectrally normalized, coverages of
the confidence intervals based on $\hat{\lambda}_{\mathrm{aML}}$ are similar
to the case of row normalization.} The nominal size is 95\%. We consider
equi-tailed two-sided confidence intervals, and right-sided confidence
intervals of the form $(-\infty,\lambda_{U})$, where $\lambda_{U}$ is a
suitably selected upper end-point.\footnote{The 95\% Wald confidence intervals
are $\hat{\lambda}\pm1.96\sqrt{\hat{v}}$ (two-sided) and $(-\infty
,\hat{\lambda}+1.645\sqrt{\hat{v}})$ (right-sided), where $\hat{\lambda}$ is
either $\hat{\lambda}_{\mathrm{LLL}}$ or $\hat{\lambda}_{\mathrm{aMLE}}$, and
$\hat{v}$ denotes the asymptotic variance given in Proposition 6.1 of
\cite{LeeLiuLin2010} and evaluated at the LLL or aMLE estimates of
$\lambda,\beta,\sigma^{2}$.} The errors $\varepsilon_{ri}$ are generated
independently from either (a) a standard normal distribution, (b) a gamma
distribution with shape parameter 1 and scale parameter 1, demeaned by the
population mean. Mean, variance, skewness, and kurtosis are $0,1,0,3$ in case
(a) and $0,1,2,9$ in case (b). The Wald confidence intervals based on
$\hat{\lambda}_{\mathrm{aML}}$ offer an improvement over Wald confidence
intervals based on $\hat{\lambda}_{\mathrm{LLL}}$ when $\tilde{k}$ is not too
small, and particularly in the case of right sided confidence intervals. The
Lugannani--Rice approximation delivers a further improvement, and indeed the
coverages of the saddlepoint confidence intervals based on $\hat{\lambda
}_{\mathrm{aML}}$ are excellent in all cases considered in the table, even
under the gamma distribution. Conversely, the empirical coverage of the Wald
confidence intervals based on $\hat{\lambda}_{\mathrm{LLL}}$ is acceptable in
the two-sided case when $R=m=30$, but quickly deteriorates as $\tilde{k}$
increases or $n$ decreases, and is considerably worse in the right-sided case.
Of course, one minus a coverage in the table gives the size of the test for
$\lambda=0$ obtained by inverting the confidence interval. Table
\ref{Table other distrib} in the Supplement reports coverages under other
severely non-normal distributions; again, the saddlepoint confidence intervals
based on the adjusted QMLE are very accurate in all cases considered.

\begin{table}[tbh]
\caption{Empirical coverages of 95\% confidence intervals in model
(\ref{network model}) with $\lambda=0$, $h=5$, and $p=0.2$. The error
distribution is either a standard normal or a centered $\mathrm{gamma}(1,1)$,
and $W$ is row normalized.}%
\label{Table network fe CI}%
\captionsetup{width=23.5cm}
\begin{adjustbox}{width=.9\textwidth,center=\textwidth}
		$
		\begin{array}[c]{lllcccccccccccccccc}\hline
		&  &  && \multicolumn{7}{l}{\text{Normal}} &  & \multicolumn{7}{l}{\text{Gamma}}\bigstrut[b]\bigstrut[t]\\
		\cline{5-11}\cline{13-19}
		&  &  && \multicolumn{3}{l}{\text{Two-sided}} &  & \multicolumn{3}{l}{\text{Right-sided}}&  & \multicolumn{3}{l}{\text{Two-sided}} &  & \multicolumn{3}{l}{\text{Right-sided}}\bigstrut[b]\bigstrut[t]\\
		
		\cline{5-7}	\cline{9-11}\cline{13-15}\cline{17-19}
		\tilde{k} & R & m &&
		\mathrm{W_{LLL}} & \mathrm{W_{aML}} & \mathrm{s_{aML}} && \mathrm{W_{LLL}} & \mathrm{W_{aML}} & \mathrm{s_{aML}}&&\mathrm{W_{LLL}} & \mathrm{W_{aML}} & \mathrm{s_{aML}}&& \mathrm{W_{LLL}} & \mathrm{W_{aML}} & \mathrm{s_{aML}}\bigstrut[b]\bigstrut[t]\\\hline
		\multicolumn{1}{l}{2}& \multicolumn{1}{l}{10} & \multicolumn{1}{l}{20} &&0.942&	0.941&	0.949&&		0.935&	0.941&	0.949&&0.945&	0.944&	0.951&&		0.939&	0.943&	0.951     \bigstrut[t]\\
		\multicolumn{1}{l}{} & \multicolumn{1}{l}{}   & \multicolumn{1}{l}{30} &&0.946&	0.945&	0.950&&		0.938&	0.945&	0.949&&0.946&	0.945&	0.950&&		0.940&	0.946&	0.949   \\
		\multicolumn{1}{l}{} & \multicolumn{1}{l}{20} & \multicolumn{1}{l}{20} &&0.946&	0.946&	0.950&&		0.940&	0.944&	0.950&&0.948&	0.948&	0.952&&		0.942&	0.946&	0.951   \\
		\multicolumn{1}{l}{} & \multicolumn{1}{l}{}   & \multicolumn{1}{l}{30} &&0.947&	0.947&	0.950&&		0.942&	0.946&	0.950&&0.949&	0.949&	0.951&&		0.943&	0.948&	0.950  \\
		\multicolumn{1}{l}{} & \multicolumn{1}{l}{30} & \multicolumn{1}{l}{20} &&0.947&	0.947&	0.950&&		0.943&	0.946&	0.950&&0.949&	0.949&	0.951&&		0.945&	0.948&	0.950    \\
		\multicolumn{1}{l}{} & \multicolumn{1}{l}{}   & \multicolumn{1}{l}{30} &&0.948&	0.947&	0.949&&		0.944&	0.948&	0.949&&0.949&	0.949&	0.951&&		0.945&	0.949&	0.949    \\
		\\[-2.5mm]
		\multicolumn{1}{l}{6}& \multicolumn{1}{l}{10} & \multicolumn{1}{l}{20} &&0.933&	0.933&	0.948&&		0.916&	0.934&	0.948&&0.935&	0.935&	0.950&&		0.916&	0.936&	0.951   \\
		\multicolumn{1}{l}{} & \multicolumn{1}{l}{}   & \multicolumn{1}{l}{30} &&0.937&	0.938&	0.948&&		0.919&	0.941&	0.949&&0.939&	0.940&	0.949&&		0.924&	0.943&	0.950   \\
		\multicolumn{1}{l}{} & \multicolumn{1}{l}{20} & \multicolumn{1}{l}{20} &&0.943&	0.942&	0.949&&		0.932&	0.942&	0.949&&0.943&	0.943&	0.950&&		0.930&	0.942&	0.949     \\
		\multicolumn{1}{l}{} & \multicolumn{1}{l}{}   & \multicolumn{1}{l}{30} &&0.944&	0.944&	0.949&&		0.931&	0.945&	0.949&&0.944&	0.945&	0.950&&		0.930&	0.945&	0.950   \\
		\multicolumn{1}{l}{} & \multicolumn{1}{l}{30} & \multicolumn{1}{l}{20} &&0.944&	0.944&	0.949&&		0.934&	0.944&	0.949&&0.946&	0.945&	0.950&&		0.934&	0.944&	0.951     \\
		\multicolumn{1}{l}{} & \multicolumn{1}{l}{}   & \multicolumn{1}{l}{30} &&0.945&	0.945&	0.948&&		0.934&	0.945&	0.949&&0.946&	0.947&	0.950&&		0.934&	0.947&	0.950    \\
		\\[-2.5mm]
		\multicolumn{1}{l}{10}& \multicolumn{1}{l}{10}& \multicolumn{1}{l}{20} &&0.921&	0.924&	0.946&&		0.898&	0.928&	0.947&&0.922&	0.925&	0.948&&		0.897&	0.930&	0.948     \\
		\multicolumn{1}{l}{} & \multicolumn{1}{l}{}   & \multicolumn{1}{l}{30} &&0.925&	0.932&	0.947&&		0.895&	0.936&	0.948&&0.928&	0.934&	0.948&&		0.899&	0.938&	0.947     \\
		\multicolumn{1}{l}{} & \multicolumn{1}{l}{20} & \multicolumn{1}{l}{20} &&0.935&	0.937&	0.948&&		0.916&	0.939&	0.948&&0.936&	0.938&	0.949&&		0.918&	0.940&	0.948    \\
		\multicolumn{1}{l}{} & \multicolumn{1}{l}{}   & \multicolumn{1}{l}{30} &&0.938&	0.941&	0.948&&		0.919&	0.943&	0.948&&0.939&	0.942&	0.949&&		0.920&	0.943&	0.949    \\
		\multicolumn{1}{l}{} & \multicolumn{1}{l}{30} & \multicolumn{1}{l}{20} &&0.940&	0.943&	0.949&&		0.922&	0.942&	0.950&&0.942&	0.943&	0.949&&		0.925&	0.943&	0.949     \\
		\multicolumn{1}{l}{} & \multicolumn{1}{l}{}   & \multicolumn{1}{l}{30} &&0.943&	0.945&	0.949&&		0.925&	0.945&	0.949&&0.942&	0.944&	0.949&&		0.926&	0.945&	0.949     \bigstrut[b]\\
		\hline
		\end{array}
		$
	\end{adjustbox}
\end{table}

\begin{mycomment}
	results in table \ref{Table network fe CI} are actually for 200000 reps (see results from saddlepoint ci simulation network fix eff adj.docx )
\end{mycomment}

\begin{mycomment}
	as p decreases coverages of the asy-LLL CI get slightly worse (see the results from saddlepoint ci simulation network fix eff adj p varies.m in some results from saddlepoint ci simulation network fix eff adj.docx)
\end{mycomment}

\section{\label{sec concl}Conclusions}

Recentering the profile score for a parameter of interest is one possible way
to deal with nuisance parameters. In this paper, we have applied this general
principle to the estimation of the autoregressive parameter $\lambda$ in a
spatial autoregression. The resulting adjusted QMLE for $\lambda$ successfully
reduces the bias in the QMLE, provides confidence intervals with excellent
coverage properties (and hence tests for $\lambda$ with excellent size
properties) even when the dimension of the nuisance parameter is large, and is
as straightforward to compute as the original QMLE. The adjusted QMLE can also
solve the incidental parameter problem that occur, for example, in social
network models with network fixed effects. However, due to the fact that the
parameter space for $\lambda$ is usually restricted to a certain interval, the
spatial autoregressive setting presents challenges for the score adjustment
procedure that do not arise in other models. Namely, the distributions of the
QMLE and of its adjusted version can be supported on different intervals,
which means that a comparison between the two estimators is not
straightforward. Our simulations suggest that the adjusted QMLE generally
performs better than the QMLE in terms of RMSE, particularly in models with a
large number of covariates.

This paper has focused on a simple version of a spatial autoregressive model.
In empirical applications, it is typically desirable to extend the model in
various directions. For example, one may want to allow for endogeneity coming
from $X$ or $W$, or for model errors that are subject to a spatial
autoregressive structure themselves. Such extensions would not preclude the
use of the score adjustment procedure, but would typically imply that the
expectation of the profile score for $\lambda$ depends on nuisance parameters.
In that case, as discussed in \cite{McCullagh1990}, the expectation would need
to be obtained numerically, rather than analytically, and the resulting
estimating equation would be unbiased up to some order, rather than being
exactly unbiased.

It is also worth mentioning that it should be possible to use the score
adjustment procedure in conjunction with modifications to the QMLE that allow
for unknown heteroskedasticity \citep[see, e.g.,][]{LiuYang2015}. Finally, the
adjusted QMLE should be effective also in models where the number of
regressors $k$ is increasing with the sample size
\citep[see, e.g.][]{GuptaRobinson2016}. If $k$ increases sufficiently quickly
with $n$, then the adjusted QMLE should have advantages, with respect to the
QMLE, even in terms of first-order asymptotics.

\begin{mycomment}
\begin{enumerate}
\item Note that the possible presence of unobserved factors responsible for
network endogeneity could be treated by network fixed effects (see, e.g., Lee,
2007; Bramoull\`{E} et al., 2009; Calvo Armengol et al., 2009; Lee et al.,
2010; Liu and Lee, 2010 The network fixed effect serves as a remedy for the
selection bias that originates from the possible sorting of individuals with
similar unobserved characteristics into a network (see Calvo Armengol et al.,
2009 peer effects education )
\item one limitation of the MLE is that it is not consistent under
heterosk.........
\item ....thus extending the applicability of ML inference for spatial models
with network fixed effects.
\item one could try other modifications of the profile lik (see in particular
I barndorff nielsen 95 which tries to produce accurate approximation of a
marginal or a conditional log likelihood, when either exists - see bias\_SLM.m
for evaluation and plot\_lik\_score\_SLM\_with\_adj: seems like BN95 behaves
very similarly to our adjustment; or another one is Cox Reid 1987)
\item extension to SARAR(p,q) should be straightforward see Liu Yang 2015 RSUE
\end{enumerate}
\end{mycomment}

\appendix%

\makeatletter\def\@seccntformat#1{Appendix\ \csname the#1\endcsname\quad}%

\makeatother



\section{\label{app add aux}Auxiliary results}

\begin{lemma}
\label{lemma null set}Let $W$ be a weights matrix, and $\omega$ a semisimple
real eigenvalue $\omega$ of $W$. The set of full column rank matrices $X$ such
that $\mathrm{tr}(M_{X}Q_{\omega})=0$ is a $\mu_{\mathbb{R}^{n\times k}}$-null set.
\end{lemma}

\begin{pff}
We need to show that $\mathcal{A}\coloneqq \{X\in\mathbb{R}^{n\times k}:\mathrm{rank}%
(X)=k$ and $\mathrm{tr}(M_{X}Q_{\omega})=0\}$ is a $\mu_{\mathbb{R}^{n\times
k}}$-null set. This is equivalent to showing that $\mathcal{B}\coloneqq \{X\in
\mathbb{R}^{n\times k}:\det(X^{\prime}X)\mathrm{tr}(M_{X}Q_{\omega})=0\}$ is a
$\mu_{\mathbb{R}^{n\times k}}$-null set, because $\mathcal{A=}\{X\in
\mathbb{R}^{n\times k}:\mathrm{rank}(X)=k\}\cap\mathcal{B}$. But
$\det(X^{\prime}X)\mathrm{tr}(M_{X}Q_{\omega})$ is a polynomial in the entries
of $X$, as it is clear from writing $\det(X^{\prime}X)\mathrm{tr}%
(M_{X}Q_{\omega})=\det(X^{\prime}X)\mathrm{tr}(Q_{\omega})-\mathrm{tr}$%
($X$\textrm{adj}$\left(  X^{\prime}X\right)  X^{\prime}Q_{\omega})$. Hence
$\mathcal{B}$ is an algebraic variety, and as such it is either a
$\mu_{\mathbb{R}^{n\times k}}$-null set or the whole $\mathbb{R}^{n\times k}$.
The latter case is easily ruled out.
\end{pff}

\begin{mycomment}
\begin{itemize}
	\item could maybe doagonalize Q (this is always possible for a projectior) and then choose $M_X$ that selects the 1's on the diagonal
	\item 	...need to exhibit an $X$ (as in potscher prein. "on size and power") s.t.
	$\mathrm{tr}(M_{X}Q_{\omega})\neq0,$ might try to take col(X)=$\mathrm{null}%
	(W-\omega^{\ast}I_{n})$ for some $\omega^{\ast}\neq\omega$ DO\ THIS
	\item	one thing one could try is to select X s.t. $M_{X}$ and $Q_{\omega}$ commute (is this possible?). the their product is a proj and therefore its trace is $\neq 0$
	\item either show that there is one X s.t. $\mathrm{tr}%
	(M_{X}Q_{\omega})\neq0$, or assume $\mathcal{B}%
	=\mathbb{R}^{n\times k}$ and try to show this leads to a contrad
\end{itemize}
\end{mycomment}

\begin{mycomment}
	An alternative statement would be: Let $\mathcal{X}\coloneqq \{X\in\mathbb{R}^{n\times
		k}:\mathrm{rank}(X)=k\}$, and consider an arbitrary $k\geq1$, and an arbitrary
	semisimple real eigenvalue $\omega$ of $W$. Then, $\mathrm{tr}(M_{X}Q_{\omega
	})\neq0$ for $\mu_{\mathbb{R}^{n\times k}}$-almost every $X\in\mathcal{X}$,
\end{mycomment}

\begin{lemma}
\label{lemma delta_a<0}If $W$ is symmetric, condition (\ref{C1}) is satisfied
for any $X$.
\end{lemma}

\begin{pff}
If $W$ is symmetric, $v^{\prime}(G(\lambda)-tI_{n})^{2}v\geq0$ for any
$t\in\mathbb{R}$, any $v\in\mathbb{R}^{n}$, and any $\lambda$ such that
$\det(S(\lambda))\neq0$. It follows that, for any $t\in\mathbb{R}$, any
$u\in\mathbb{R}^{m}$, any $m\times n$ matrix $C$, and any $\lambda$ such that
$\det(S(\lambda))\neq0$, $u^{\prime}C(G(\lambda)-tI_{n})^{2}C^{\prime}u\geq0$
and hence $\mathrm{tr} \left(  C(G(\lambda)-tI_{n})^{2}C^{\prime}\right)
\geq0$. Let us now choose $C$ to be an $\left(  n-k\right)  \times n$ matrix
such that $CC^{\prime}=I_{n-k}$ and $C^{\prime}C=M_{X}$, and $t$ to be
$\mathrm{tr}(M_{X} G(\lambda))/(n-k)$. Then, $\mathrm{tr}\left(
C(G(\lambda)-tI_{n} )^{2}C^{\prime}\right)  =\mathrm{tr}(M_{X}G^{2}%
(\lambda))-\left[  \mathrm{tr}\left(  M_{X}G(\lambda)\right)  \right]
^{2}/(n-k)=-(n-k)\delta_{\mathrm{a}}(\lambda)$. The proof is completed on
noting that $\mathrm{tr} \left(  C(G(\lambda)-tI_{n})^{2}C^{\prime}\right)
=0$ if and only if $G(\lambda)$, and hence $W$, is a scalar multiple of
$I_{n},$ which is ruled out by the maintained assumption that $W$ has at least
one negative and at least one positive eigenvalue.
\end{pff}

\begin{lemma}
\label{lemma trMG}For any semisimple nonzero real eigenvalue $\omega$ of $W$,

\begin{enumerate}
\item[(i)] if $\mathrm{tr}(M_{X}Q_{\omega})>0$ then $\lim_{\lambda
\uparrow\omega^{-1}}\mathrm{tr}(M_{X}G(\lambda))=+\infty$ and $\lim
_{\lambda\downarrow\omega^{-1}}\mathrm{tr}(M_{X}G(\lambda))=-\infty$;

\item[(ii)] if $\mathrm{tr}(M_{X}Q_{\omega})=0$ then $\lim_{\lambda
\rightarrow\omega^{-1}}\mathrm{tr}(M_{X}G(\lambda))$ is bounded;

\item[(iii)] if $\mathrm{tr}(M_{X}Q_{\omega})<0$ then $\lim_{\lambda
\uparrow\omega^{-1}}\mathrm{tr}(M_{X}G(\lambda))=-\infty$ and $\lim
_{\lambda\downarrow\omega^{-1}}\mathrm{tr}(M_{X}G(\lambda))=+\infty$.
\end{enumerate}
\end{lemma}

\begin{pff}
Let \textrm{Sp}$(W)$ denote the spectrum (defined as the set of distinct
eigenvalues) of $W$. For any $\lambda\in\mathbb{R}\setminus\mathrm{Sp}(W)$,
consider the function $f(z)\coloneqq z\left(  1-\lambda z\right)  ^{-1}$ from
$\mathbb{C}\setminus\{\lambda^{-1}\}$ to $\mathbb{C}$, and let $f^{(i)}(z)$
denote its $i$-th order derivative. Since $f(z)$ is defined at each eigenvalue
of $W$, the matrix $G(\lambda)$, for any $\lambda\in\mathbb{R}\setminus
\mathrm{Sp}(W)$, admits the spectral resolution
\citep[e.g.,][p. 603]{Meyer2000}
\begin{equation}
G(\lambda)=\sum_{\chi\in\mathrm{Sp}(W)}\sum_{i=0}^{k_{\chi}-1}\frac
{f^{(i)}(\chi)}{j!}(W-\chi I_{n})^{i}T_{\chi}, \label{sp resol G}%
\end{equation}
where $k_{\chi}$ denotes the index of an eigenvalue $\chi$, and $T_{\chi}$
denotes the projector onto the generalized eigenspace $\mathrm{null}\left(
(W-\chi I_{n})^{k_{\chi}}\right)  $ along $\operatorname{col}\left(  (W-\chi
I_{n})^{k_{\chi}}\right)  $. If an eigenvalue $\omega$ of $W$ is semisimple
(and only in that case), then $k_{\omega}=1$, and hence
\begin{equation}
\mathrm{tr}(M_{X}G(\lambda))=f(\omega)\mathrm{tr}\left(  M_{X}Q_{\omega
}\right)  +\sum_{\chi\in\mathrm{Sp}(W)\setminus\{\omega\}}\sum_{i=0}^{k_{\chi
}-1}\frac{f^{(i)}(\chi)}{j!}\mathrm{tr}\left(  M_{X}(W-\chi I_{n})^{i}T_{\chi
}\right)  . \label{trMG2}%
\end{equation}
The stated result obtains, because all derivatives $f^{(i)}(\chi)$ are bounded
for $\lambda\neq\chi^{-1}$.
\end{pff}

\begin{lemma}
\label{lemma extremes}For any semisimple nonzero real eigenvalue $\omega$ of
$W$, if $y\notin\mathrm{null}(M_{X}S(\omega^{-1}))$, then $\lim_{\lambda
\rightarrow\omega^{-1}}l_{\mathrm{a}}(\lambda)$ is

\begin{enumerate}
\item[(i)] $-\infty$ if $\mathrm{tr}(M_{X}Q_{\omega})>0$;

\item[(ii)] bounded if $\mathrm{tr}(M_{X}Q_{\omega})=0;$

\item[(iii)] $+\infty$ if $\mathrm{tr}(M_{X}Q_{\omega})<0$.
\end{enumerate}
\end{lemma}

\begin{pff}
The result can be proved by looking at the two terms that make up
$s_{\mathrm{a2}}(\lambda)$ in expression (\ref{sa2}): $(n-k)(y^{\prime
}W^{\prime}M_{X}S(\lambda)y)/(y^{\prime}S(\lambda)^{\prime}M_{X}S(\lambda)y)$
and $-\mathrm{tr}(M_{X}G(\lambda))$. Consider an arbitrary nonzero real
eigenvalue $\omega$ of $W$. If $y\notin\mathrm{null}(M_{X}S(\omega^{-1}))$,
the function $\lambda\mapsto(n-k)(y^{\prime}W^{\prime}M_{X}S(\lambda
)y)/(y^{\prime}S(\lambda)^{\prime}M_{X}S(\lambda)y)$ is continuous at
$\lambda=\omega^{-1}$ because it is well defined at $\lambda=\omega^{-1}$, and
is well defined in a neighborhood of $\lambda=\omega^{-1}$
\citep[the last claim
follows by Lemma S.1.1 in the online supplement to][]{Hillier2017}. The proof
is completed on using Lemma \ref{lemma trMG} to establish the limiting
behavior of the term $-\mathrm{tr}(M_{X}G(\lambda))$ as $\lambda
\rightarrow\omega^{-1}$.
\end{pff}

\begin{lemma}
\label{lemma limdelta}For any semisimple nonzero real eigenvalue $\omega$ of
$W$, $\left[  \mathrm{tr}\left(  M_{X}G(\lambda)\right)  \right]
^{2}-(n-k)\mathrm{tr}(M_{X}G^{2}(\lambda))\rightarrow+\infty$ as
$\lambda\rightarrow\omega^{-1}$ if $\mathrm{tr}(M_{X}Q_{\omega})<0$.
\end{lemma}

\begin{pff}
This proof is based on the proof of Lemma \ref{lemma trMG}. For any
$\lambda\in\mathbb{R}\setminus\mathrm{Sp}(W)$, consider the function
$g(z)=f^{2}(z)=z^{2}\left(  1-\lambda z\right)  ^{-2}$ from $\mathbb{C}%
\setminus\{\lambda^{-1}\}$ to $\mathbb{C}$, and let $g^{(i)}(z)$ denote its
$i$-th order derivative. Then, for any $\lambda\in\mathbb{R}\setminus
\mathrm{Sp}(W)$, a spectral resolution of $G^{2}(\lambda)$ is given by the
right hand side of equation (\ref{sp resol G}) with all $f^{(i)}(\cdot)$
replaced by $g^{(i)}(\cdot)$. Hence, for an arbitrary semisimple nonzero real
eigenvalue $\omega$ of $W$, $\mathrm{tr}(M_{X}G^{2}(\lambda))$ can be
expressed as in the right hand side of (\ref{trMG2}), with $f$ replaced by
$g$. Since all derivatives $g^{(i)}(\chi)$ are bounded for $\lambda\neq
\chi^{-1}$, $\lim_{\lambda\rightarrow\omega^{-1}}\mathrm{tr}(M_{X}%
G^{2}(\lambda))=-\infty$ if $\mathrm{tr}(M_{X}Q_{\omega})<0$. The proof is
completed on observing that, by Lemma \ref{lemma trMG}, $\lim_{\lambda
\rightarrow\omega^{-1}}\left[  \mathrm{tr}\left(  M_{X}G(\lambda)\right)
\right]  ^{2}=+\infty$ if $\mathrm{tr}(M_{X}Q_{\omega})<0$.
\end{pff}

\begin{lemma}
\label{lemma inv sub aML}Suppose that, in a SAR model, $\mathrm{col}(X)$ is an
invariant subspace of $W$. Then the adjusted log-likelihood function
$l_{\mathrm{a}}(\sigma^{2},\lambda)$ is the same as the quasi log-likelihood
function for $(\sigma^{2},\lambda)$ based on $Dy$, for any full rank $\left(
n-k\right)  \times n$ matrix $D$ such that $DX=0$, and for any $y\in
\mathbb{R}^{n}$.
\end{lemma}

\begin{pff}
We only provide a sketch of the proof here; full details are given in Section
\ref{sec suppl proof aux lemma} of the Supplement. If $\operatorname{col}(X)$
is an invariant subspace $W$, there exists a unique $k\times k$ matrix $A$
such that $WX=XA$, and hence $S^{-1}(\lambda)X=X(I_{k}-\lambda A)^{-1}$, for
any $\lambda$ such that $S(\lambda)$ is invertible. Thus, when
$\operatorname{col}(X)$ is an invariant subspace $W$, the SAR model
$y=S^{-1}(\lambda)X\beta+\sigma S^{-1}(\lambda)\varepsilon$ can be written as
$y=X(I_{k}-\lambda A)^{-1}\beta+\sigma S^{-1}(\lambda)\varepsilon$, which
corresponds to the spatial error model (\ref{SEM}) with $\beta$ replaced by
$(I_{k}-\lambda A)^{-1}\beta$. Intuitively, the lemma then follows from the
fact that, for a spatial error model, the adjusted (quasi) log-likelihood
$l_{\mathrm{a}}(\sigma^{2},\lambda)$ is the same as the restricted, or
residual, quasi log-likelihood for $(\sigma^{2},\lambda)$ (see Section
\ref{sec suppl reml} in the Supplement).
\end{pff}

\begin{mycomment}
	OLD:Then $(\hat{\lambda}_{\mathrm{aML}},\hat{\sigma}_{\mathrm{aML}}^{2})$
coincide with the estimator that maximizes the likelihood for $(\lambda
,\sigma)$ based on $Dy$ for any full rank $\left(  n-k\right)  \times n$
matrix $D$ such that $DX=0$.\bigskip
\end{mycomment}

\begin{mycomment}
I wouldn't call the estimator that maximizes the likelihood based on $Dy$ REML
because the model is not really a SEM is a SEM with a param restriction (see
Remark \ref{rem SEM-like} in the Suppl and also from MLE paper: The condition
that $\operatorname{col}(X)$ is an invariant subspace of $W$ is equivalent to
the existence of a $k\times k$ matrix $A$ such that $WX=XA$. Since the
eigenvalues of $A$ must be a subset of the eigenvalues of $W$, $WX=XA$ if and
only if $S^{-1}(\lambda)X=X(I_{k}-\lambda A)^{-1}$, for any $\lambda$ such
that $S_{\lambda}$ is invertible.
\end{mycomment}

\begin{mycomment}
when $\mathrm{col}(X)$ is an invariant subspace of $W$ SAR is equiv to a
restricted SEM, and the two models have same prof lik for $(\lambda,\sigma)$
\end{mycomment}

\section{\label{app proofs}Proofs}

\begin{pff}
[Proof of Proposition \ref{prop l_a SLM}]The adjusted likelihood
$l_{\mathrm{a}}(\sigma^{2},\lambda)$ is defined by the property that its
gradient is the score $s_{\mathrm{a}}(\sigma^{2},\lambda)$ given in equation
(\ref{sa sig lam}), for any $\sigma>0$ and any $\lambda\in\Lambda_{u}$, where,
recall, $\Lambda_{u}\coloneqq \left\{  \lambda\in\mathbb{R}:\det\left(  S(\lambda
)\right)  \neq0\right\}  $. Solving the two equations
\begin{align*}
\frac{\partial l_{\mathrm{a}}(\sigma^{2},\lambda)}{\partial\sigma^{2}}  &
=-\frac{n-k}{2\sigma^{2}}+\frac{1}{2\sigma^{4}}y^{\prime}S^{\prime}%
(\lambda)M_{X}S(\lambda)y,\\
\frac{\partial l_{\mathrm{a}}(\sigma^{2},\lambda)}{\partial\lambda}  &
=\frac{1}{\sigma^{2}}y^{\prime}W^{\prime}M_{X}S(\lambda)y-\mathrm{tr}%
(M_{X}G(\lambda)),
\end{align*}
gives, up to an additive constant,%
\[
l_{\mathrm{a}}(\sigma^{2},\lambda)=-\frac{n-k}{2}\log(\sigma^{2})-\frac
{1}{2\sigma^{2}}y^{\prime}S^{\prime}(\lambda)M_{X}S(\lambda)y+\int%
\mathrm{tr}(M_{X}G(\lambda))\diff\lambda.
\]
Using elementary complex analysis, the integral $\int\mathrm{tr}%
(M_{X}G(\lambda))\diff\lambda$ can be expressed in terms of the matrix
logarithm. For $\lambda\in(-1,1)$, $\log S(\lambda)$ admits the convergent
power series representation $-\sum_{k=1}^{\infty}\frac{1}{k}\lambda^{k}W^{k}$,
from which it is immediately clear that $\frac{\diff}{\diff\lambda}\log
S(\lambda)=-G(\lambda)$. This expression for $\frac{\diff}{\diff\lambda}\log
S(\lambda)$ can be extended over the set $\Lambda_{u}$, by selecting a
suitable branch of the matrix logarithm. Let $\Xi\coloneqq \left\{  1-\lambda
\omega:\lambda\in\mathbb{R}\text{, }\omega\in\mathrm{Sp}(W)\right\}  $, where
\textrm{Sp}($W$) denotes the set of eigenvalues of $W$, be the subset of
$\mathbb{C}$ formed by the eigenvalues of $S(\lambda)$ as $\lambda$ ranges
over $\mathbb{R}$. Since $\Xi$ is formed by a finite number of lines in
$\mathbb{C}$ going through $1$, there must exist a half line $l$ in
$\mathbb{C}$ starting from the origin that does not intersect $\Xi.$ In the
rest of the proof, $\log$ denotes the matrix logarithm associated to the
branch cut $l$. Then, $\log S(\lambda)$, and hence its first derivative, is
holomorphic over $\mathbb{C}\backslash l$. Since $\mathbb{C}\backslash l$ is a
connected set and $\Lambda_{u}\subset\mathbb{C}\backslash l$,$\ $the identity
theorem for holomorphic functions implies that $\frac{\diff}{\diff\lambda}\log
S(\lambda)=-G(\lambda)$ for any $\lambda\in\Lambda_{u}$. It follows that
$\mathrm{tr}(M_{X}\log S(\lambda))$ is an antiderivative of $-\mathrm{tr}%
(M_{X}G(\lambda))$, and hence that $\operatorname{Re}\left[  \mathrm{tr}%
(M_{X}\log S(\lambda))\right]  $ is a real antiderivative of $-\mathrm{tr}%
(M_{X}G(\lambda))$, for any $\lambda\in\Lambda_{u}$, which completes the proof.
\end{pff}

\begin{pff}
[Proof of Theorem \ref{lemma lim gen W}]According to Assumption \ref{assum id}%
, $M_{X}S(\omega^{-1})\neq0$, for any nonzero real eigenvalue $\omega$ of $W$.
Hence $\mathrm{null}(M_{X}S(\omega^{-1}))$ is a $\mu_{\mathbb{R}^{n}}$-null
set, for any nonzero real eigenvalue $\omega$ of $W$. The result follows by
Lemma \ref{lemma extremes}.
\end{pff}

\begin{pff}
[Proof of Lemma \ref{lemma null}](i) For any semisimple eigenvalue $\omega$ of
$W$, $Q_{\omega}$ is a projector onto $\operatorname{col}(Q_{\omega
})=\mathrm{null}(W-\omega I_{n})$, so $\mathrm{null}(W-\omega I_{n}
)\subseteq\operatorname{col}(X)$ if and only if $M_{X}Q_{\omega}=0$, which is
obviously sufficient for $\mathrm{tr}(M_{X}Q_{\omega})=0$. (ii) Let
$H_{\omega}$ be a matrix whose columns form an orthonormal basis for
$\mathrm{null}(W-\omega I_{n})$. If $W$ is symmetric, $Q_{\omega}=H_{\omega
}H_{\omega}^{\prime}$ (the orthogonal projector onto $\mathrm{null}(W-\omega
I_{n})$), and hence $\mathrm{tr}(M_{X}Q_{\omega})=\mathrm{tr}(H_{\omega
}^{\prime}M_{X}H_{\omega})$. Observe now that, since $M_{X}$ is positive
semidefinite, all diagonal entries of $H_{\omega}^{\prime}M_{X}H_{\omega}$ are
nonnegative. Thus, $\mathrm{tr}(M_{X}Q_{\omega})\geq0$, and the desired
conclusion follows from (i).
\end{pff}

\begin{pff}
[Proof of Proposition \ref{theo cdf MLE}]According to Lemma
\ref{lemma limdelta}, condition (\ref{C1}) implies that $\mathrm{tr}%
(M_{X}Q_{\omega})\geq0$ for all semisimple eigenvalues $\omega$ such that
$\omega^{-1}\in\Lambda_{\mathrm{a}}$. Hence, by Theorem \ref{lemma lim gen W},
as $\lambda$ approaches the reciprocals of such eigenvalues, $l_{\mathrm{a}%
}(\lambda)$ either a.s.\ diverges to $-\infty$ or is a.s.\ bounded, provided
that Assumption \ref{assum id} holds. The former case is impossible by the
definition of $\Lambda_{\mathrm{a}}$, so $l_{\mathrm{a}}(\lambda)$ must be
a.s.\ continuous on the whole $\Lambda_{\mathrm{a}}$ (after extension of the
domain of $l_{\mathrm{a}}(\lambda)$ to include any zeros of $\det(S(\lambda))$
in $\Lambda_{\mathrm{a}}$). By the definition of $\Lambda_{\mathrm{a}}$ we
also have that $l_{\mathrm{a}}(\lambda)\rightarrow-\infty$ a.s.\ at the
extremes of $\Lambda_{\mathrm{a}}$. We now show that the second derivative
$\ddot{l}_{\mathrm{a}}(\lambda)$ of $l_{\mathrm{a}}(\lambda)$ is negative at
any critical point of $l_{\mathrm{a}}(\lambda)$ in $\Lambda_{\mathrm{a}}$.
This implies that $l_{\mathrm{a}}(\lambda)$ a.s.\ has a single critical point
in $\Lambda_{\mathrm{a}}$, corresponding to a maximum (that is, it is
single-peaked with no stationary inflection points). Write
\[
\ddot{l}_{\mathrm{a}}(\lambda)=\left(  n-k\right)  \left(  -\frac{(ac-b^{2}%
)}{(a\lambda^{2}-2b\lambda+c)^{2}}+\frac{(b-a\lambda)^{2}}{(a\lambda
^{2}-2b\lambda+c)^{2}}\right)  -\mathrm{tr}(M_{X}G^{2}(\lambda)),
\]
where $a\coloneqq y^{\prime}W^{\prime}M_{X}Wy\text{, }b\coloneqq y^{\prime}W^{\prime}%
M_{X}y\text{, }c\coloneqq y^{\prime}M_{X}y.$ The first order condition $s_{\mathrm{a2}%
}(\lambda)=0$ implies
\[
\frac{(b-a\lambda)^{2}}{(a\lambda^{2}-2b\lambda+c)^{2}}=\frac{1}{\left(
n-k\right)  ^{2}}\left[  \mathrm{tr}\left(  M_{X}G(\lambda)\right)  \right]
^{2},
\]
so that, at any critical point of $l_{\mathrm{a}}(\lambda)$,
\begin{equation}
\ddot{l}_{\mathrm{a}}(\lambda)=-\left(  n-k\right)  \frac{(ac-b^{2}
)}{(a\lambda^{2}-2b\lambda+c)^{2}}+\frac{1}{n-k}\left[  \mathrm{tr}\left(
M_{X}G(\lambda)\right)  \right]  ^{2}-\mathrm{tr}(M_{X}G^{2}(\lambda)).
\label{l2 at l1=0}%
\end{equation}
By the Cauchy-Schwarz inequality the first of the three terms on the right
hand side of (\ref{l2 at l1=0}) is nonpositive. Hence condition (\ref{C1}) is
sufficient for $\ddot{l}_{\mathrm{a}}(\lambda)$ to be negative at any zero of
$s_{\mathrm{a2}}(\lambda)$ in $\Lambda_{\mathrm{a}}$, which completes the proof.
\end{pff}

\begin{pff}
[Proof of Proposition \ref{prop aML=LLL}]If $W_{r}\iota_{m_{r}}=\iota_{m_{r}}%
$, for all $r=1,\ldots,R$, then $\operatorname{col}(\bigoplus_{r=1}^{R}%
\iota_{m_{r}})$ is an invariant subspace of $W=\bigoplus_{r=1}^{R}W_{r}$.
Since $F_{r}\iota_{m_{r}}=0$, for all $r=1,\ldots,R$, the desired claim follows
by Lemma \ref{lemma inv sub aML}.
\end{pff}

\begin{mycomment}
OLD PROOF, not convincing as \ref{NFE SEM} is not really a SEM. The network
fixed effects model without covariates is
\begin{equation}
y_{r}=\lambda W_{r}y_{r}+\alpha_{r}\iota_{m_{r}}+\sigma\varepsilon_{r},\text{
}r=1,\ldots,R. \label{nfe model}%
\end{equation}
Define the orthogonal projector $M_{r}\coloneqq I_{m_{r}}-\frac{1}{m_{r}}\iota_{m_{r}%
}\iota_{m_{r}}^{^{\prime}}$, and let $F_{r}$ be the $m_{r}\times\left(
m_{r}-1\right)  $ matrix of orthonormal eigenvectors of $M_{r}$ corresponding
to the eigenvalue $1$. Under the condition $W_{r}\iota_{m_{r}}=\iota_{m_{r}}$,
the LLL estimator $(\hat{\lambda}_{\mathrm{LLL}},\hat{\sigma}_{\mathrm{LLL}%
}^{2})$ is the MLE obtained after premultiplying each equation in
(\ref{nfe model}) by $F_{r}$. We now show that, in this context,
$(\hat{\lambda}_{\mathrm{aML}},\hat{\sigma}_{\mathrm{aML}}^{2})$ is equivalent
to $(\hat{\lambda}_{\mathrm{LLL}},\hat{\sigma}_{\mathrm{LLL}}^{2})$. Provided
that each $I_{m_{r}}-\lambda W_{r}$ is nonsingular, the reduced form for model
(\ref{nfe model}) is
\begin{equation}
y_{r}=\alpha_{r}(I_{m_{r}}-\lambda W_{r})^{-1}\iota_{m_{r}}+\sigma(I_{m_{r}%
}-\lambda W_{r})^{-1}\varepsilon_{r},\text{ }r=1,\ldots,R. \label{netsar}%
\end{equation}
If $W_{r}\iota_{m_{r}}=\iota_{m_{r}}$, equation (\ref{netsar}) becomes
\begin{equation}
y_{r}=\frac{\alpha_{r}}{1-\lambda}\iota_{m_{r}}+\sigma(I_{m_{r}}-\lambda
W_{r})^{-1}\varepsilon_{r},\text{ }r=1,\ldots,R. \label{NFE SEM}%
\end{equation}
This is a spatial error model as defined in Section \ref{sec SEM}, and
therefore $(\hat{\lambda}_{\mathrm{aML}},\hat{\sigma}_{\mathrm{aML}}^{2})$
must be equivalent to the REML estimator of $(\lambda,\sigma^{2})$; see
Section \ref{sec suppl reml} of the Supplement. The REML estimator maximizes
the likelihood obtained on multiplying each equation in (\ref{NFE SEM}) by any
matrix $D_{r}$ such that $D_{r}\iota_{m_{r}}=0$. Choosing $D_{i}=F_{i}$ shows
that $(\hat{\lambda}_{\mathrm{aML}},\hat{\sigma}_{\mathrm{aML}}^{2})$ is the
same as $(\hat{\lambda}_{\mathrm{LLL}},\hat{\sigma}_{\mathrm{LLL}}^{2})$.
\end{mycomment}

\newpage

\numberwithin{equation}{section}
\renewcommand{\thesection}{S.\arabic{section}}

\renewcommand{\thetable}{S.\arabic{table}}

\setcounter{equation}{0}
\setcounter{figure}{0}
\setcounter{table}{0}
\setcounter{section}{0}
\setcounter{page}{1}

	\enlargethispage{2\baselineskip}
	
	\begin{center}
		\renewcommand{\thefootnote}{\fnsymbol{footnote}} {\Large \textbf{Supplement to
				\textquotedblleft Adjusted QMLE for the Spatial Autoregressive
				Parameter\textquotedblright}}
		
		\vspace{.5cm}
		
		\begin{savenotes}
			\begin{tabular}
				[c]{ccc}%
				Federico Martellosio & \hspace{0.54cm} & Grant Hillier\\
				University of Surrey, UK &  & CeMMAP and\\
				&  & University
				of Southampton, UK%
			\end{tabular}
		\end{savenotes}\vspace{0.5cm} \setcounter{footnote}{0}
		
		February 28, 2019 \vspace{0.2cm}

	\end{center}
	
	\noindent This Supplement contains technical material related to the paper
	``Adjusted QMLE for the spatial autoregressive parameter''. All numbers of
	equations, lemmas, etc., in the Supplement are preceded by `S.'.
	
	\section{\label{sec suppl Assum1} Details on Assumption \ref{assum id}}
	
	The following lemma provides an analysis of Assumption \ref{assum id}.
	
	\begin{lemma}
		\label{lemma remarks}\leavevmode

		\begin{enumerate}
			\item[(i)] If $M_{X}(zI_{n}-W)=0$ then $z$ is a real eigenvalue of $W$.
			
			\item[(ii)] For fixed $W$ and $X$, the condition $M_{X}(\omega I_{n}-W)=0$ can
			be satisfied at most by one eigenvalue $\omega$ of $W$.
			
			\item[(iii)] For any real eigenvalue $\omega$ of $W$, $M_{X}(\omega
			I_{n}-W)=0$ if and only if $\operatorname{col}^{\perp}(X)\subseteq
			\mathrm{null}(W^{\prime}-\omega I_{n})$.
			
			\item[(iv)] For any real eigenvalue $\omega$ of $W$, if $M_{X}(\omega
			I_{n}-W)=0$ then $k\geq n-g_{\omega}$, where $g_{\omega}$ denotes the
			geometric multiplicity of $\omega$.
			
			\item[(v)] For any real eigenvalue $\omega$ of $W$, a necessary condition for
			$M_{X}(\omega I_{n}-W)=0$ is that all eigenvectors of $W$ associated with
			eigenvalues other than $\omega$ are in $\operatorname{col}(X)$.\footnote{When
				saying that an eigenvector of $W$ is in $\operatorname{col}(X)$, we take it as
				given that the vector space field used to define $\operatorname{col}(X)$ is
				$\mathbb{C}^{n}$, not $\mathbb{R}^{n}$. If the field was $\mathbb{R}^{n}$, the
				condition should read \textquotedblleft the real parts and imaginary parts of
				all eigenvectors of $W$ associated with eigenvalues other than $\omega$ are in
				$\operatorname{col}(X)$\textquotedblright.} The condition is also sufficient
			if $W$ is diagonalizable.
		\end{enumerate}
	\end{lemma}
	
	\begin{pff}
		(i) Clearly $z$ must be real, because if $M_{X}(zI_{n}-W)=0$ then
		$(\operatorname{Re}(z)+\operatorname{Im}(z))M_{X}=M_{X}W$, which implies
		$\operatorname{Im}(z)=0$. If $z$ is not an eigenvalue of $W$ then
		$\mathrm{rank}\left(  M_{X}(zI_{n}-W)\right)  =n-k$ and hence $M_{X}%
		(zI_{n}-W)$ cannot be $0$, since $k<n$. (ii) Straightforward, because
		$M_{X}(\omega_{1}I_{n}-W)=M_{X}(\omega_{2}I_{n}-W)$ implies $\omega_{1}%
		=\omega_{2}$. (iii) Follows from $\mathrm{\operatorname{col}}(\omega
		I_{n}-W)=\mathrm{null}^{\perp}(\omega I_{n}-W^{\prime})$, and the fact that
		$A\subseteq B$ if and only if $B^{\perp}\subseteq A^{\perp}$, for any two
		subspaces $A$ and $B$. (iv) In order for $\mathrm{\operatorname{col}}(\omega
		I_{n}-W)$ to be a subset of $\operatorname{col}(X)$, the dimension of
		$\mathrm{\operatorname{col}}(X)$ must not be smaller than that of
		$\operatorname{col}(\omega I_{n}-W)$, that is, it must hold that
		$k\geq\mathrm{rank}(\omega I_{n}-W)=n-\mathrm{nullity}(\omega I_{n}%
		-W)=n-g_{\omega}$ (the nullity of a matrix being the dimension of its null
		space). (v) If $M_{X}(\omega I_{n}-W)=0$ then $M_{X}(\omega I_{n}-W)v=0$ for
		any $v\in\mathbb{C}^{n}$, that is,
		\begin{equation}
		\omega M_{X}v=M_{X}Wv. \label{1}%
		\end{equation}
		Suppose now $v$ is an eigenvector of $W$ associated to the eigenvalue
		$\varkappa$. Then equation (\ref{1}) gives $\omega M_{X}v=\varkappa M_{X}v$,
		which implies that $v$ must be in $\operatorname{col}(X)$ if $\varkappa
		\neq\omega$. If $W$ is diagonalizable, we can write $W=HDH^{-1}$, where
		$D\coloneqq \mathrm{diag}(\omega I_{g_{\omega}},\omega\in\mathrm{Sp}(W))$ is the
		diagonal matrix containing the eigenvalues (ordered in some arbitrary manner)
		of $W$, and $H$ is the matrix of eigenvectors. Sufficiency of the condition
		follows, because $M_{X}(\omega I_{n}-W)=M_{X}H\mathrm{diag}\left(  \left(
		\omega-\chi\right)  I_{n_{\chi}},\chi\in\mathrm{Sp}(W)\right)  H^{-1}$.
	\end{pff}
	
	In view of part (i) of Lemma \ref{lemma remarks}, Assumption \ref{assum id}
	could be reformulated as $M_{X}(zI_{n}-W)\neq0$ for any $z$ in $\mathbb{R}$
	(or in $\mathbb{C}$). Part (iii) shows that Assumption \ref{assum id} is the
	same as the condition used in \cite{Preinerstorfer2015} for identifiability of
	the spatial autoregressive parameter in a spatial error model
	\citep[as it should
	be, because the SAR and spatial error models have the same
	profile log-likelihood $l(\lambda)$ under Assumption \ref{assum id}; see][]{Martellosio2018}.
	Part (iv) indicates how large $k$ must be in order for the identifiability
	issue related to Assumption \ref{assum id} to arise.
	
	Note that the set of (full rank) matrices $X$ such that Assumption 1 is
	violated for a given $W$ is a null set with respect to any absolutely
	continuous distribution of $X$ on $\mathbb{R}^{n\times k}$. The condition that
	the distribution of $X$ is absolutely continuous would of course not be
	satisfied if some of the columns of $X$ were fixed (e.g., there are group
	intercepts as in Example \ref{ex BGI}). In that case, Assumption 1 may be
	violated for any draw of the remaining columns.
	
	\begin{mycomment}
		for prev paragraph don;t really need to specify that we are treating X as random there
	\end{mycomment}

	The next result states some consequences of a violation of Assumption
	\ref{assum id} on the profile log-likelihood $l(\lambda)$; see also
	\cite{Martellosio2018}.
	
	\begin{lemma}
		\label{lemma l(lambda) violation}Suppose Assumption \ref{assum id} is violated
		for some eigenvalue $\omega$ of $W$. In both the SAR and the spatial error model:
		
		\begin{enumerate}
			\item[(i)] if $\omega\neq0$, the profile log-likelihood $l(\lambda)$ is
			a.s.\ unbounded from above in a neighborhood of $\omega^{-1}$;
			
			\item[(ii)] if $\omega=0$, then, up to additive constants and a.s.,
			$l(\lambda)=\log\left\vert \det\left(  S(\lambda)\right)  \right\vert $, for
			any $\lambda$ such that $\det(S(\lambda))\neq0$.
		\end{enumerate}
	\end{lemma}
	
	\begin{pff}
		(i) For any $\lambda$ such that $\mathrm{rank}\left(  S(\lambda)\right)  =n$,
		and for any $y\notin\mathrm{null}(M_{X}S(\lambda))$, the profile
		log-likelihood $l(\lambda)$ for a SAR model is given by equation
		(\ref{prof lik}). Note that equation (\ref{prof lik}) holds a.s.\ for any
		fixed $\lambda$ such that \textrm{rank} $\left(  S(\lambda)\right)  =n$,
		because $\mathrm{null}(M_{X}S(\lambda))$ is a $\mu_{\mathbb{R}^{n}}$-null set
		when \textrm{rank}$\left(  S(\lambda)\right)  =n$ (since $k<n$). If Assumption
		\ref{assum id} is violated for an eigenvalue $\omega$ (which must be real and
		unique by parts (i) and (ii) of Lemma \ref{lemma remarks}) of $W$, then
		$M_{X}(\omega I_{n}-W)=0$ and hence $M_{X}S(\lambda)=(1-\lambda\omega)M_{X}$.
		Substituting this last equation into (\ref{prof lik}) gives, for any
		$y\notin\operatorname{col}(X)$,
		\begin{align}
		l(\lambda)  &  =\log\left\vert \det\left(  S(\lambda)\right)  \right\vert
		-n\log\left\vert 1-\lambda\omega\right\vert -\frac{n}{2}\log(y^{\prime}%
		M_{X}y)\label{prof lik ASSUM1 violated}\\
		&  =\log\left(  \frac{\left\vert \prod_{\varkappa\in\mathrm{Sp}(W)\setminus
				\{\omega\}}\left(  1-\lambda\varkappa\right)  ^{n_{\varkappa}}\right\vert
		}{\left(  y^{\prime}M_{X}y\right)  ^{\frac{n}{2}}}\right)  -(n-n_{\omega}%
		)\log(\left\vert 1-\lambda\omega\right\vert ), \label{two terms}%
		\end{align}
		where $n_{\varkappa}$ denotes the algebraic multiplicity of an eigenvalue
		$\varkappa$. The first term in equation (\ref{two terms}) is a.s.\ bounded as
		$\lambda\rightarrow\omega^{-1}$. The second term goes to $+\infty$ as
		$\lambda\rightarrow\omega^{-1}$, because $n_{\omega}<n$ by the assumption that
		$W$ has at least one positive and one negative eigenvalues. Thus,
		$\lim_{\lambda\rightarrow\omega^{-1}}l(\lambda)=+\infty$ a.s.\ (ii) If
		Assumption \ref{assum id} is violated for $\omega=0$, equation
		(\ref{prof lik ASSUM1 violated}) gives $l(\lambda)=\log\left\vert \det\left(
		S(\lambda)\right)  \right\vert -\frac{n}{2}\log(y^{\prime}M_{X}y)$, for any
		$y\notin\operatorname{col}(X).$
	\end{pff}
	
	\begin{rem}
		More can be said about $\hat{\lambda}_{\mathrm{ML}}$ in case (ii) of Lemma
		\ref{lemma l(lambda) violation}. In that case, $\hat{\lambda}_{\mathrm{ML}}$
		is a zero of $\mathrm{tr}(G(\lambda))$, because $\frac{\diff}{\diff\lambda
		}\log\left\vert \det\left(  S(\lambda)\right)  \right\vert =-\mathrm{tr}%
		(G(\lambda))$ and $\log\left\vert \det\left(  S(\lambda)\right)  \right\vert
		\rightarrow-\infty$ as $\lambda$ approaches any real zero of $\det
		(S(\lambda))$. For many weights matrices $W$, $\mathrm{tr}(G(\lambda))$ has
		the same sign as $\lambda$ on $\Lambda$, which implies that $\log\left\vert
		\det\left(  S(\lambda)\right)  \right\vert $ is single-peaked on $\Lambda$
		with peak at $0$, and hence $\hat{\lambda}_{\mathrm{ML}}=0$. For example,
		$\mathrm{tr}(G(\lambda))$ has the same sign as $\lambda$ on $\Lambda$ whenever
		all eigenvalues of $W$ are real and $\mathrm{tr}(W)=0$.
	\end{rem}
	
	\begin{mycomment}
		for
		myself: under the condition that $\mathrm{tr}(G(\lambda))$ has the same sign
		as $\lambda$ (see OLS notes), the max is zero, but it can be some other zero
		of $\mathrm{tr}(G(\lambda))$ otherwise, see
		example\_MW\_equal\_zero.docx)
	\end{mycomment}

	\section{\label{app smooth} The profile log-likelihood $l(\lambda)$}
	
	This section establishes that the (quasi) profile log-likelihood function
	$l(\lambda)$, given in equation (\ref{prof lik}), is a.s.\ well defined. Let
	us start by considering a fixed $y\in\mathbb{R}^{n}$. For any $\lambda$ such
	that $S(\lambda)$ is nonsingular, $\beta$ and $\sigma^{2}$ can be concentrated
	out of the log-likelihood (\ref{loglik}) if and only if there is no $\beta
	\in\mathbb{R}^{k}$ such that $S(\lambda)y-X\beta\neq0$, or equivalently, if
	$M_{X}S(\lambda)y\neq0$. If $M_{X}S(\lambda)y=0$ for some $\lambda$ such that
	$S(\lambda)$ is nonsingular, then $y$ is fitted perfectly (zero residuals) by
	the SAR model and the profile log-likelihood is not defined for that value of
	$\lambda$. More precisely, if $M_{X}S(\lambda)y=0$ is satisfied by a unique
	value $\bar{\lambda}$ of $\lambda$, then the profile log-likelihood function
	approaches $+\infty$ as $\lambda$ approaches $\bar{\lambda}$. If
	$M_{X}S(\lambda)y=0$ for more than one value of $\lambda$, then $M_{X}%
	S(\lambda)y=0$ for any $\lambda$
	\citep[see Lemma S.1.1 in the online supplement to][]{Hillier2017}, and hence
	the whole profile log-likelihood function is undefined.
	
	Let us now turn to the case of random $y$. Provided that the distribution of
	$y$ is absolutely continuous with respect to $\mu_{\mathbb{R}^{n}}$, the set
	of $y$'s such that $M_{X}S(\lambda)y=0$, for a fixed $\lambda$ such that
	$S(\lambda)$ is nonsingular, is $\mu_{\mathbb{R}^{n}}$-null (because
	$\mathrm{null}\left(  M_{X}S(\lambda)\right)  $ has dimension $k$ if
	$\mathrm{rank}\left(  S(\lambda)\right)  =n$, and $k<n$ by assumption). That
	is, the profile log-likelihood $l(\lambda)$ is a.s.\ well defined at any fixed
	value of $\lambda$ such that $S(\lambda)$ is nonsingular. This does not imply
	that the \textit{function} $\lambda\mapsto l(\lambda)$ is a.s.\ well defined,
	because the fact that $\mathrm{null}\left(  M_{X}S(\lambda)\right)  $ is
	$\mu_{\mathbb{R}^{n}}$-null for any $\lambda$ such that $\det\left(
	S(\lambda)\right)  \neq0$ does not guarantee that the uncountable union of
	$\mathrm{null}\left(  M_{X}S(\lambda)\right)  $ over the set $\{\lambda
	\in\mathbb{R}^{n}:\det\left(  S(\lambda)\right)  \neq0\}$ is $\mu
	_{\mathbb{R}^{n}}$-null. That the function $l(\lambda)$ is a.s.\ well defined
	on its domain is established by the following result.
	
	\begin{lemma}
		\label{lemma lp smooth SLM}In a SAR model, the profile log-likelihood function
		$l(\lambda)$ is a.s.\ well defined for any $\lambda$ such that $S(\lambda)$ is nonsingular.
	\end{lemma}
	
	\begin{pff}
		We need to show that the set $A\coloneqq \{y\in\mathbb{R}^{n}:M_{X}S(\lambda)y=0$ for
		some $\lambda$ such that $\det(S(\lambda))\neq0\}$ is a $\mu_{\mathbb{R}^{n}}%
		$-null set. Suppose first that Assumption \ref{assum id} is violated for an
		eigenvalue $\omega$, which must be real and unique by parts (i) and (ii) of
		Lemma \ref{lemma remarks}, of $W$. That is, $M_{X}(\omega I_{n}-W)=0$, which
		implies $M_{X}S(\lambda)=(1-\lambda\omega)M_{X}$, and hence
		$A=\operatorname{col}(X)$, which is clearly $\mu_{\mathbb{R}^{n}}$-null since
		$k<n$ by assumption. For the rest of the proof, suppose Assumption
		\ref{assum id} holds. We establish that $A$ is a $\mu_{\mathbb{R}^{n}}$-null
		set by showing that the larger set $B\coloneqq \{y\in\mathbb{R}^{n}:M_{X}%
		S(\lambda)y=0$ for some $\lambda\in\mathbb{R}\}$ is a $\mu_{\mathbb{R}^{n}}%
		$-null set. Note that $B=\{y\in\mathbb{R}^{n}:\mathrm{rank}(y,Wy,X)<k+2\}$,
		and that $\mathrm{rank}(y,Wy,X)<k+2$ if and only if the determinants of all
		$\left(  k+2\right)  \times\left(  k+2\right)  $ submatrices of $(y,Wy,X)$ are
		zero (or, equivalently, if and only if $\mathrm{\det}\left(  (y,Wy,X)^{\prime
		}(y,Wy,X)\right)  =0$). Hence, $B$ is an algebraic variety and as such it is
		either a $\mu_{\mathbb{R}^{n}}$-null set or the whole $\mathbb{R}^{n}$. We now
		show that $B$ cannot be $\mathbb{R}^{n}$ under Assumption \ref{assum id}.
		Noting that the problem is invariant to a change of basis for $\mathbb{R}^{n}%
		$, we can take $X=(e_{n-k+1},\ldots,e_{n})=\left(  0,I_{k}\right)  ^{\prime}$,
		where $e_{i}$ is the $i$-th standard unit vector. In order for $B=\mathbb{R}%
		^{n}$, the determinants of all $\left(  k+2\right)  \times\left(  k+2\right)
		$ submatrices of $(y,Wy,X)$ must be zero for all $y\in\mathbb{R}^{n}$.
		Recalling that $n\geq k+2$ by assumption, this requires, in particular, that
		the matrix
		\[
		\left[
		\begin{array}
		[c]{cc}%
		y_{i} & \left(  Wy\right)  _{i}\\
		y_{j} & \left(  Wy\right)  _{j}%
		\end{array}
		\right]
		\]
		is singular for any $i,j\leq n-k$, and for any $y\in\mathbb{R}^{n}$ (note that
		if the assumption $n\geq k+2$ was violated, then $B$ would be equal to
		$\mathbb{R}^{n}$). Now, $y_{i}\left(  Wy\right)  _{j}-y_{j}(Wy)_{i}=0$ for any
		$y\in\mathbb{R}^{n}$ implies $\left(  Wy\right)  _{i}=\alpha y_{i}$ and
		$\left(  Wy\right)  _{j}=\alpha y_{j}$, for some scalar $\alpha$. Hence, in
		order for $B=\mathbb{R}^{n}$ there must be a scalar $\alpha$ such that
		$y-\alpha Wy\in\operatorname{col}(X)$ for any $y$, which contradicts
		Assumption \ref{assum id}. Thus, $B$, and hence $A$, must be a $\mu
		_{\mathbb{R}^{n}}$-null set.
	\end{pff}
	
	\begin{mycomment}
		alternatively, to prove the lemma, having noted that $B$ is an algebraic
		variety and as such it is either a $\mu_{\mathbb{R}^{n}}$-null set or the
		whole $\mathbb{R}^{n}$, it would suffice to exhibit one $y$ s.t.
		$M_{X}S(\lambda)y\neq0$ for some $\lambda\in\mathbb{R}$. I tried that but
		didn't manage
	\end{mycomment}

	\begin{mycomment}
		for myself: $A=\bigcup_{\lambda:\det(S(\lambda))\neq0}\mathrm{null}%
		(M_{X}S(\lambda))$, $B=\bigcup_{\lambda\in\mathbb{R}}\mathrm{null}%
		(M_{X}S(\lambda))$
		wtp the "for any $y\in\mathbb{R}^{n}$ there is $\lambda\in\mathbb{R}$ such
		that $S(\lambda)y\in\operatorname{col}(X)$" (i.e., A=$\mathbb{R}^{n}$)
		implies $\operatorname{col}(\omega I_{n}-W)\subseteq\operatorname{col}(X)$
		that is, I need to prove that if $\operatorname{col}(\omega I_{n}%
		-W)\nsubseteq\operatorname{col}(X)$ then there are some y's such that there is
		no $\lambda$ s.t. $M_{X}S(\lambda)y=0$ (some y's that cannot be perfectly fitted)
		suppose $\operatorname{col}(\omega I_{n}-W)\nsubseteq\operatorname{col}(X)$
		for $\omega\neq0$ then there are y's s.t. $M_{X}S(\omega^{-1})y\neq0$ and
		some s.t. $M_{X}S(\omega^{-1})y=0$
		OR wtp there is at least one y s.t. $\mathrm{rank}(y,Wy,X)=k+2$, i.e. $y,Wy,X$
		are l.i., i.e. a y s.t. there is no $\lambda$ s.t. $S(\lambda)y\in
		\operatorname{col}(X)$
		show it is impossible that for any $y$ $S(\lambda)y\in\operatorname{col}(X)$
		for some $\lambda\in R$ s.t. $S(\lambda)$ is nonsing. y and Wy span a plane,
		so unless $\operatorname{col}(S(\lambda))\subset\operatorname{col}(X)$
		$\operatorname{col}(S(\lambda))\nsubseteq\operatorname{col}(X)$ for some
		$\lambda$ s.t. $S(\lambda)$ is nonsing. hence there are y's s.t. $S_{\lambda
		}y\notin\operatorname{col}(X)$. what is A$_{\lambda}\coloneqq $\{y$\in R^{n}%
		:S(\lambda)y\in\operatorname{col}(X)$\}? It's $null(M_{X}S(\lambda))$
		we want to show there is at least one y that is not in $\bigcup_{\lambda\in
			R}\mathrm{null}(M_{X}S(\lambda))$
		$M_{X}S(\lambda)y=0$ means there is a l.c. of $y$ and $Wy$ that belongs to
		$\operatorname{col}(X)$
		$\operatorname{col}(\omega I_{n}-W)\subseteq\operatorname{col}(X)$ for
		$\omega\neq0$ iff for any $y\in\mathbb{R}^{n}$ $(I_{n}-\omega^{-1}%
		W)y\in\operatorname{col}(X)$ (or $M_{X}S(\omega^{-1})y=0$). and hence
		A=$\mathbb{R}^{n}$
		$%
		\begin{array}
		[c]{cccc}%
		a & e & 0 & 0\\
		b & f & 0 & 0\\
		c & g & 1 & 0\\
		d & h & 0 & 1
		\end{array}
		$, determinant: $af-be$
		\bigskip
		$%
		\begin{array}
		[c]{cccc}%
		a & c\ast a & 0 & 0\\
		b & c\ast b & 0 & 0\\
		c & g & 1 & 0\\
		d & h & 0 & 1
		\end{array}
		$, rank: $3$
	\end{mycomment}

	\begin{mycomment}
		When Assumption \ref{assum id} is satisfied, $l(\lambda)$ a.s.\ goes to $-\infty$ at each zero of $\det(S(\lambda))$ (see Proposition 1
		in \cite{Hillier2017}), and hence Lemma \ref{lemma lp smooth SLM} implies that
		$l(\lambda)$ a.s.\ has at least one critical point corresponding to a maximum
		between any two consecutive real zeros of $\det(S(\lambda))$.
	\end{mycomment}

	It is worth noting that, provided it is well defined, $l(\lambda)$ is
	$C^{\infty}$ between any two consecutive real zeros of $\det(S(\lambda))$
	(because the term $\log\left\vert \det\left(  S(\lambda)\right)  \right\vert $
	in equation (\ref{prof lik}) is). Also, on comparing equation (\ref{prof lik})
	for $l(\lambda)$ and equation (\ref{l a lam}) for $l_{\mathrm{a}}(\lambda)$,
	it is clear that the statement of Lemma \ref{lemma lp smooth SLM} also applies
	to $l_{\mathrm{a}}(\lambda)$. Finally, recall from above that the set
	$\mathrm{null}\left(  M_{X}S(z)\right)  $ (the set of $y$'s that are perfectly
	fitted by a SAR model with $\lambda=z$) is $\mu_{\mathbb{R}^{n}}$-null for any
	$z$ such $S(z)$ is nonsingular. For the values of $z$ such that $S(z)$ is
	singular (i.e., for the values $z=\omega^{-1}$, for the nonzero real
	eigenvalues $\omega$ of $W$), $\mathrm{null}\left(  M_{X}S(z)\right)  $ is a
	$\mu_{\mathbb{R}^{n}}$-null set if Assumption \ref{assum id} holds, whereas
	$\mathrm{null}\left(  M_{X}S(z)\right)  $ is the whole $\mathbb{R}^{n}$ if
	Assumption \ref{assum id} is violated. In the next section we will prove a
	result, Lemma \ref{lemma extr zero prob}, that establishes the limiting
	behavior of $l_{\mathrm{a}}(\lambda)$ as $\lambda\rightarrow\omega^{-1}$ when
	$y\in\mathrm{null}(M_{X}S(\omega^{-1}))$.
	
	\section{\label{sec suppl analytical expr} Remarks on Proposition
		\ref{prop l_a SLM}}
	
	\begin{rem}
		From the proof of Proposition \ref{prop l_a SLM} it is clear that if
		$S(\lambda)$ does not have any real nonpositive eigenvalues, then the
		\textit{principal matrix logarithm} (that is, the logarithm associated to the
		branch cut $(-\infty,0]$) can be used in equation (\ref{l_a(sig,lambda)}).
		This is certainly the case for any $\lambda\in\Lambda$, because $S(\lambda)$
		is positive definite on $\Lambda$. A different branch cut is required to
		evaluate $l_{\mathrm{a}}{(\sigma^{2},\lambda)}$ outside $\Lambda$.
	\end{rem}
	
	\begin{rem}
		Proposition \ref{prop l_a SLM} shows that the effect of the score adjustment
		on the profile log-likelihood (\ref{lik sig lam SAR}) is to replace $n$ with
		$n-k$ and the term $\log\left\vert \det\left(  S(\lambda)\right)  \right\vert
		=\operatorname{Re}\left[  \mathrm{tr}\left(  \log S(\lambda)\right)  \right]
		$ with $\operatorname{Re}\left[  \mathrm{tr}(M_{X}\log S(\lambda))\right]  $,
		for any $\lambda$ such that $S(\lambda)$ is invertible. If we restrict
		attention to $\lambda\in\Lambda$, then $S(\lambda)$ is positive definite, and
		the adjustment amounts to replacing $\log\left(  \det\left(  S(\lambda
		)\right)  \right)  =\mathrm{tr}(\log S(\lambda))$ with $\mathrm{tr}(M_{X}\log
		S(\lambda))$ (the matrix logarithm here can be taken to be the principal one,
		by the previous remark).
	\end{rem}
	
	\section{\label{sec more} Further results related to Theorem
		\ref{lemma lim gen W}}
	
	The following lemma complements Lemma \ref{lemma null} in the main text by
	giving a condition for $\mathrm{tr}(M_{X}Q_{\omega})=0$ in terms of right and
	left eigenvectors of $W$ associated to $\omega$.
	
	\begin{lemma}
		\label{lemma left right}For any simple eigenvalue $\omega$ of $W$,
		$\mathrm{tr}(M_{X}Q_{\omega})=0$ if and only if $l^{\prime}M_{X}h=0$, for some
		(and hence all) $h\in\mathrm{null}(W-\omega I_{n})$ and for some (and hence
		all) $l\in\mathrm{null}(W^{\prime}-\omega I_{n})$.
	\end{lemma}
	
	\begin{pff}
		[Proof of Lemma \ref{lemma left right}]If $\omega$ is simple, $Q_{\omega
		}=hl^{\prime}/l^{\prime}h$ for some (and hence all) $h\in\mathrm{null}
		(W-\omega I_{n})$ and for some (and hence all) $l\in\mathrm{null}(W^{\prime
		}-\omega I_{n})$. It follows that $\mathrm{tr}(M_{X}Q_{\omega})=0$ if and only
		if $\mathrm{tr}(l^{\prime}M_{X}h)=0$, or equivalently $l^{\prime}M_{X}h=0$.
	\end{pff}
	
	It is a well known fact in linear algebra that left and right eigenvectors
	corresponding to the same eigenvalue cannot be orthogonal. Lemma
	\ref{lemma left right} establishes that $\mathrm{tr}(M_{X}Q_{\omega})=0$ if
	and only if left and right eigenvectors corresponding to $\omega$ are
	orthogonal after orthogonal projection onto $\operatorname{col}^{\perp}(X)$.
	
	Next, recall that the conclusions in Theorem \ref{lemma lim gen W} require
	$y\notin\mathrm{null}(M_{X}S(\omega^{-1}))$, with $\mathrm{null}(M_{X}%
	S(\omega^{-1}))$ a $\mu_{\mathbb{R}^{n}}$-null set under Assumption
	\ref{assum id}. It is therefore of interest to elucidate what happens when
	$y\in\mathrm{null}(M_{X}S(\omega^{-1}))$. Note that if $y\in\mathrm{null}%
	(M_{X}S(\omega^{-1}))$ and $y\in\mathrm{null}(M_{X}S(\lambda))$ for some
	$\lambda\in\mathbb{R}/\{\omega^{-1}\}$, then, by Lemma S.1.1 in the online
	supplement to \cite{Hillier2017}, $y\in\mathrm{null}(M_{X}S(\lambda))$ for
	every $\lambda$, and hence $l_{\mathrm{a}}(\lambda)$ is undefined for every
	$\lambda$.
	
	\begin{lemma}
		\label{lemma extr zero prob}Consider an arbitrary semisimple nonzero real
		eigenvalue $\omega$ of $W$. If $y\in\mathrm{null}(M_{X}S(\omega^{-1}))$, and
		$y\notin\mathrm{null}(M_{X}S(\lambda))$ for any $\lambda\in\mathbb{R}%
		/\{\omega^{-1}\}$, then $\lim_{\lambda\rightarrow\omega^{-1}}l_{\mathrm{a}%
		}(\lambda)$ is
		
		\begin{enumerate}
			\item[(i)] $-\infty$ if $\mathrm{tr}(M_{X}Q_{\omega})>n-k$;
			
			\item[(ii)] bounded if $\mathrm{tr}(M_{X}Q_{\omega})=n-k;$
			
			\item[(iii)] $+\infty$ if $\mathrm{tr}(M_{X}Q_{\omega})<n-k$.
		\end{enumerate}
	\end{lemma}
	
	\begin{pff}
		Assume $y\in\mathrm{null}(M_{X}S(\omega^{-1}))$, and $y\notin\mathrm{null}%
		(M_{X}S(\lambda))$ for any $\lambda\in\mathbb{R}/\{\omega^{-1}\}$. Then,
		$s_{\mathrm{a}2}(\lambda)$ in equation (\ref{sa2}) is well defined for every
		$\lambda\ $such that $S(\lambda)$ is non singular. Since $y\in\mathrm{null}%
		(M_{X}S(\omega^{-1}))$, we have $M_{X}Wy=\omega M_{X}y$ and hence
		$M_{X}S(\lambda)y=(1-\lambda\omega)M_{X}y$. Plugging this last expression into
		equation (\ref{sa2}), we obtain
		\begin{align*}
		s_{\mathrm{a}2}(\lambda)  &  =(n-k)\frac{\omega(1-\lambda\omega)y^{\prime
			}M_{X}y}{(1-\lambda\omega)^{2}y^{\prime}M_{X}y}-\mathrm{tr}(M_{X}G(\lambda))\\
		&  =(n-k)\frac{\omega}{1-\lambda\omega}-\mathrm{tr}(M_{X}G(\lambda)).
		\end{align*}
		Equation (\ref{trMG2}) now gives
		\[
		s_{\mathrm{a}2}(\lambda)=\left(  (n-k)-\mathrm{tr}(M_{X}Q_{\omega})\right)
		\frac{\omega}{1-\lambda\omega}-\sum_{\chi\in\mathrm{Sp}(W)\setminus\{\omega
			\}}\sum_{i=0}^{k_{\chi}-1}\frac{f^{(i)}(\chi)}{j!}\mathrm{tr}\left(
		M_{X}(W-\chi I_{n})^{i}T_{\chi}\right)  .
		\]
		Since all derivatives $f^{(i)}(\chi)$ are bounded for $\lambda\neq\chi^{-1}$,
		it follows that: if $\mathrm{tr}(M_{X}Q_{\omega})<n-k$, then $\lim
		_{\lambda\uparrow\omega^{-1}}s_{\mathrm{a2}}(\lambda)=+\infty$ and
		$\lim_{\lambda\downarrow\omega^{-1}}s_{\mathrm{a2}}(\lambda)=-\infty$, that
		is, $\lim_{\lambda\rightarrow\omega^{-1}}l_{\mathrm{a}}(\lambda)=+\infty$; if
		$\mathrm{tr}(M_{X}Q_{\omega})>n-k$, then $\lim_{\lambda\rightarrow\omega^{-1}%
		}l_{\mathrm{a}}(\lambda)=-\infty$; if $\mathrm{tr}(M_{X}Q_{\omega})=n-k$, then
		$\lim_{\lambda\rightarrow\omega^{-1}}l_{\mathrm{a}}(\lambda)$ is bounded.
	\end{pff}
	
	Finally, we remark that, for a fixed eigenvalue $\omega$, the statement of
	Theorem \ref{lemma lim gen W} does not require the full Assumption
	\ref{assum id}, but only that $\operatorname{col}(\omega I_{n}-W)\nsubseteq
	\operatorname{col}(X)$ for that specific $\omega$. However, the result is
	trivial whenever Assumption \ref{assum id} fails, because in that case
	$l_{\mathrm{a}}(\lambda)$ is flat (and hence $\mathrm{tr}(M_{X}Q_{\omega}%
	)=0$); see \cite{Martellosio2018}.
	
	\section{\label{sec further simul} Additional simulation evidence}
	
	\subsection{\label{sec sup unr}Relationship between the adjusted MLE and the
		unrestricted MLE}
	
	We report some simulation evidence on the relationship between the adjusted
	MLE $\hat{\lambda}_{\mathrm{aML}}$ and the unrestricted MLE $\hat{\lambda
	}_{\mathrm{uML}}$. We employ the simulation design that is used in Section
	\ref{sec cross} for Table \ref{Table cross landscape}, with $W$ row
	normalized, and $\lambda=0.5$.
	
	In Table \ref{Table unr}, the columns headed by \textquotedblleft$\hat
	{\lambda}_{\mathrm{uML}}>1$\textquotedblright\ and \textquotedblleft%
	$\hat{\lambda}_{\mathrm{aML}}>1$\textquotedblright\ report the percentage of
	Monte Carlo repetitions in which, respectively, $\hat{\lambda}_{\mathrm{uML}%
	}>1$ and $\hat{\lambda}_{\mathrm{aML}}>1$; the columns headed by
	\textquotedblleft$\mathrm{ua}$\textquotedblright\ report the percentage of
	times that $\hat{\lambda}_{\mathrm{uML}}>1$\ out of all repetitions in which
	$\hat{\lambda}_{\mathrm{aML}}>1$; the columns headed by \textquotedblleft%
	$\mathrm{au}$\textquotedblright\ report the percentage of times that
	$\hat{\lambda}_{\mathrm{aML}}>1$\ out of all repetitions in which
	$\hat{\lambda}_{\mathrm{uML}}>1$.
	
	The main conclusions are as follows. Firstly, the probabilities that
	$\hat{\lambda}_{\mathrm{uML}}$ and $\hat{\lambda}_{\mathrm{aML}}$ are greater
	than $1$ increase with the density of $W$ and with the rewiring probability
	$p$. The table reports the estimate of these probabilities only for
	$\lambda=0.5$; of course, the probabilities would be larger for larger values
	of $\lambda$. Secondly, when $\hat{\lambda}_{\mathrm{uML}}$ is greater than 1,
	$\hat{\lambda}_{\mathrm{aML}}$ is almost always greater than 1 too, and, on
	the other hand, the probability that $\hat{\lambda}_{\mathrm{uML}}$ is greater
	than 1 given that $\hat{\lambda}_{\mathrm{aML}}$ is greater than 1 is
	decreasing in $\tilde{k}$ and increasing in $p$.
	
	\begin{table}[h]
		\caption{Study of the probabilities that $\hat{\lambda}_{\mathrm{uML}}$ and
			$\hat{\lambda}_{\mathrm{aML}}$ are greater than 1, for a SAR model on a
			Watts-Strogatz network of size $n=200$, when $W$ is row normalized,
			$\lambda=0.5$, and the errors are i.i.d. $\mathrm{N}(0,1)$ }%
		\label{Table unr}
		\begin{adjustbox}{width=1\linewidth}		
			$
			\begin{array}[c]{lllLLLLlLLLLlLLLL}
			\hline
			&  & & \multicolumn{4}{l}{\tilde{k}=1} &  & \multicolumn{4}{l}{\tilde{k}=3}&  & \multicolumn{4}{l}{\tilde{k}=5}\bigstrut[t]\\
			\cline{4-7}\cline{9-12}\cline{14-17}

			p & h &  & \mc{\hat{\lambda}_{\mathrm{uML}}>1} & \mc{\hat{\lambda}_{\mathrm{aML}}>1} & \mc{\mathrm{ua}} & \mc{\mathrm{au}} &&
			\mc{\hat{\lambda}_{\mathrm{uML}}>1} & \mc{\hat{\lambda}_{\mathrm{aML}}>1} & \mc{\mathrm{ua}} & \mc{\mathrm{au}} &&\mc{\hat{\lambda}_{\mathrm{uML}}>1} & \mc{\hat{\lambda}_{\mathrm{aML}}>1} & \mc{\mathrm{ua}} & \mc{\mathrm{au}}\bigstrut[t]\bigstrut[b]\\
			\hline
			
			\multicolumn{1}{l}{0} & \multicolumn{1}{l}{5} &&  0		& 0		&\mc{-}	&\mc{-}&&	0		& 0		&\mc{-}	&\mc{-}&&0		& 0		&\mc{-}	&\mc{-}\bigstrut[t] \\
			\multicolumn{1}{l}{} & \multicolumn{1}{l}{10} &&  0		& 0		&\mc{-}	&\mc{-}&&	0		& 0		&\mc{-}	&\mc{-}&&0		& 0		&\mc{-}	&\mc{-} \\
			\multicolumn{1}{l}{} & \multicolumn{1}{l}{50} && 2.64	&3.45		&76.43	&100.00&&0.57		&11.18		&5.12	&100.00&&0.12&21.13		&0.58	&100.00	\\
			\multicolumn{1}{l}{} & \multicolumn{1}{l}{75} && 24.09	&26.28		&91.63	&100.00&&5.38		&29.74		&51.59	&99.73&&9.69		&33.20		&29.20	&100.00	\\
			\\[-2.5mm]
			
			\multicolumn{1}{l}{0.2} & \multicolumn{1}{l}{5}  &&  0		& 0		&\mc{-}	&\mc{-}&&	0		& 0		&\mc{-}	&\mc{-}&&0		& 0		&\mc{-}	&\mc{-}\\
			\multicolumn{1}{l}{} & \multicolumn{1}{l}{10} &&	0.00	&0.00		&100.00	&100.00&&0.00		& 0.00		&50.00	&100.00&&0		& 0		&\mc{-}	&\mc{-}\\
			\multicolumn{1}{l}{} & \multicolumn{1}{l}{50} &&	14.76	&16.16		&91.34	&100.00&&10.84		&18.09		&59.91	&100.00&&8.15		&19.68		&41.43	&100.00\\
			\multicolumn{1}{l}{} & \multicolumn{1}{l}{75} &&	33.15	&33.41		&99.23	&100.00&&32.24		&33.79		&95.28	&99.86&&30.91		&34.07		&90.70	&99.96\\
			\\[-2.5mm]
			\multicolumn{1}{l}{0.5} & \multicolumn{1}{l}{5} && 0.00		&0.00		&94.12	&100.00&&0.00		&0.00		&100.00	&100.00&&0.00		&0.00		&66.67	&100.00\\
			\multicolumn{1}{l}{} & \multicolumn{1}{l}{10} &&	0.49	&0.51		&96.01	&100.00&&0.27		&0.35		&76.98	&100.00&&0.22		&0.33		&65.49	&100.00\\
			\multicolumn{1}{l}{} & \multicolumn{1}{l}{50} &&	22.95	&23.11		&99.30	&100.00&&22.22		&23.33		&95.21	&99.96&&21.33		&23.52		&90.68	&100.00\\
			\multicolumn{1}{l}{} & \multicolumn{1}{l}{75} &&	34.13	&34.19		&99.84	&100.00&&34.02		&34.36		&98.84	&99.83&&33.97		&34.55		&98.17	&99.86\\
			\\[-2.5mm]
			\multicolumn{1}{l}{1} & \multicolumn{1}{l}{5} && 0.01		&0.01		&94.83	&100.00&&0.00		&0.01		&80.00	&100.00&&0.00		&0.00		&65.91	&100.00\\
			\multicolumn{1}{l}{} & \multicolumn{1}{l}{10} &&	1.09	&1.10		&98.77	&100.00&&0.95		&1.01		&93.55	&100.00&&0.79		&0.90		&88.23	&100.00\\
			\multicolumn{1}{l}{} & \multicolumn{1}{l}{50} &&	23.83	&23.88		&99.78	&100.00&&23.89		&24.16		&98.73	&99.86&&23.96		&24.31		&98.26	&99.70\\
			\multicolumn{1}{l}{} & \multicolumn{1}{l}{75} &&	34.16	&34.21		&99.84	&100.00&&33.91		&34.32		&98.65	&99.82&&34.08		&34.55		&98.39	&99.75\bigstrut[b]\\
			\hline
			
		\end{array}
		$
		
	\end{adjustbox}
\end{table}

\begin{mycomment}
	Table \ref{Table unr} produced using proportion_unrestricted_MLE_greater_1.m (results available in proportion_unrestricted_MLE_greater_1_results.docx)
	note that some 0.00 are zeros some are v small numbers
\end{mycomment}

\subsection{Large $R$}

Table \ref{Table panel R=100} extends the results reported in Table
\ref{Table panel} in the paper to the case when $R=100$ and $W$ is row
normalized. At the sample sizes considered in the table, the bias of both the
LLL estimator and the adjusted QMLE is small, but, as expected, the adjusted
QMLE still offers an improvement especially when $\tilde{k}$ is large.
\begin{table}[tbh]
	\caption{Model (\ref{network model}) with $R=100$, $\lambda=0.5$, $h=5$ and
		$p=0.2$, when $W$ is row normalized.}%
	\label{Table panel R=100}%
	\captionsetup{width=14cm} \begin{threeparttable}	
		\begin{adjustbox}{width=.58\textwidth,center=\textwidth}
			$		
			\begin{array}[c]{llcccL}\hline
			
			&   & \hat{\lambda}_{\mathrm{LLL}} &
			\hat{\lambda}_{\mathrm{aML}} &	&\bigstrut[t]\\
			\tilde{k} &  m & \text{bias(s.d.)} &
			\text{bias(s.d.)} & \Delta\%\left\vert \text{bias}\right\vert
			& \mc{\Delta\%\text{RMSE}}\\\hline
			
			\multicolumn{1}{l}{2} & \multicolumn{1}{l}{20} &-0.003(0.068)&    -0.001(0.069)&        -53.61&       -0.01  \bigstrut[t]\\
			\multicolumn{1}{l}{}   & \multicolumn{1}{l}{30} & -0.003(0.043)&    -0.001(0.043)&        -44.20&       -0.01\\
			
			\\[-2.5mm]
			\multicolumn{1}{l}{10} & \multicolumn{1}{l}{20} & -0.009(0.067)&	-0.002(0.067)&           -82.70&      -0.57\\
			\multicolumn{1}{l}{}   & \multicolumn{1}{l}{30} &-0.006(0.041)&    -0.001(0.041)&           -81.93&      -1.30 \\
			
			\\[-2.5mm]
			\multicolumn{1}{l}{20}& \multicolumn{1}{l}{20} & -0.015(0.065)&    -0.001(0.066)&        -91.46&      -2.07\\
			\multicolumn{1}{l}{}   & \multicolumn{1}{l}{30} &-0.010(0.039)&    -0.001(0.039)&       -89.12  &    -3.36\\
			\\[-2.5mm]
			\multicolumn{1}{l}{30}& \multicolumn{1}{l}{20} & -0.022(0.064)&    -0.002(0.064)&        -93.17&      -5.16\\
			\multicolumn{1}{l}{}   & \multicolumn{1}{l}{30} &-0.014(0.037)&    -0.001(0.037)&        -92.37 &     -5.94			\bigstrut[b]\\
			\hline
		\end{array}
		$
		\end{adjustbox}
		\end{threeparttable}
		\end{table}\begin{mycomment}
		from results bias_MLE_panel_table_X_in_each_rep_k_varies r100.docx
		\end{mycomment}

		Table \ref{Table network fe CI r100} extends Table \ref{Table network fe CI}
		in the paper to the case when $R=100$, for various values of
		$\tilde{k}$. Even at these sample sizes, the coverage of the Wald confidence
		intervals deteriorates significantly as $\tilde{k}$ increases (particularly
		for the case of the confidence intervals based on the QMLE), but that is not
		the case for the saddlepoint confidence intervals based on the adjusted QMLE.
		
		\begin{table}[tbh]
			\caption{Empirical coverages of 95\% confidence intervals in model
				(\ref{network model}) with $R=100$, $\lambda=0$, $h=5$, and $p=0.2$. The error
				distribution is a standard normal, and $W$ is row normalized.}%
			\label{Table network fe CI r100}
			\captionsetup{width=14cm}
			\par
			\begin{adjustbox}{width=.51\textwidth,center=\textwidth}
				$
				\begin{array}[c]{llcccccccc}\hline
				&&& \multicolumn{3}{l}{\text{Two-sided}} &  & \multicolumn{3}{l}{\text{Right-sided}}\bigstrut[b]\bigstrut[t]\\
				
				\cline{4-6}	\cline{8-10}
				\tilde{k} &m&&
				\mathrm{W_{LLL}} & \mathrm{W_{aML}} & \mathrm{s_{aML}} && \mathrm{W_{LLL}} & \mathrm{W_{aML}} & \mathrm{s_{aML}}\bigstrut[b]\bigstrut[t]\\\hline
				\multicolumn{1}{l}{2} & \multicolumn{1}{l}{20}&& 0.950	&0.950&0.951&&			0.945&0.947&0.951 \bigstrut[t]\\
				
				& \multicolumn{1}{l}{30}&&   0.949&	0.949&	0.950&&		0.947&	0.949&	0.950  \\
				\\[-2.5mm]
				
				\multicolumn{1}{l}{10} & \multicolumn{1}{l}{20}&& 0.947	&0.948&0.950&&			0.939&0.948&0.950  \\
				
				& \multicolumn{1}{l}{30}&&   0.948&	0.948&	0.950&&		0.936&	0.947&	0.949  \\
				\\[-2.5mm]
				\multicolumn{1}{l}{20} & \multicolumn{1}{l}{20}&& 0.945&   0.947&  0.950&& 			0.928&0.947&0.950       \\
				
				& \multicolumn{1}{l}{30}&&   0.944&	0.946&	0.948&&		0.928&	0.947&	0.949 \\
				\\[-2.5mm]
				\multicolumn{1}{l}{30} & \multicolumn{1}{l}{20}&& 0.939&  0.944&  0.948&&     0.917&  0.944&  0.950  \\
				
				& \multicolumn{1}{l}{30}&&  0.941&  0.947&  0.949&&     0.919&  0.947&  0.949  \\
				\\[-2.5mm]
				\multicolumn{1}{l}{50} & \multicolumn{1}{l}{20}&& 0.928&  0.942&  0.947&&     0.896&  0.943&  0.949   \\
				
				& \multicolumn{1}{l}{30}&&   0.934&  0.946&  0.949&&     0.901&  0.948&  0.949  \bigstrut[b] \\
				
				\hline
			\end{array}
			$
			\end{adjustbox}
			\end{table}
			
			\subsection{Non-normality}
			
			Table \ref{Table other distrib} extends Table \ref{Table network fe CI} in the
			paper to other non-normal distributions. More specifically, the errors
			$\varepsilon_{ri}$ are generated independently from either: (a) a
			$\mathrm{Laplace}(0,2^{-1/2}$) distribution, (b) a $\chi_{3}$ distribution,
			standardized so that its mean is 0 and its variance is 1, (c) a gamma
			distribution with shape parameter 1/2 and scale parameter 1, standardized so
			that its mean is 0 and its variance is 1. Skewness and kurtosis are,
			respectively, $0$ and $6$ in case (a); $\sqrt{8/3}=1.63$ and $7$ in case (b);
			$2^{3/2}=2.83$ and $15$ in case (c).
			
			\begin{mycomment}
				for plots of the pdf's of these distributions, see distributions_mean_var_hist.m
			\end{mycomment}

			\section{\label{sec suppl ci} The Lugannani--Rice approximation}
			
			This section provides additional details about the approximation needed for
			the construction of the confidence intervals in Section \ref{sec ci} of the
			paper. The first-order Lugannani--Rice approximation to the cdf $\Pr(V\leq v)$
			of a random variable $V$ having cumulant generating function (cgf)
			$K_{V}(\cdot)$ is%
			
			\begin{equation}
			\widetilde{\Pr}(V\leq v)\coloneqq \left\{
			\begin{array}
			[c]{cl}%
			\Phi(\hat{w})+\phi(\hat{w})\left(  \frac{1}{\hat{w}}-\frac{1}{\hat{u}}\right)
			, & \text{if }v\neq\mathrm{E}\left(  V\right)  ,\\
			\frac{1}{2}+\frac{K_{V}^{\prime\prime\prime}(0)}{6\sqrt{2\pi}(K_{V}%
				^{^{\prime\prime}}(0))^{\frac{3}{2}}}, & \text{if }v=\mathrm{E}\left(
			V\right)  ,
			\end{array}
			\right.  \label{lug-rice}%
			\end{equation}
			where $\Phi(\cdot)$ and $\phi(\cdot)$ denote the cdf and pdf of the standard
			normal distribution, respectively, primes denote derivatives, and%
			\begin{equation}
			\hat{w}\coloneqq \mathrm{sgn}(\hat{s})\sqrt{-2K_{V}(\hat{s})},\text{ }\hat{u}\coloneqq \hat
			{s}\sqrt{K_{V}^{\prime\prime}(\hat{s})},
			\end{equation}
			where the saddlepoint value $\hat{s}$ is determined by $K_{V}^{\prime}(\hat
			{s})=v$ \citep[see, e.g.,][p. 12]{Butler2007}. This is a normal-based
			approximation. For extensions to non-normal bases, see \cite{Wood1993}.
			
			Formula (\ref{lug-rice}) can be used to derive an approximation to the cdf of
			$\hat{\lambda}_{\mathrm{aML}}$ based on the equality $\Pr(\hat{\lambda
			}_{\mathrm{aML}}\leq z;\beta,\sigma^{2},\lambda)=\Pr(y^{\prime}S^{\prime
			}(z)R(z)S(z)y\leq0)$ (equation ({\ref{cdf}}) in the paper), which, remember,
			requires $l_{\mathrm{a}}(\lambda)$ to be single-peaked. More precisely,
			formula (\ref{lug-rice}) can be applied with $V=\frac{1}{\sigma^{2}}y^{\prime
			}S^{\prime}(z)R(z)S(z)y$ and $v=0$. The cgf of $\frac{1}{\sigma^{2}}y^{\prime
			}S^{\prime}(z)R(z)S(z)y$ when $y\sim\mathrm{N}(S^{-1}(\lambda)X\beta
			,\sigma^{2}\left[  S^{\prime}(\lambda)S(\lambda)\right]  ^{-1})$ is
			\begin{equation}
			K_{V}(s)=-\frac{1}{2}\log\left(  \det\left(  I_{n}-2sB(z,\lambda)\right)
			\right)  -\frac{1}{2\sigma^{2}}\beta^{\prime}X^{\prime}\left(  I_{n}-\left(
			I_{n}-2sB(z,\lambda)\right)  ^{-1}\right)  X\beta,\label{cgf Z}%
			\end{equation}
			for $s\in\left(  1/(2b_{\min}),1/(2b_{\max})\right)  $, where
			\[
			B(z,\lambda)\coloneqq \frac{1}{2}\left[  S(z)S^{-1}(\lambda)\right]  ^{\prime}\left[
			R(z)+R^{\prime}(z)\right]  \left[  S(z)S^{-1}(\lambda)\right]  ,
			\]
			and $b_{\min}$ and $b_{\max}$ denote, respectively, the smallest and the
			largest eigenvalues of $B(z,\lambda)$.\footnote{It is easily verified that the
				cgf of $y^{\prime}Ay$, for a symmetric $A$ and when $y\sim\mathrm{N}%
				(\mu,\Sigma)$ is $-\frac{1}{2}\log\left(  \det\left(  I_{n}-2sA\Sigma\right)
				\right)  -\frac{1}{2}\mu^{\prime}\left(  I_{n}-\left(  I_{n}-2sA\Sigma\right)
				^{-1}\right)  \Sigma^{-1}\mu$. Equation (\ref{cgf Z}) obtains immediately from
				this expression on rewriting the quadratic form $V=\frac{1}{\sigma^{2}%
				}y^{\prime}S^{\prime}(z)R(z)S(z)y$ as $V=\tilde{y}^{\prime}B(z,\lambda
				)\tilde{y}$, with $\tilde{y}\coloneqq \frac{1}{\sigma}S(\lambda)y\sim\mathrm{N}%
				(\frac{1}{\sigma}X\beta,I_{n}).$}
			
			We define $\widetilde{\Pr}(\hat{\lambda}_{\mathrm{aML}}\leq z;\beta,\sigma
			^{2},\lambda)\coloneqq \widetilde{\Pr}(y^{\prime}S^{\prime}(z)R(z)S(z)y\leq0)$. It is
			interesting to note that the approximation is available in closed form in the
			case $z=\lambda$. This is because $\mathrm{E}\left(  y^{\prime}S^{\prime
			}(\lambda)R(\lambda)S(\lambda)y\right)  =0$ (see Section
			\ref{subsec SLM adj prof lik} in the paper), and therefore the expression at
			the bottom of equation (\ref{lug-rice}) applies when approximating
			$\Pr(y^{\prime}S^{\prime}(z)R(z)S(z)y\leq0)$. Specifically, the probability
			that $\hat{\lambda}_{\mathrm{aML}}$ underestimates $\lambda$ has the closed
			form expression
			\[
			\widetilde{\Pr}(\hat{\lambda}_{\mathrm{aML}}\leq\lambda;\beta,\sigma
			^{2},\lambda)=\frac{1}{2}+\frac{K_{V}^{\prime\prime\prime\prime}(0)}%
			{6\sqrt{2\pi}(K_{V}^{^{\prime\prime}}(0))^{\frac{3}{2}}}.
			\]

			\begin{mycomment}
				workings:
				using mathai provost the cgf of $Z=y^{\prime}S^{\prime}(z)R(z)S(z)y=\tilde
				{y}^{\prime}B(z,\lambda)\tilde{y}$ assuming $\tilde{y}\sim\mathrm{N}%
				(X\beta,\sigma^{2}I_{n})$ is given by
				\begin{equation}
				K_{Z}(s;\beta,\sigma^{2},\lambda)=-\frac{1}{2}\log\left(  \det\left(
				I_{n}-2s\sigma^{2}B(z,\lambda)\right)  \right)  -\frac{1}{2\sigma^{2}}%
				\beta^{\prime}X^{\prime}\left(  I_{n}-\left(  I_{n}-2s\sigma^{2}%
				B(z,\lambda)\right)  ^{-1}\right)  X\beta,
				\end{equation}
				\bigskip
				the cgf of $Z=y^{\prime}S^{\prime}(z)R(z)S(z)y=y^{\ast\prime}\sigma
				^{2}B(z,\lambda)y^{\ast}$ assuming $y^{\ast}=\frac{1}{\sigma}\tilde{y}%
				\sim\mathrm{N}(\frac{1}{\sigma}X\beta,I_{n})$ is given by
				\begin{equation}
				K_{Z}(s;\beta,\sigma^{2},\lambda)=-\frac{1}{2}\log\left(  \det\left(
				I_{n}-2s\sigma^{2}B(z,\lambda)\right)  \right)  -\frac{1}{2\sigma^{2}}%
				\beta^{\prime}X^{\prime}\left(  I_{n}-\left(  I_{n}-2s\sigma^{2}%
				B(z,\lambda)\right)  ^{-1}\right)  X\beta,
				\end{equation}
				\bigskip
				so
				\begin{equation}
				K_{Z}(s;\beta,\sigma^{2},\lambda)=-\frac{1}{2}\log\left(  \det\left(
				I_{n}-2sB(z,\lambda)\right)  \right)  -\frac{1}{2\sigma^{2}}\beta^{\prime
				}X^{\prime}\left(  I_{n}-\left(  I_{n}-2sB(z,\lambda)\right)  ^{-1}\right)
				X\beta,
				\end{equation}
				is the cgf of $Z=\frac{1}{\sigma^{2}}y^{\prime}S^{\prime}(z)R(z)S(z)y=y^{\ast
					\prime}B(z,\lambda)y^{\ast}$ assuming $y^{\ast}=\frac{1}{\sigma}\tilde{y}%
				\sim\mathrm{N}(\frac{1}{\sigma}X\beta,I_{n})$ is given by
			\end{mycomment}

			\afterpage{
				\thispagestyle{empty}
				\newgeometry{left=6cm, right=0cm, top=1cm, bottom=-1.2cm}
				\begin{landscape}
					\begin{table}[ptbh]
						\captionsetup{width=17.5cm,margin={3.25cm,3.4cm}}
						
						\caption{Empirical coverages of 95\% confidence intervals in model
							(\ref{network model}) with $\lambda=0$, $h=5$, and $p=0.2$. The error distribution is either a $\mathrm{Laplace}(0,2^{-1/2}$), a $\chi_3$, or $\mathrm{gamma}(1/2,1)$ (with the $\chi_3$ and the gamma distributions standardized to have mean zero and variance 1), and $W$ is row normalized.}			\label{Table other distrib}
						
						\begin{adjustbox}{width=.78\linewidth,center=\linewidth}
							$
							\begin{array}[c]{lllcccccccccccccccccccccccc}\hline
							&  &  && \multicolumn{7}{l}{\text{Laplace}} &  & \multicolumn{7}{l}{\text{Chi-square}}&  & \multicolumn{7}{l}{\text{Gamma}}\bigstrut[b]\bigstrut[t]\\
							\cline{5-11}\cline{13-19}\cline{21-27}
							&  &  && \multicolumn{3}{l}{\text{Two-sided}} &  & \multicolumn{3}{l}{\text{Right-sided}}&  & \multicolumn{3}{l}{\text{Two-sided}} &  & \multicolumn{3}{l}{\text{Right-sided}}&  & \multicolumn{3}{l}{\text{Two-sided}} &  & \multicolumn{3}{l}{\text{Right-sided}}\bigstrut[b]\bigstrut[t]\\
							
							\cline{5-7}	\cline{9-11}\cline{13-15}\cline{17-19}\cline{21-23}\cline{25-27}
							\tilde{k} & R & m &&
							\mathrm{W_{LLL}} & \mathrm{W_{aML}} & \mathrm{s_{aML}} && \mathrm{W_{LLL}} & \mathrm{W_{aML}} & \mathrm{s_{aML}}&&\mathrm{W_{LLL}} & \mathrm{W_{aML}} & \mathrm{s_{aML}}&& \mathrm{W_{LLL}} & \mathrm{W_{aML}} & \mathrm{s_{aML}}&&\mathrm{W_{LLL}} & \mathrm{W_{aML}} & \mathrm{s_{aML}}&& \mathrm{W_{LLL}} & \mathrm{W_{aML}} & \mathrm{s_{aML}}\bigstrut[b]\bigstrut[t]\\\hline
							\multicolumn{1}{l}{2}& \multicolumn{1}{l}{10} & \multicolumn{1}{l}{20} && 0.945&	0.944&	0.951&&	0.938&	0.942&	0.950&&0.944&	0.943&	0.950&&	0.938&	0.943&	0.950&&	0.946&	0.945&	0.953&&	0.938&	0.944&	0.952     \bigstrut[t]\\
							\multicolumn{1}{l}{} & \multicolumn{1}{l}{}   & \multicolumn{1}{l}{30} && 0.947&	0.946&	0.951&&	0.940&	0.945&	0.951&&0.945&	0.944&	0.950&&	0.940&	0.945&	0.950&&	0.949&	0.948&	0.953&&	0.943&	0.949&	0.951          \\
							\multicolumn{1}{l}{} & \multicolumn{1}{l}{20} & \multicolumn{1}{l}{20} && 0.947&	0.946&	0.950&&	0.942&	0.945&	0.950&&0.947&	0.946&	0.949&&	0.942&	0.945&	0.950&&	0.948&	0.948&	0.951&&	0.943&	0.947&	0.952       \\
							\multicolumn{1}{l}{} & \multicolumn{1}{l}{}   & \multicolumn{1}{l}{30} && 0.948&	0.947&	0.950&&	0.942&	0.947&	0.950&&0.948&	0.947&	0.950&&	0.942&	0.947&	0.949&&	0.949&	0.948&	0.951&&	0.944&	0.949&	0.949      \\
							\multicolumn{1}{l}{} & \multicolumn{1}{l}{30} & \multicolumn{1}{l}{20} && 0.948&	0.948&	0.951&&	0.944&	0.946&	0.949&&0.948&	0.947&	0.950&&	0.943&	0.947&	0.950&&	0.948&	0.948&	0.951&&	0.943&	0.947&	0.951  \\
							\multicolumn{1}{l}{} & \multicolumn{1}{l}{}   & \multicolumn{1}{l}{30} && 0.949&	0.949&	0.950&&	0.944&	0.948&	0.950&&0.949&	0.949&	0.950&&	0.945&	0.949&	0.949&&	0.949&	0.949&	0.951&&	0.946&	0.950&	0.949  \\
							\\[-2.5mm]
							\multicolumn{1}{l}{6}& \multicolumn{1}{l}{10} & \multicolumn{1}{l}{20} && 0.935&	0.935&	0.949&&	0.920&	0.936&	0.949&&0.932&	0.933&	0.949&&	0.912&	0.934&	0.949&&	0.938&	0.936&	0.951&&	0.922&	0.937&	0.951         \\
							\multicolumn{1}{l}{} & \multicolumn{1}{l}{}   & \multicolumn{1}{l}{30} && 0.937&	0.939&	0.949&&	0.920&	0.940&	0.950&&0.937&	0.939&	0.950&&	0.917&	0.941&	0.949&&	0.939&	0.941&	0.951&&	0.918&	0.943&	0.949	     \\
							\multicolumn{1}{l}{} & \multicolumn{1}{l}{20} & \multicolumn{1}{l}{20} && 0.943&	0.943&	0.950&&	0.931&	0.942&	0.950&&0.942&	0.942&	0.949&&	0.929&	0.943&	0.949&&	0.942&	0.942&	0.950&&	0.929&	0.942&	0.950         \\
							\multicolumn{1}{l}{} & \multicolumn{1}{l}{}   & \multicolumn{1}{l}{30} && 0.944&	0.945&	0.949&&	0.932&	0.945&	0.950&&0.945&	0.945&	0.950&&	0.931&	0.946&	0.950&&	0.945&	0.946&	0.951&&	0.932&	0.946&	0.949            \\
							\multicolumn{1}{l}{} & \multicolumn{1}{l}{30} & \multicolumn{1}{l}{20} && 0.944&	0.945&	0.949&&	0.935&	0.944&	0.950&&0.944&	0.945&	0.949&&	0.933&	0.944&	0.949&&	0.945&	0.946&	0.951&&	0.934&	0.945&	0.950          \\
							\multicolumn{1}{l}{} & \multicolumn{1}{l}{}   & \multicolumn{1}{l}{30} && 0.946&	0.946&	0.949&&	0.935&	0.946&	0.949&&0.946&	0.947&	0.950&&	0.935&	0.946&	0.950&&	0.947&	0.948&	0.951&&	0.937&	0.948&	0.950           \\
							\\[-2.5mm]
							\multicolumn{1}{l}{10}& \multicolumn{1}{l}{10}& \multicolumn{1}{l}{20} && 0.921&	0.927&	0.948&&	0.894&	0.932&	0.949&&0.918&	0.924&	0.947&&	0.890&	0.929&	0.949&& 0.920&	0.927&	0.950&&	0.889&	0.931&	0.950   \\
							\multicolumn{1}{l}{} & \multicolumn{1}{l}{}   & \multicolumn{1}{l}{30} && 0.930&	0.937&	0.949&&	0.905&	0.939&	0.949&&0.928&	0.933&	0.948&&	0.901&	0.937&	0.949&&	0.928&	0.935&	0.949&&	0.899&	0.939&	0.949        \\
							\multicolumn{1}{l}{} & \multicolumn{1}{l}{20} & \multicolumn{1}{l}{20} && 0.935&	0.939&	0.949&&	0.914&	0.940&	0.949&&0.937&	0.939&	0.949&&	0.918&	0.940&	0.949&&	0.937&	0.939&	0.950&&	0.917&	0.941&	0.949       \\
							\multicolumn{1}{l}{} & \multicolumn{1}{l}{}   & \multicolumn{1}{l}{30} && 0.940&	0.944&	0.950&&	0.920&	0.944&	0.950&&0.938&	0.942&	0.949&&	0.916&	0.943&	0.949&&	0.939&	0.942&	0.949&&	0.918&	0.944&	0.948       \\
							\multicolumn{1}{l}{} & \multicolumn{1}{l}{30} & \multicolumn{1}{l}{20} && 0.942&	0.943&	0.949&&	0.930&	0.944&	0.949&&0.940&	0.942&	0.949&&	0.924&	0.942&	0.950&&	0.942&	0.944&	0.950&&	0.925&	0.943&	0.951    \\
							\multicolumn{1}{l}{} & \multicolumn{1}{l}{}   & \multicolumn{1}{l}{30} && 0.943&	0.945&	0.949&&	0.928&	0.945&	0.950&&0.943&	0.945&	0.949&&	0.927&	0.945&	0.949&&	0.943&	0.945&	0.949&&	0.925&	0.946&	0.949	    \bigstrut[b]\\
							\hline
						\end{array}
						$
						\end{adjustbox}
						\end{table}
						
					\end{landscape}
					\restoregeometry
				}
				
				\restoregeometry

				\begin{mycomment}
					results in table \ref{network model} are actually for 200000 reps (see results from saddlepoint ci simulation network fix eff adj.docx )
				\end{mycomment}

				\section{\label{sec suppl s(lambda)} Recentering the score for $\lambda$
					alone}
				
				The adjusted QMLE $\lambda_{\mathrm{aML}}$ is obtained by recentering the
				profile score $s(\sigma^{2},\lambda)$; see Section
				\ref{subsec SLM adj prof lik}. It is natural to wonder what would happen if,
				instead, one were to treat $\sigma^{2}$ as a nuisance parameter, and recenter
				the profile score for $\lambda$ alone, say $s(\lambda)$ (i.e., the score
				associated to the log-likelihood (\ref{prof lik})). To investigate, first we
				recenter $s(\lambda)$ under a stronger assumption than the assumption we used to recenter $s(\sigma
				^{2},\lambda)$, the latter assumption being $\mathrm{E}(\varepsilon)=0$ and
				$\mathrm{var}(\varepsilon)=I_{n}$.\footnote{That the assumption $\mathrm{E}%
					(\varepsilon)=0$ and $\mathrm{var}(\varepsilon)=I_{n}$ is not sufficient for
					an exact recentering of $s(\lambda)$ could also be seen by noticing that the
					assumption is not sufficient for the profile adjusted score $s_{\mathrm{a2}%
					}(\lambda)$ given in equation (\ref{sa2}) to be (exactly) unbiased (even
					though it is sufficient for the estimating equation (\ref{unbesteq}) to be
					unbiased).} Under the stronger assumption, recentering $s(\lambda)$ produces
				the same estimator for $\lambda$ as recentering $s(\sigma^{2},\lambda)$. But,
				of course, recentering $s(\lambda)$ does not automatically deliver an adjusted
				estimator of $\sigma^{2}$, which is needed for inference on $\lambda$. Then,
				we show how a small modification of the procedure produces the same unbiased
				estimating equation for $\lambda$ derived in the paper (equation
				(\ref{unbesteq})), without the need for the stronger distributional assumption.
				
				The log-likelihood obtained after profiling out both $\beta$ and $\sigma^{2}$
				from the (quasi) log-likelihood $l(\beta,\sigma^{2},\theta)$ is given in
				equation (\ref{prof lik}), and the associated profile score is%
				\begin{equation}
				s(\lambda)\coloneqq n\frac{y^{\prime}W^{\prime}M_{X}S(\lambda)y}{y^{\prime}S^{\prime
					}(\lambda)M_{X}S(\lambda)y}-\mathrm{tr}(G(\lambda)). \label{score}%
				\end{equation}
				We now show that $s(\lambda)$ can be recentered under the assumption that the
				distribution of $y$ is a scale-mixture of the $\mathrm{N}(S^{-1}%
				(\lambda)X\beta,\sigma^{2}(S^{\prime}(\lambda)S(\lambda))^{-1})$ distribution.
				By Lemma \ref{lemma exp} given at the end of this subsection, the expectation
				of $s(\lambda)$ when $y\sim\mathrm{N}(S^{-1}(\lambda)X\beta,\sigma
				^{2}(S^{\prime}(\lambda)S(\lambda))^{-1}$ (or, equivalently, $\varepsilon
				\sim\mathrm{N}(0,I)$) is given by%
				\begin{equation}
				\mathrm{E}(s(\lambda))=\frac{n}{n-k}\mathrm{tr}(M_{X}G(\lambda))-\mathrm{tr}%
				(G(\lambda)), \label{exp adj score}%
				\end{equation}
				which is generally nonzero. The adjusted profile score for $\lambda$ is
				therefore
				\begin{equation}
				s_{\mathrm{a}}(\lambda)\coloneqq s(\lambda)-\mathrm{E}(s(\lambda))=n\frac{y^{\prime
					}W^{\prime}M_{X}S(\lambda)y}{y^{\prime}S^{\prime}(\lambda)M_{X}S(\lambda
					)y}-\frac{n}{n-k}\mathrm{tr}(M_{X}G(\lambda)), \label{adj score}%
				\end{equation}
				with associated profile likelihood $l_{\mathrm{a}}^{\ast}(\lambda)\coloneqq \int
				s_{\mathrm{a}}(\lambda)\diff\lambda$. In terms of the score $s_{\mathrm{a}%
					2}(\lambda)$ in equation (\ref{sa2}), $s_{\mathrm{a}}(\lambda)=\frac{n}%
				{n-k}s_{\mathrm{a}2}(\lambda)$, which implies that the associated estimating
				equations are identical. Of course, up to additive constants, we also have
				$l_{\mathrm{a}}^{\ast}(\lambda)=\frac{n}{n-k}l_{\mathrm{a}}(\lambda)$,
				$l_{\mathrm{a}}(\lambda)$ being the log-likelihood given in equation
				(\ref{l a lam}). In fact, equation (\ref{exp adj score}) is quite robust to
				deviations from normality. For one thing, it is evident that, as a function of
				$y,$ $s(\lambda)$ is scale-invariant, which implies that its properties under
				Gaussian assumptions will be retained under all scale-mixtures of the
				$\mathrm{N}(S^{-1}(\lambda)X\beta,\sigma^{2}(S^{\prime}(\lambda)S(\lambda
				))^{-1})$ distribution for $y.$ This is a much broader class of distributions
				for $y.$
				
				\begin{mycomment}
					\emph{question for Grant: what is the exact relationship between the
						assumption that the distrib of }$y$\emph{ is a scale-mixture of }%
					$N(S^{-1}(\lambda)X\beta,\sigma^{2}(S^{\prime}(\lambda)S(\lambda))^{-1}%
					)$\emph{ and the assumption that }$\varepsilon$\emph{ has a spherically
						symmetric distribution?}\textrm{\{\{\{Why do we need to discuss this??
						The only issue it seems to me is the one just stated - results that hold under
						normality also hold for a broader class of distributions. You could discuss
						the general question you ask, but not here. I don't think the spherical
						symmetry of }$\varepsilon$\textrm{ needs to be mentioned.\}\}\}}
					F: for example saying that epsilon is sperically symm means we don;t require
					indep errors
				\end{mycomment}

				Also, the result holds approximately in much greater generality, since the
				first term on the right hand side of equation (\ref{exp adj score}) can be interpreted as a
				Laplace approximation to the expectation of the ratio of quadratic forms in
				equation (\ref{score}); see \cite{Lieberman94}.
				
				Thus, so far we have shown that the estimator $\hat{\lambda}_{\mathrm{aML}}$,
				which has been produced in the paper by recentering $s(\sigma^{2},\lambda)$,
				can be also obtained by recentering $s(\lambda)$ under a stronger
				distributional assumption. There is an alternative way to produce
				$\hat{\lambda}_{\mathrm{aML}}$ starting from the score $s(\lambda)$, without
				the stronger assumption. The denominator $y^{\prime}S^{\prime}(\lambda
				)M_{X}S(\lambda)y$ in expression (\ref{score}) for $s(\lambda)$ is a.s.
				positive. Hence, the score equation $s(\lambda)=0$ is a.s. equivalent to the
				equation
				\begin{equation}
				y^{\prime}W^{\prime}M_{X}S(\lambda)y-\frac{\mathrm{tr}(G(\lambda))}%
				{n}y^{\prime}S^{\prime}(\lambda)M_{X}S(\lambda)y=0. \label{est eq score}%
				\end{equation}
				This estimating equation can be recentered assuming only that $\mathrm{E}%
				(S(\lambda)y)=X\beta$ and $\mathrm{var}(S(\lambda)y)=\sigma^{2}I_{n}$ (or,
				equivalently, $\mathrm{E}(\varepsilon)=0$ and $\mathrm{var}(\varepsilon
				)=I_{n}$). Under such assumptions,
				\[
				\mathrm{E}\left(  y^{\prime}W^{\prime}M_{X}S(\lambda)y-\frac{\mathrm{tr}%
					(G(\lambda))}{n}y^{\prime}S^{\prime}(\lambda)M_{X}S(\lambda)y\right)
				=\sigma^{2}\left(  \mathrm{tr}(M_{X}G(\lambda))-\frac{n-k}{n}\mathrm{tr}%
				(G(\lambda))\right)  ,
				\]
				which is generally different from $0$. The recentered version of equation
				(\ref{est eq score}) is therefore
				\[
				y^{\prime}W^{\prime}M_{X}S(\lambda)y-\frac{\mathrm{tr}(G(\lambda))}%
				{n}y^{\prime}S^{\prime}(\lambda)M_{X}S(\lambda)y-\sigma^{2}\left(
				\mathrm{tr}(M_{X}G(\lambda))-\frac{n-k}{n}\mathrm{tr}(G(\lambda))\right)  =0,
				\]
				which now involves the unknown $\sigma^{2}$. But, replacing $\sigma^{2}$ with
				its unbiased estimator given $\lambda$, $\frac{1}{n-k}y^{\prime}S^{\prime
				}(\lambda)M_{X}S(\lambda)y$ gives the equation%
				\[
				y^{\prime}W^{\prime}M_{X}S(\lambda)y-\frac{\mathrm{tr}(M_{X}G(\lambda))}%
				{n-k}y^{\prime}S^{\prime}(\lambda)M_{X}S(\lambda)y=0,
				\]
				which is the same as the estimating equation (\ref{unbesteq}) in the paper.
				
				\begin{mycomment}
					laplace approx to $\mathrm{E}\left(  \frac{1}{2}\frac{y^{\prime}(W^{\prime
						}M_{X}S(\lambda)+S^{\prime}(\lambda)M_{X}W)y}{y^{\prime}S^{\prime}%
						(\lambda)M_{X}S(\lambda)y}\right)  $ is $\frac{1}{2}\frac{\mathrm{E}\left(
						y^{\prime}(W^{\prime}M_{X}S(\lambda)+S^{\prime}(\lambda)M_{X}W)y\right)
					}{\mathrm{E}\left(  y^{\prime}S^{\prime}(\lambda)M_{X}S(\lambda)y\right)  }$
					and this, assuming $E(S(\lambda)y)=X\beta$ and $var(S(\lambda)y)=\sigma^{2}I$,
					is
					\[
					\frac{1}{2}\frac{\mathrm{tr}\left(  \left(  W^{\prime}M_{X}S(\lambda
						)+S^{\prime}(\lambda)M_{X}W\right)  \left(  S^{\prime}S\right)  ^{-1}\right)
						+\beta^{\prime}X^{\prime}S^{\prime-1}(W^{\prime}M_{X}S(\lambda)+S^{\prime
						}(\lambda)M_{X}W)S^{-1}X\beta}{n-k}=\frac{1}{2}\frac{\mathrm{tr}\left(
						S^{\prime-1}\left(  W^{\prime}M_{X}S(\lambda)+S^{\prime}(\lambda
						)M_{X}W\right)  \left(  S\right)  ^{-1}\right)  }{n-k}=\frac{\mathrm{tr}%
						\left(  M_{X}WS^{-1}\right)  }{n-k}%
					\]
				\end{mycomment}

				\begin{mycomment}
					By definition, $l_{\mathrm{a}}(\lambda)$ is obtained by integrating the
					adjusted score (\ref{adj score}). That is, up to an additive constant,
					\[
					l_{\mathrm{a}}(\lambda)\coloneqq \int s_{\mathrm{a}}(\lambda)\diff\lambda=-\frac{n}%
					{2}\log\left(  y^{\prime}S^{\prime}(\lambda)M_{X}S(\lambda)y\right)  -\frac
					{n}{n-k}\int\mathrm{tr}(M_{X}G(\lambda))\diff\lambda,
					\]
				\end{mycomment}

				\begin{mycomment}
					recentering $s(\sigma^{2},\lambda)$ and then concentrating out $\lambda$ does
					not give same score as recentering $s(\lambda)$ (although zeros are the same).
					see check\_transformation\_approach\_panel\_indiv\_fixed\_effects.m,
					spatial\_panel\_two\_way\_fixed\_effects.m,
					check\_transformation\_approach\_network\_fixed\_effects\_unbalanced. maybe a
					little unexpected as $s_{2}(\hat{\sigma}_{ML}^{2}(\lambda),\lambda
					)=s(\lambda)$
				\end{mycomment}

				Finally, we observe that, depending on $W$ and $X$, $\mathrm{E}(s(\lambda))$
				in equation (\ref{exp adj score}) can be positive, zero, or negative for a given
				$\lambda$. Unreported simulations suggest that, in most cases of interest for
				applications, $\mathrm{E}(s(\lambda))<0$ for $\lambda\in\Lambda,\ $or at least
				for $\lambda\in(-1,1)$, which typically induces a negative bias in
				$\hat{\lambda}_{\mathrm{ML}}$. Note that if $\mathrm{E}(s(\lambda))<0$ for all
				$\lambda\in\Lambda$ and $s(\lambda)$ has a unique zero on $\Lambda$ (see
				Proposition \ref{theo cdf MLE}), then $\hat{\lambda}_{\mathrm{aML}}%
				>\hat{\lambda}_{\mathrm{ML}}$ for any $y$.
				
				The lemma required to obtain equation (\ref{exp adj score}) under normality is:
				
				\begin{lemma}
					\label{lemma exp}For any $n\times n$ nonrandom matrix $A$, and for
					$z\sim\mathrm{N}(X\beta,\sigma^{2}I_{n})$,
					\[
					\mathrm{E}\left(  \frac{z^{\prime}AM_{X}z}{z^{\prime}M_{X}z}\right)
					=\frac{\mathrm{tr}(AM_{X})}{n-k}.
					\]
					
				\end{lemma}
				
				\begin{pff}
					Let $C$ be a matrix such that $CC^{\prime}=I_{n-k}$ and $C^{\prime}C=M_{X}.$
					We can write, uniquely, $z=Vz_{1}+C^{\prime}z_{2}$, where $V\coloneqq X(X^{\prime
					}X)^{-1/2}$, $z_{1}\coloneqq (X^{\prime}X)^{-1/2}X^{\prime}z$, and $z_{2}=Cz$. Thus
					\[
					\frac{z^{\prime}AM_{X}z}{z^{\prime}M_{X}z}=\frac{z_{2}^{\prime}CAC^{\prime
						}z_{2}}{z_{2}^{\prime}z_{2}}+\frac{z_{1}^{\prime}V^{\prime}AC^{\prime}z_{2}%
					}{z_{2}^{\prime}z_{2}}.
					\]
					Note that $z_{1}\sim\mathrm{N}((X^{\prime}X)^{1/2}\beta,\sigma^{2}I_{n})$,
					$z_{2}\sim\mathrm{N}(0,\sigma^{2}I_{n-k}),$ and $z_{1}$ and $z_{2}$ are
					independent. The expectation of $z_{1}^{\prime}V^{\prime}AC^{\prime}%
					z_{2}/z_{2}^{\prime}z_{2}$ with respect to $z_{1}$ is $\beta^{\prime
					}(X^{\prime}X)^{1/2}V^{\prime}AC^{\prime}z_{2}/z_{2}^{\prime}z_{2}$, and the
					expectation with respect to $z_{2}$ is then zero, because the distribution of
					$z_{2}$ is invariant under $z_{2}\rightarrow-z_{2}$. The proof is completed on
					noting that the ratio $z_{2}^{\prime}CAC^{\prime}z_{2}/z_{2}^{\prime}z_{2}$ is
					independent of its denominator \citep{Pitman1937}, and hence has expected
					value equal to $\mathrm{tr}(AM_{X})/(n-k).$
				\end{pff}
				
				\begin{mycomment}
					(it is not necessary that $\mathrm{E}_{\lambda}(s(\lambda))<0$ for $\lambda
					\in\Lambda_{\mathrm{a}}$)
				\end{mycomment}

				\begin{mycomment}
					\begin{align}
					s_{\mathrm{a}}(\lambda)  &  \coloneqq s-\mathrm{E}_{\lambda}(s(\lambda))\\
					&  =s(\lambda)-\frac{n}{n-k}\mathrm{tr}(M_{X}G(\lambda))+\mathrm{tr}%
					(G(\lambda))
					\end{align}
					What we can establish is that when $\mathrm{tr}(M_{X}Q_{\omega})=0$ for a
					semisimple nonzero real eigenvalue $\omega$ of $W$, $\frac{n}{n-k}%
					\mathrm{tr}(M_{X}G(\lambda))<\mathrm{tr}(G(\lambda))$ in a neighborhood of
					$\omega^{-1}$. When $\mathrm{tr}(M_{X}Q_{\omega})=0$ for a semisimple nonzero
					real eigenvalue $\omega$ of $W$ (so that $l_{\mathrm{a}}(\lambda)$ is a.s.
					continuous at $\lambda=\omega^{-1}$ by Lemma \ref{lemma trMG}) $\lim
					_{\lambda\rightarrow\omega^{-1}}\mathrm{tr}(M_{X}G(\lambda))$ is bounded. On
					the other hand $\lim_{\lambda\rightarrow\omega^{-1}}\mathrm{tr}(G(\lambda
					))=+\infty$ for any nonzero real eigenvalue $\omega$ of $W$.
				\end{mycomment}

				\begin{mycomment}
					\[
					\frac{d}{\diff\lambda}\mathrm{tr}(G(\lambda))=\mathrm{tr}(WS^{-1}(\lambda
					)WS^{-1}(\lambda))=\mathrm{tr}(G^{2}(\lambda))
					\]
					Recall $\mathrm{tr}(C^{2}(\lambda))=\mathrm{tr}(G^{2}(\lambda))-\frac{1}%
					{n}[\mathrm{tr}(G(\lambda))]^{2}$ (with $\mathrm{tr}(C^{2}(\lambda))>0$ being
					equivalent to $\delta(\lambda)<0$). Then $\mathrm{tr}(C^{2}(\lambda))>0$ for
					all $\lambda\in\Lambda$ (which is the case in particular if all eval of $W$
					are real) iff $\mathrm{tr}(G^{2}(\lambda))>\frac{1}{n}[\mathrm{tr}%
					(G(\lambda))]^{2}$. That is $\mathrm{tr}(C^{2}(\lambda))>0$ for all
					$\lambda\in\Lambda$ implies $\log\left\vert \det\left(  S(\lambda)\right)
					\right\vert $ is concave on $\Lambda$ (the shrinking term $\mathrm{tr}%
					(G(\lambda))$ in $s(\lambda)$ is monotonically incr; same sign as $\lambda$ if
					$\mathrm{tr}(W)=0$) (if the condition $\mathrm{tr}(C^{2}(\lambda))>0$ for all
					$\lambda\in\Lambda$ is not satisfied $\log\left\vert \det\left(
					S(\lambda)\right)  \right\vert $ may be multimodal and this may cause
					multimodality of $l(\lambda)$)%
					\[
					\frac{d}{\diff\lambda}\mathrm{tr}(M_{X}G(\lambda))=\mathrm{tr}(M_{X}%
					WS^{-1}(\lambda)WS^{-1}(\lambda))=\mathrm{tr}(M_{X}G^{2}(\lambda))
					\]
					Similarly $\delta_{\mathrm{a}}(\lambda)<0$ iff $(n-k)\mathrm{tr}(M_{X}%
					G^{2}(\lambda))>\left[  \mathrm{tr}\left(  M_{X}G(\lambda)\right)  \right]
					^{2}$. Hence if $\delta_{\mathrm{a}}(\lambda)<0$ for all $\lambda\in\Lambda$
					the shrinking term $\frac{n}{n-k}\mathrm{tr}(M_{X}G(\lambda))$ in
					$s_{\mathrm{a}}(\lambda)$ is monotonically incr (with $\frac{n}{n-k}%
					\mathrm{tr}(M_{X}G_{0})=\frac{n}{n-k}\mathrm{tr}(M_{X}W)$), and the
					corresponding term in $l_{\mathrm{a}}(\lambda)$ is concave
				\end{mycomment}

				\section{\label{suppl SEM} Spatial error model}
				
				A spatial error model is a particular case of a regression model with
				correlated errors, which we write as $y=X\beta+u,$ with $\mathrm{E}(u)=0$ and
				$\mathrm{var}(u)=\sigma^{2}\Sigma(\theta)$. We assume that $X$ is fixed, but
				of course one could condition on a random $X$ if $X$ and $u$ are assumed to be
				independent. The parameter $\theta$ belongs to some subset $\Theta
				\subseteq\mathbb{R}^{p}$, and it is assumed that $\Sigma(\theta)$ is positive
				definite for any $\theta\in\Theta$ and differentiable in $\theta$. We also
				define a matrix $A(\theta)$ such that $A^{\prime}(\theta)A(\theta)=\Sigma
				^{-1}(\theta)$. The model can the be written as $y=X\beta+\sigma
				A^{-1}\varepsilon,$ with $\mathrm{E}(\varepsilon)=0$ and $\mathrm{var}%
				(\varepsilon)=I_{n}$. In the spatial error model case, we may take
				$A(\theta)=S(\lambda)$.
				
				In Section \ref{sec der} we derive expression (\ref{l_a_regr_sigma_theta}) for
				the adjusted profile log-likelihood $l_{\mathrm{a}}(\sigma^{2},\theta)$, in
				the context of a general $\Sigma(\theta)$, by recentering the profile score
				for $\theta$. Then, in Section \ref{sec other der}, we show that the
				corresponding adjusted likelihood is equivalent to the likelihood suggested by
				other inferential approaches.
				
				\subsection{\label{sec der}The adjusted profile log-likelihood}
				
				The quasi log-likelihood prevailing under the assumption $\varepsilon
				\sim\mathrm{N}(0,I_{n})$ is%
				\[
				l(\beta,\sigma^{2},\theta)\coloneqq -\frac{n}{2}\log(\sigma^{2})-\frac{1}{2}%
				\log\left(  \det\left(  \Sigma(\theta)\right)  \right)  -\frac{1}{2\sigma^{2}%
				}(y-X\beta)^{\prime}\Sigma^{-1}(\theta)(y-X\beta),
				\]
				where $(\beta,\sigma^{2},\theta)\in\mathbb{R}^{k}\times\mathbb{R}^{+}%
				\times\Theta$, and additive constants are omitted.
				
				Provided that $y\notin\operatorname{col}(X)$, $\beta$ can be concentrated out
				of $l(\beta,\sigma^{2},\theta)$.\footnote{If $y\in\operatorname{col}(X)$, the
					model provides perfect fit. Of course, $\operatorname{col}(X)$ is a
					$\mu_{\mathbb{R}^{n}}$-null set, since $k<n$ by assumption.} The MLE of
				$\beta$ given $\theta$ is $\hat{\beta}_{\mathrm{ML}}(\theta)\coloneqq \left(
				X^{\prime}\Sigma^{-1}(\theta)X\right)  ^{-1}X^{\prime}\Sigma^{-1}(\theta)y$.
				Thus, the profile log-likelihood for $(\sigma^{2},\theta)$ is%
				\begin{equation}
				l(\sigma^{2},\theta)\coloneqq l(\hat{\beta}_{\mathrm{ML}}(\theta),\sigma^{2}%
				,\theta)=-\frac{n}{2}\log(\sigma^{2})-\frac{1}{2}\log\left(  \det\left(
				\Sigma(\theta)\right)  \right)  -\frac{1}{2\sigma^{2}}y^{\prime}U(\theta)y,
				\label{l(sig,theta) SEM}%
				\end{equation}
				where
				\begin{align*}
				U(\theta)  &  \coloneqq \left(  I_{n}-X\left(  X^{\prime}\Sigma^{-1}(\theta)X\right)
				^{-1}X^{\prime}\Sigma^{-1}(\theta)\right)  ^{\prime}A^{\prime}(\theta
				)A(\theta)\left(  I_{n}-X\left(  X^{\prime}\Sigma^{-1}(\theta)X\right)
				^{-1}X^{\prime}\Sigma^{-1}(\theta)\right) \\
				&  =A^{\prime}(\theta)M_{A(\theta)X}A(\theta).
				\end{align*}

				\begin{mycomment}
					\begin{align*}
					M_{A(\theta)X}  &  =I-A(\theta)X(X^{\prime}A^{\prime}(\theta)A(\theta
					)X)^{-1}X^{\prime}A^{\prime}(\theta)\\
					A^{\prime}(\theta)M_{A(\theta)X}A(\theta)  &  =A^{\prime}(\theta)\left(
					I-A(\theta)X(X^{\prime}A^{\prime}(\theta)A(\theta)X)^{-1}X^{\prime}A^{\prime
					}(\theta)\right)  A(\theta)\\
					&  =\Sigma^{-1}(\theta)-\Sigma^{-1}(\theta)X\left(  X^{\prime}\Sigma
					^{-1}(\theta)X\right)  ^{-1}X^{\prime}\Sigma^{-1}(\theta)=\Sigma^{-1}%
					(\theta)(I-X\left(  X^{\prime}\Sigma^{-1}(\theta)X\right)  ^{-1}X^{\prime
					}\Sigma^{-1}(\theta))\\
					& \\
					&  \left(  I_{n}-X\left(  X^{\prime}\Sigma^{-1}(\theta)X\right)
					^{-1}X^{\prime}\Sigma^{-1}(\theta)\right)  ^{\prime}\Sigma^{-1}(\theta)\left(
					I_{n}-X\left(  X^{\prime}\Sigma^{-1}(\theta)X\right)  ^{-1}X^{\prime}%
					\Sigma^{-1}(\theta)\right) \\
					&  =\left(  I_{n}-\Sigma^{-1}(\theta)X\left(  X^{\prime}\Sigma^{-1}%
					(\theta)X\right)  ^{-1}X^{\prime}\right)  \left(  \Sigma^{-1}(\theta
					)-\Sigma^{-1}(\theta)X\left(  X^{\prime}\Sigma^{-1}(\theta)X\right)
					^{-1}X^{\prime}\Sigma^{-1}(\theta)\right) \\
					&  =\Sigma^{-1}(\theta)-\Sigma^{-1}(\theta)X\left(  X^{\prime}\Sigma
					^{-1}(\theta)X\right)  ^{-1}X^{\prime}\Sigma^{-1}(\theta)
					\end{align*}
				\end{mycomment}

				The score associated to the log-likelihood (\ref{l(sig,theta) SEM}) is
				\begin{equation}
				s(\sigma^{2},\theta)\coloneqq \left[
				\begin{array}
				[c]{c}%
				\frac{\diff s(\sigma^{2},\theta)}{\diff\sigma^{2}}\\
				\frac{\diff s(\sigma^{2},\theta)}{\diff\theta}%
				\end{array}
				\right]  =\left[
				\begin{array}
				[c]{c}%
				-\frac{n}{2\sigma^{2}}+\frac{1}{2\sigma^{4}}y^{\prime}U(\theta)y\\
				-\frac{1}{2}\mathrm{tr}\left(  \Sigma^{-1}(\theta)\frac{\diff\Sigma(\theta
					)}{\diff\theta}\right)  +\frac{1}{2\sigma^{2}}y^{\prime}U(\theta
				)\frac{\diff\Sigma(\theta)}{\diff\theta}U(\theta)y
				\end{array}
				\right]  . \label{s(sig,theta) SEM}%
				\end{equation}

				Due to the presence of the nuisance parameter $\beta$, the score $s(\sigma
				^{2},\theta)$ is generally biased. In order to obtain an unbiased estimating
				equation, we recenter $s(\sigma^{2},\theta)$. Using the fact that
				$U(\theta)X=0,$ we can write%
				\begin{align*}
				s(\sigma^{2},\theta)  &  =\left[
				\begin{array}
				[c]{c}%
				-\frac{n}{2\sigma^{2}}+\frac{1}{2\sigma^{4}}u^{\prime}U(\theta)u\\
				-\frac{1}{2}\mathrm{tr}\left(  \Sigma^{-1}(\theta)\frac{\diff\Sigma(\theta
					)}{\diff\theta}\right)  +\frac{1}{2\sigma^{2}}u^{\prime}U(\theta
				)\frac{\diff\Sigma(\theta)}{\diff\theta}U(\theta)u
				\end{array}
				\right] \\
				&  =\left[
				\begin{array}
				[c]{c}%
				-\frac{n}{2\sigma^{2}}+\frac{1}{2\sigma^{2}}\varepsilon^{\prime}M_{A(\theta
					)X}\varepsilon\\
				-\frac{1}{2}\mathrm{tr}\left(  \Sigma^{-1}(\theta)\frac{\diff\Sigma(\theta
					)}{\diff\theta}\right)  +\frac{1}{2}\varepsilon^{\prime}M_{A(\theta)X}%
				A(\theta)\frac{\diff\Sigma(\theta)}{\diff\theta}A^{\prime}(\theta
				)M_{A(\theta)X}\varepsilon
				\end{array}
				\right]  .
				\end{align*}

				\begin{mycomment}
					
					$\varepsilon=\frac{1}{\sigma}A(\theta)u$ $u=\sigma A^{-1}(\theta)\varepsilon$
					$-\frac{n}{2\sigma^{2}}+\frac{1}{2\sigma^{2}}\varepsilon^{\prime}\left(
					A^{\prime}(\theta)\right)  ^{-1}U(\theta)A^{-1}(\theta)\varepsilon=-\frac
					{n}{2\sigma^{2}}+\frac{1}{2\sigma^{2}}\varepsilon^{\prime}M_{A(\theta
						)X}\varepsilon$
				\end{mycomment}

				Assuming that $\mathrm{E}(\varepsilon)=0$ and $\mathrm{var}(\varepsilon
				)=I_{n}$, we have%
				\begin{equation}
				\mathrm{E}(s(\sigma^{2},\theta))=\left[
				\begin{array}
				[c]{c}%
				-\frac{n}{2\sigma^{2}}+\frac{n-k}{2\sigma^{2}}\\
				\frac{1}{2}\mathrm{tr}\left(  X(X^{\prime}\Sigma^{-1}(\theta)X)^{-1}X^{\prime
				}\frac{\diff\Sigma^{-1}(\theta)}{\diff\theta}\right)
				\end{array}
				\right]  , \label{Es SEM}%
				\end{equation}
				where we have used%
				\begin{align}
				&  \mathrm{E}\left(  \varepsilon^{\prime}M_{A(\theta)X}A(\theta)\frac
				{\diff\Sigma(\theta)}{\diff\theta}A^{\prime}(\theta)M_{A(\theta)X}%
				\varepsilon\right) \nonumber\\
				&  \hspace{3cm}=\mathrm{tr}\left(  M_{A(\theta)X}A(\theta)\frac{\diff\Sigma
					(\theta)}{\diff\theta}A^{\prime}(\theta)M_{A(\theta)X}\right) \nonumber\\
				&  \hspace{3cm}=\mathrm{tr}\left(  M_{A(\theta)X}A(\theta)\frac{\diff\Sigma
					(\theta)}{\diff\theta}A^{\prime}(\theta)\right) \nonumber\\
				&  \hspace{3cm}=\mathrm{tr}\left(  \left(  I_{n}-A(\theta)X(X^{\prime}%
				\Sigma^{-1}(\theta)X)^{-1}X^{\prime}A^{\prime}(\theta)\right)  A(\theta
				)\frac{\diff\Sigma(\theta)}{\diff\theta}A^{\prime}(\theta)\right) \nonumber\\
				&  \hspace{3cm}=\mathrm{tr}\left(  \Sigma^{-1}(\theta)\frac{\diff\Sigma
					(\theta)}{\diff\theta}\right)  -\mathrm{tr}\left(  X(X^{\prime}\Sigma
				^{-1}(\theta)X)^{-1}X\frac{\diff\Sigma^{-1}(\theta)}{\diff\theta}\right)  .
				\label{EE}%
				\end{align}
				Hence, the adjusted score is%
				\begin{align*}
				s_{\mathrm{a}}(\sigma^{2},\theta)  &  \coloneqq s(\sigma^{2},\theta)-\mathrm{E}%
				(s(\sigma^{2},\theta))=\left[
				\begin{array}
				[c]{c}%
				s_{\mathrm{a}1}(\sigma^{2},\theta)\\
				s_{\mathrm{a}2}(\sigma^{2},\theta)
				\end{array}
				\right] \\
				&  =\left[
				\begin{array}
				[c]{c}%
				\frac{1}{2\sigma^{4}}y^{\prime}U(\theta)y-\frac{n-k}{2\sigma^{2}}\\
				-\frac{1}{2}\mathrm{tr}\left(  \Sigma^{-1}(\theta)\frac{\diff\Sigma(\theta
					)}{\diff\theta}\right)  +\frac{1}{2\sigma^{2}}y^{\prime}U(\theta
				)\frac{\diff\Sigma(\theta)}{\diff\theta}U(\theta)y-\frac{1}{2}\mathrm{tr}%
				\left(  X(X^{\prime}\Sigma^{-1}(\theta)X)^{-1}X^{\prime}\frac{\diff\Sigma
					^{-1}(\theta)}{\diff\theta}\right)
				\end{array}
				\right]  .
				\end{align*}
				Note that setting $s_{\mathrm{a}1}(\sigma^{2},\theta)=0$ gives
				\[
				\hat{\sigma}_{\mathrm{a}\mathrm{ML}}^{2}(\theta)=\frac{y^{\prime}U(\theta
					)y}{n-k},
				\]
				the adjusted QMLE of $\sigma^{2}$ for given $\theta$. The likelihood
				corresponding to $s_{\mathrm{a}}(\sigma^{2},\theta)$ is
				\begin{equation}
				l_{\mathrm{a}}(\sigma^{2},\theta)=-\frac{n-k}{2}\log(\sigma^{2})-\frac
				{1}{2\sigma^{2}}y^{\prime}U(\theta)y-\frac{1}{2}\log\left(  \det\left(
				\Sigma(\theta)\right)  \right)  -\frac{1}{2}\log\left(  \det\left(  X^{\prime
				}\Sigma^{-1}(\theta)X\right)  \right)  , \label{la sig lam}%
				\end{equation}
				and hence the profile adjusted likelihood for $\theta$ only is%
				\begin{align}
				l_{\mathrm{a}}(\theta)  &  \coloneqq l_{\mathrm{a}}(\hat{\sigma}_{\mathrm{a}%
					\mathrm{ML}}^{2}(\theta),\theta)\nonumber\\
				&  =-\frac{n-k}{2}\log(y^{\prime}U(\theta)y)-\frac{1}{2}\log\left(
				\det\left(  \Sigma(\theta)\right)  \right)  -\frac{1}{2}\log\left(
				\det\left(  X^{\prime}\Sigma^{-1}(\theta)X\right)  \right)  . \label{la theta}%
				\end{align}

				\begin{mycomment}
					$\frac{\diff\log\left(  \det\left(  X^{\prime}\Sigma^{-1}(\theta)X\right)
						\right)  }{\diff\theta}=\mathrm{tr}\left(  \left(  X^{\prime}\Sigma
					^{-1}(\theta)X\right)  ^{-1}\frac{\diff\left(  X^{\prime}\Sigma^{-1}%
						(\theta)X\right)  }{\diff\theta}\right)  =\mathrm{tr}\left(  \left(
					X^{\prime}\Sigma^{-1}(\theta)X\right)  ^{-1}X^{\prime}\frac{\diff\Sigma
						^{-1}(\theta)}{\diff\theta}X\right)  =$
				\end{mycomment}

				\begin{mycomment}
					$\frac{\partial l_{\mathrm{a}}(\sigma^{2},\theta)}{\partial\sigma^{2}}%
					=-\frac{n-k}{2\sigma^{2}}+\frac{1}{2\sigma^{4}}y^{\prime}U(\theta)y=0$
					$-\left(  n-k\right)  +\frac{1}{\sigma^{2}}y^{\prime}U(\theta)y=0$ $\sigma
					^{2}=\frac{y^{\prime}U(\theta)y}{n-k}$ $\frac{\partial l_{\mathrm{a}}%
						(\sigma^{2},\theta)}{\partial\theta}=...$
				\end{mycomment}

				\begin{mycomment}
					$loglik=k_{1}+k_{2}\log f(y)$
					$lik=e^{k_{1}+\log f(y)^{k_{2}}}=e^{k_{1}}f(y)^{k_{2}}$
					$lik=k_{1}+k_{2}f(y)$
					$\log lik=k_{1}+k_{2}f(y)$
					$d\log\left(  \det\left(  X^{\prime}\Sigma^{-1}(\theta)X\right)  \right)
					=-tr\left(  \left(  X^{\prime}\Sigma^{-1}(\theta)X\right)  ^{-1}d\left(
					X^{\prime}\Sigma^{-1}(\theta)X\right)  \right)  =-tr\left(  \left(  X^{\prime
					}\Sigma^{-1}(\theta)X\right)  ^{-1}X^{\prime}d\Sigma^{-1}(\theta)X\right)
					=tr\left(  X\left(  X^{\prime}\Sigma^{-1}(\theta)X\right)  ^{-1}X^{\prime
					}\Sigma(\theta)d\Sigma^{-1}(\theta)\Sigma(\theta)\right)  =$
					$d\log\left(  \det\left(  \Sigma(\theta)\right)  \right)  =\mathrm{tr}%
					(\Sigma^{-1}(\theta)d\Sigma(\theta))$
					$\Sigma^{-1}(\theta)=A^{\prime}(\theta)S_{\theta}$
					$d\Sigma^{-1}(\theta)=-A^{\prime}(\theta)S_{\theta}d\Sigma(\theta)S_{\theta
					}^{\prime}S_{\theta}$
				\end{mycomment}

				In the same way as for the SAR model (see Section
				\ref{subsec SLM adj prof lik}), an alternative to recentering the score
				$s(\sigma^{2},\theta)$ would be to profile out both $\beta$ and $\sigma^{2}$
				from the (quasi) log-likelihood $l(\beta,\sigma^{2},\theta)$ (provided that
				$y\notin\operatorname{col}(X)$), and then recenter the score $s(\theta)$. We
				now report the related calculations, for completeness. Note that
				recentering $s(\theta)$ produces the same likelihood as (\ref{la theta}), but
				under a normality assumption, and does not produce a corrected estimator for
				$\sigma^{2}$.
				
				The MLE's of $\beta$ and $\sigma^{2}$ given $\theta$ are $\hat{\beta
				}_{\mathrm{ML}}(\theta)\coloneqq \left(  X^{\prime}\Sigma^{-1}(\theta)X\right)
				^{-1}X^{\prime}\Sigma^{-1}(\theta)y$ and $\hat{\sigma}_{\mathrm{ML}}%
				^{2}(\theta)\coloneqq \frac{y^{\prime}U(\theta)y}{n}$. Thus the profile log-likelihood
				for $\theta$ only is%
				\begin{equation}
				l(\theta)\coloneqq l(\hat{\beta}_{\mathrm{ML}}(\theta),\hat{\sigma}_{\mathrm{ML}}%
				^{2}(\theta),\theta)=-\frac{n}{2}\log\left(  y^{\prime}U(\theta)y\right)
				-\frac{1}{2}\log\left(  \det\left(  \Sigma(\theta)\right)  \right)  .
				\label{l_theta}%
				\end{equation}

				\begin{mycomment}
					$\hat{\sigma}_{\mathrm{ML}}^{2}(\theta)\coloneqq \frac{y^{\prime}U(\theta)y}{n}$,
					where
					\[
					U(\theta)\coloneqq \Sigma^{-1}(\theta)-\Sigma^{-1}(\theta)X(X^{\prime}\Sigma
					^{-1}(\theta)X)^{-1}X^{\prime}\Sigma^{-1}(\theta).
					\]
					In terms of the matrix $A(\theta)$,
					\begin{equation}
					U(\theta)=A^{\prime}(\theta)M_{A(\theta)X}A(\theta),
					\end{equation}
					\bigskip
				\end{mycomment}

				The score associated to the log-likelihood (\ref{l_theta}) is%
				\begin{align}
				s(\theta)  &  \coloneqq \frac{\diff l(\theta)}{\diff\theta}=\frac{n}{2}\frac
				{y^{\prime}U(\theta)\frac{\diff\Sigma(\theta)}{\diff\theta}U(\theta
					)y}{y^{\prime}U(\theta)y}-\frac{1}{2}\mathrm{tr}\left(  \Sigma^{-1}%
				(\theta)\frac{\diff\Sigma(\theta)}{\diff\theta}\right) \nonumber\\
				&  =\frac{n}{2}\frac{u^{\prime}U(\theta)\frac{\diff\Sigma(\theta)}%
					{\diff\theta}U(\theta)u}{u^{\prime}U(\theta)u}-\frac{1}{2}\mathrm{tr}\left(
				\Sigma^{-1}(\theta)\frac{\diff\Sigma(\theta)}{\diff\theta}\right) \nonumber\\
				&  =\frac{n}{2}\frac{\varepsilon^{\prime}M_{A(\theta)X}A(\theta)\frac
					{\diff\Sigma(\theta)}{\diff\theta}A^{\prime}(\theta)M_{A(\theta)X}\varepsilon
				}{\varepsilon^{\prime}M_{A(\theta)X}\varepsilon}-\frac{1}{2}\mathrm{tr}\left(
				\Sigma^{-1}(\theta)\frac{\diff\Sigma(\theta)}{\diff\theta}\right)  ,
				\label{dltheta}%
				\end{align}
				where in the second line we have used $U(\theta)X=0$, and in the third the
				definition $U(\theta)\coloneqq A^{\prime}(\theta)M_{A(\theta)X}A(\theta)$ and the fact
				that $\varepsilon=\frac{1}{\sigma}A(\theta)u$.
				
				The ratio of quadratic forms in (\ref{dltheta}) is independent of its
				denominator \citep{Pitman1937}. Hence, its expectation under the assumption
				underlying (\ref{l_theta}) (that is, $y\sim\mathrm{N}(X\beta,\sigma^{2}%
				\Sigma(\theta))$, or equivalently $\varepsilon\sim\mathrm{N}(0,\sigma^{2}%
				I_{n})$) is%
				\begin{align*}
				&  \frac{\mathrm{E}\left(  \varepsilon^{\prime}M_{A(\theta)X}A(\theta
					)\frac{\diff\Sigma(\theta)}{\diff\theta}A^{\prime}(\theta)M_{A(\theta
						)X}\varepsilon\right)  }{\mathrm{E}\left(  \varepsilon^{\prime}M_{A(\theta
						)X}\varepsilon\right)  }\\
				&  \hspace{1cm}=\frac{1}{n-k}\left\{  \mathrm{tr}\left(  \Sigma^{-1}%
				(\theta)\frac{\diff\Sigma(\theta)}{\diff\theta}\right)  +\mathrm{tr}\left(
				X(X^{\prime}\Sigma^{-1}(\theta)X)^{-1}X^{\prime}\frac{\diff\Sigma^{-1}%
					(\theta)}{\diff\theta}\right)  \right\}  ,
				\end{align*}
				where we have used expression (\ref{EE}). The adjusted score $s_{\mathrm{a}%
				}(\theta)\coloneqq s(\theta)-\mathrm{E}(s(\theta))$ is therefore%
				\begin{multline*}
				\frac{n}{2}\left\{  \frac{y^{\prime}U(\theta)\frac{\diff\Sigma(\theta
						)}{\diff\theta}U(\theta)y}{y^{\prime}U(\theta)y}-\frac{1}{n-k}\left[
				\mathrm{tr}\left(  \frac{\diff\Sigma(\theta)}{\diff\theta}\Sigma^{-1}%
				(\theta)\right)  +\mathrm{tr}\left(  X(X^{\prime}\Sigma^{-1}(\theta
				)X)^{-1}X^{\prime}\frac{\diff\Sigma^{-1}(\theta)}{\diff\theta}\right)
				\right]  \right\} \\
				=-\frac{n}{2(n-k)}\left\{  -\left(  n-k\right)  \frac{y^{\prime}U(\theta
					)\frac{\diff\Sigma(\theta)}{\diff\theta}U(\theta)y}{y^{\prime}U(\theta
					)y}+\mathrm{tr}(d\Sigma(\theta)\Sigma^{-1}(\theta))\right. \\
				+\biggl.\mathrm{tr}\left(  X(X^{\prime}\Sigma^{-1}(\theta)X)^{-1}X^{\prime
				}\frac{\diff\Sigma^{-1}(\theta)}{\diff\theta}\right)  \biggr\}.
				\end{multline*}
				Integration of $s_{\mathrm{a}}(\theta)$ gives%
				\begin{align*}
				l_{\mathrm{a}}^{\ast}(\theta)  &  \coloneqq \frac{n}{n-k}\left\{  -\frac{n}{2}%
				\log\left(  y^{\prime}U(\theta)y\right)  -\frac{1}{2}\log\left(  \det\left(
				\Sigma(\theta)\right)  \right)  -\frac{1}{2}\log\left(  \det\left(  X^{\prime
				}\Sigma^{-1}(\theta)X\right)  \right)  \right\} \\
				&  =\frac{n}{n-k}l_{\mathrm{a}}(\theta).
				\end{align*}

				\subsection{\label{sec other der}Other interpretations of $l_{\mathrm{a}%
					}(\sigma^{2},\theta)$ and $l_{\mathrm{a}}(\theta)$}
				
				\subsubsection{\label{subsec inv}$\mathcal{G}_{X}$-invariance}
				
				The adjusted log-likelihoods $l_{\mathrm{a}}(\sigma^{2},\theta)$ and
				$l_{\mathrm{a}}(\theta)$, given respectively in equation (\ref{la sig lam})
				and (\ref{la theta}), are genuine likelihoods. The next lemma shows that
				$l_{\mathrm{a}}(\theta)$ corresponds to the density of $v\coloneqq Cy/\left\Vert
				Cy\right\Vert $, where $C$ is an $\left(  n-k\right)  \times n$ matrix such
				that $CC^{\prime}=I_{n-k}$ and $C^{\prime}C=M_{X}$ (that is, the columns of
				$C$ form an orthonormal basis for col$^{\perp}(X)$), and with the convention
				that $v\coloneqq 0$ if $Cy=0$. The vector $v$ is a maximal invariant under the group
				$\mathcal{G}_{X}$ defined in Section \ref{sec SEM}, so in the present context
				recentering the profile score for $\theta$ amounts to imposing $\mathcal{G}%
				_{X}$-invariance. The lemma is established under the assumption that $u$ has an elliptically symmetric
				distribution with no atom at the origin. We note that a similar
				result to Lemma \ref{lemma l_RE} is stated in \cite{RahmanKing1997}.
				
				\begin{lemma}
					\label{lemma l_RE}Assume that, in a spatial error model, $u$ has an
					elliptically symmetric distribution with $\mathrm{Pr}(u=0)=0$. Then,
					$l_{\mathrm{a}}(\theta)$ corresponds to the density of the maximal invariant
					$v$.
				\end{lemma}
				
				\begin{pff}
					[Proof of Lemma \ref{lemma l_RE}](i) Under the stated assumption, the density
					function of the maximal invariant $v$ with respect to the normalized invariant
					measure on the sphere $\{v\in\mathbb{R}^{n-k}:\left\Vert v\right\Vert =1\}$ is%
					\begin{equation}
					f(v;\theta)=2\det(C\Sigma(\theta)C^{\prime})^{-\frac{1}{2}}\left(  v^{\prime
					}\left(  C\Sigma(\theta)C^{\prime}\right)  ^{-1}v\right)  ^{-\frac{n-k}{2}};
					\label{pdf v}%
					\end{equation}
					see, e.g., \cite{kariya1980}, who considers the case when $u$ has a density,
					and note that existence of the density of $u$ is not required for
					(\ref{pdf v}), because the density of $v$ is the same for any elliptically
					symmetric distribution of $u$. Now,
					\[
					\det(C\Sigma(\theta)C^{\prime})=\det(\Sigma(\theta))\det\left(  X^{\prime
					}\Sigma^{-1}(\theta)X\right)  \left(  \det(X^{\prime}X)\right)  ^{-1}%
					\]
					\citep[e.g.,][]{Verbyla90}, and, since the columns of $C$ span
					$\operatorname{col}^{\perp}(X)$, it is easily seen that $C^{\prime}\left(
					C\Sigma(\theta)C^{\prime}\right)  ^{-1}C=U(\theta)$. It follows that the
					likelihood of $\theta$ based on $v$ is proportional to%
					\[
					\left(  \det\left(  \Sigma(\theta)\right)  \right)  ^{-\frac{1}{2}}\left(
					\det\left(  X^{\prime}\Sigma^{-1}(\theta)X\right)  \right)  ^{-\frac{1}{2}%
					}\left(  \frac{y^{\prime}U(\theta)y}{y^{\prime}M_{X}y}\right)  ^{-\frac
						{n-k}{2}},
					\]
					which is equivalent to the log-likelihood (\ref{la theta}).
				\end{pff}
				
				Similarly, it is easily seen that, under the assumption in Lemma
				\ref{lemma l_RE}, $l_{\mathrm{a}}(\sigma^{2},\theta)$ corresponds to the
				density of a maximal invariant under the group of transformations
				$y\rightarrow y+X\delta$.
				
				\begin{mycomment}
					density not needed see king' thesis lemma 8.1
				\end{mycomment}

				\begin{mycomment}
					- cf rhaman and king 1997, where they say this is shown in a wp by Ara and King
				\end{mycomment}

				\begin{mycomment}
					For a spatial error model this is the case if $\varepsilon$ has a spherical
					distribution with $\mathrm{Pr}(\varepsilon=0)=0$ as $u=\sigma S^{-1}%
					(\lambda)\varepsilon$
				\end{mycomment}

				\begin{mycomment}
					(the density would be $\det\left[  \Sigma(\theta)\right]  ^{-1/2}q(u^{\prime
					}\Sigma^{-1}(\theta)u)$ but again this is not needed as in king' thesis lemma
					8.1) with a density I could also say the density function of $\varepsilon$
					depends on $\varepsilon$ only through $\varepsilon^{\prime}\varepsilon$, and
					hence $u=\sigma S^{-1}(\lambda)\varepsilon$ has density function $\det
					(\sigma^{2}\Sigma(\theta))^{-1/2}q(\sigma^{-2}u^{\prime}\Sigma^{-1}(\theta)u)$
					for some function $q$ (see, e.g., Muirhead , p. 34) but this is not needed
				\end{mycomment}

				\subsubsection{\label{sec suppl reml}Restricted likelihood}
				
				The adjusted log-likelihood $l_{\mathrm{a}}(\sigma^{2},\theta)$ given in
				equation (\ref{la sig lam}) is also equivalent to the restricted, or residual,
				log-likelihood for $(\sigma^{2},\theta)$ \citep{Thompson1962, Patterson1971}.
				
				Let $D$ be a full rank $(n-k)\times n$ matrix, independent of $\beta$, and
				such that $DX=0$. Then, the distribution of $Dy$ does not depend on $\beta$.
				The Gaussian restricted log-likelihood for $(\sigma^{2},\theta)$ is the
				log-likelihood based on $Dy$. It can be immediately verified that this
				log-likelihood does not depend on the specific $D$ that is chosen, and is
				equivalent to the adjusted log-likelihood $l_{\mathrm{a}}(\sigma^{2},\theta)$.
				The procedure can also be taken a step further, to produce a likelihood for
				$\theta$ only. The distribution of $Dy/\left\Vert Dy\right\Vert $ does not
				depend on $\beta$ and $\sigma^{2}$, and it is easily seen that the
				corresponding log-likelihood is the same as $l_{\mathrm{a}}(\theta)$ in
				(\ref{la theta}). Contrary to the profile log-likelihood $l(\theta)$ in
				(\ref{l_theta}), the log-likelihoods (\ref{la sig lam}) and (\ref{la theta})
				are genuine likelihoods (because they correspond to the density of observable
				random variables, $Dy$ and $Dy/\left\Vert Dy\right\Vert $, respectively).
				Hence, they provide unbiased and information unbiased estimating equations.
				There is a large literature on estimating covariance parameters by maximizing
				the restricted likelihood, particularly for variance components models. Often
				the resulting estimator has a distribution that is closer to its asymptotic
				distribution, and has important robustness properties
				\citep[e.g.,][]{Verbyla93}. Also, the estimator can be consistent even when
				$k$ increases at same rate as $n$ \citep[e.g.,][]{SmythVerbyla99}. Note that
				equations (\ref{la sig lam}) and (\ref{la theta}) do not depend on $D$. Taking
				$D=C$ shows the equivalence to the maximal invariant approach described in
				Section \ref{subsec inv}.
				
				Finally, it is worth pointing out that the log-likelihood $l_{\mathrm{a}%
				}(\sigma^{2},\theta)$ can also be interpreted as a \cite{CoxReid1987}
				approximate conditional likelihood; see, e.g., \cite{Bellhouse1990}.
				
				\begin{mycomment}
					FOR MYSELF: \emph{I have not proved yet }$l(\lambda)\rightarrow+\infty$\emph{
						in these cases, but see plot\_lik\_SEM\_with\_first\_term.m}. If a failure of
					Assumption \ref{assum id SEM} implied a failure of Assumption \ref{assum id}
					(I don;t think that's the case), we could have used Lemma
					\ref{lemma l(lambda) violation}
				\end{mycomment}

				\begin{mycomment}
					I had the following lemma, but It's wrong I think that neither $M_{X}%
					S(\omega^{-1})=0$ then $M_{S(\omega^{-1})X}S(\omega^{-1})=0$, nor the reverse.
					Finally, it is possible to relate Assumption \ref{assum id SEM} to the
					corresponding assumption for a SAR model, Assumption \ref{assum id}.
					\begin{lemma}
						\label{lemma colcol implies colcol}For any nonzero real eigenvalue $\omega$ of
						$W$, if $M_{X}S(\omega^{-1})=0$ then $M_{S(\omega^{-1})X}S(\omega^{-1})=0.$
					\end{lemma}
					\begin{pff}
						[Proof of Lemma \ref{lemma colcol implies colcol}]$M_{X}S(\omega^{-1})=0$ is
						equiv to $\operatorname{col}(S(\omega^{-1}))\subseteq\operatorname{col}(X)$
						but $\operatorname{col}(S(\omega^{-1})X)\subseteq\operatorname{col}%
						(S(\omega^{-1}))$ (because $\operatorname{col}(AB)\subseteq\operatorname{col}%
						(A)$ for any two conformable matrices $A$ and $B$) so if $M_{X}S(\omega
						^{-1})=0$ then $\operatorname{col}(S(\omega^{-1})X)\subseteq\operatorname{col}%
						(X)$ (i.e. $M_{X}S(\omega^{-1})X=0$ ) $\operatorname{col}(S(\omega
						^{-1})X)\subseteq\operatorname{col}(X)$ if and only if $\operatorname{col}%
						^{\perp}(X)\subseteq\operatorname{col}^{\perp}(S(\omega^{-1})X)$ if
						$\operatorname{col}^{\perp}(X)\subseteq\operatorname{col}^{\perp}%
						(S(\omega^{-1})X)$ $M_{X}A=M_{S(\omega^{-1})X}A$ desired:\ then
						$\operatorname{col}(S(\omega^{-1}))\subseteq\operatorname{col}(S(\omega
						^{-1})X);$ recall $\operatorname{col}(S(\omega^{-1}))\subseteq
						\operatorname{col}(S(\omega^{-1})X)$ is the same as $\operatorname{col}%
						(S(\omega^{-1}))=\operatorname{col}(S(\omega^{-1})X)$, because
						$\operatorname{col}(S(\omega^{-1})X)$ is obviously a subset of
						$\operatorname{col}(S(\omega^{-1}))$ WTS\ $\operatorname{col}(S(\omega
						^{-1}))\subseteq\operatorname{col}(X)$ implies $\operatorname{col}%
						(S(\omega^{-1}))\subseteq\operatorname{col}(S(\omega^{-1})X)$ but
						$\operatorname{col}(S(\omega^{-1})X)\subseteq\operatorname{col}(S(\omega
						^{-1}))$, so if $\operatorname{col}(S(\omega^{-1}))\subseteq\operatorname{col}%
						(X)$ then $\operatorname{col}(S(\omega^{-1})X)\subseteq\operatorname{col}(X)$
						$\left(  I_{n}-X(X^{\prime}X)^{-1}X^{\prime}\right)  S(\omega^{-1})=0$
						$\left(  I_{n}-S(\omega^{-1})X(X^{\prime}S^{\prime}(\omega^{-1})S^{\prime
						}(\omega^{-1})X)^{-1}X^{\prime}S^{\prime}(\omega^{-1})\right)  S(\omega
						^{-1})=0$ UPDATE\ THIS\ PROOF\ TO\ NEW\ FORMULATION: Since, clearly,
						$\operatorname{col}(AB)\subseteq\operatorname{col}(A)$ for any two conformable
						matrices $A$ and $B$, it follows that $\operatorname{col}(S(\omega
						^{-1}))\subseteq\operatorname{col}(X)$ implies $\operatorname{col}%
						(S(\omega^{-1})X)\subseteq\operatorname{col}(X)$, for any nonzero real
						eigenvalue $\omega$ of $W$. But $\operatorname{col}(S(\omega^{-1}%
						)X)\subseteq\operatorname{col}(X)$ if and only if $\operatorname{col}^{\perp
						}(X)\subseteq\operatorname{col}^{\perp}(S(\omega^{-1})X)$, and therefore
						$M_{X}S(\omega^{-1})=0$ implies $M_{S(\omega^{-1})X}S(\omega^{-1})=0$
						(NOOOOOOOOOO!). The desired statement follows, because $M_{X}S(\omega^{-1})=0$
						is equivalent to $\operatorname{col}(S(\omega^{-1}))\subseteq
						\operatorname{col}(X)$, and, by Lemma \ref{lemma SSX}, $M_{S(\omega^{-1}%
							)X}S(\omega^{-1})=0$ is equivalent to $\operatorname{col}(S(\omega
						^{-1}))=\operatorname{col}(S(\omega^{-1})X)$. OLD Since, clearly,
						$\operatorname{col}(AB)\subseteq\operatorname{col}(A)$ for any two conformable
						matrices $A$ and $B$, it follows that $\operatorname{col}(S(\omega
						^{-1}))\subseteq\operatorname{col}(X)$ implies $\operatorname{col}%
						(S(\omega^{-1})X)\subseteq\operatorname{col}(X)$, for any nonzero real
						eigenvalue $\omega$ of $W$. But $\operatorname{col}(S(\omega^{-1}%
						)X)\subseteq\operatorname{col}(X)$ if and only if $\operatorname{col}^{\perp
						}(X)\subseteq\operatorname{col}^{\perp}(S(\omega^{-1})X)$, and therefore
						$M_{X}S(\omega^{-1})=0$ implies $M_{S(\omega^{-1})X}S(\omega^{-1})=0$. The
						desired statement follows, because $M_{X}S(\omega^{-1})=0$ is equivalent to
						$\operatorname{col}(S(\omega^{-1}))\subseteq\operatorname{col}(X)$, and, by
						Lemma \ref{lemma SSX}, $M_{S(\omega^{-1})X}S(\omega^{-1})=0$ is equivalent to
						$\operatorname{col}(S(\omega^{-1}))=\operatorname{col}(S(\omega^{-1})X)$.
					\end{pff}
					It is worth noting that the reverse of Lemma \ref{lemma colcol implies colcol}
					does not hold; in other words, Assumption \ref{assum id SEM} is stronger than
					Assumption \ref{assum id}. \bigskip
				\end{mycomment}

				\begin{mycomment}
					\bigskip
					so so far the required modification of Assumption \ref{assum id} seems to be
					(any relation between the 2 conditions in the assumption \ref{assum id SEM}?)
					for a spatial error model there are more cases such that lik is unbounded from
					above (\textbf{but REML is not flat})
					if $\operatorname{col}(S(\omega^{-1}))\subseteq\operatorname{col}(X)$ REML
					flat. if $\operatorname{col}(S(\omega^{-1}))=\operatorname{col}(S(\omega
					^{-1})X)$ (which is more general) then from simul it looks like REML a.s.
					continuous at $\omega^{-1}$ (need to check)
					$y=S^{-1}(\lambda)X\beta+\sigma S^{-1}(\lambda)\varepsilon,$
					$\operatorname{col}(S^{-1}(\lambda)X)=\operatorname{col}(X)$ then this is like
					a spatial error model (see section on invariance)\bigskip
					SEM
					$y=X\beta+\sigma S^{-1}(\lambda)\varepsilon$
					$S(\lambda)y=S(\lambda)X\beta+\sigma\varepsilon$
					$y=\lambda Wy+S(\lambda)X\beta+\sigma\varepsilon$
					$y=\lambda Wy+X\beta-\lambda WX\beta+\sigma\varepsilon$ (*)
				\end{mycomment}

				\begin{mycomment}
					Also it is clear from (*) that if $\operatorname{col}(W)\subseteq
					\operatorname{col}(X)$ (which requires $rank(W)\leq k$) then $\lambda$ and
					$\beta$ are not separately identifiable\footnote{If $\operatorname{col}%
						(W)\subseteq\operatorname{col}(X)$, then for any $y\in\mathbb{R}^{n}$ there
						exists a $\bar{\beta}$ such that $Wy=X\bar{\beta}$, and hence the spatial
						error model becomes $y=X(\lambda\bar{\beta}+\beta)-\lambda WX\beta
						+\varepsilon$.
						\par
						but if $\operatorname{col}(W)\subseteq\operatorname{col}(X)$ then col(X) is
						inv subsp of $W$, so there is $k$ by $k$ $A$ such that $WX=XA$ so
						\par
						$y=X(\lambda\bar{\beta}+\beta)-\lambda XA\beta+\varepsilon$
						\par
						$=X(\lambda\bar{\beta}+\beta-\lambda A\beta)+\varepsilon$}\emph{check this in
						view of PP identif result}
					note that if $\operatorname{col}(W)\subseteq\operatorname{col}(X)$ the prof
					score does not depend on $y$ and adj prof lik is flat
					also recall if $\operatorname{col}(W)\subseteq\operatorname{col}(X)$ then
					col(X) is inv subs of $W$ so prof lik for SEM and SLM are the same
					- anything else?
					We can NOT join $\operatorname{col}(S(\omega^{-1}))=\operatorname{col}%
					(S(\omega^{-1})X)$ and $\operatorname{col}(W)\subseteq\operatorname{col}%
					(X)$\bigskip
				\end{mycomment}

				\begin{mycomment}
					For any nonzero real eigenvalue $\omega$ of $W$, $\operatorname{col}%
					(S(\omega^{-1}))=\operatorname{col}(S(\omega^{-1})X)$ creates identif prob in
					SEM. The case $\omega=0$ would give $\operatorname{col}(W)=\operatorname{col}%
					(WX)$ but this we don;t need to rule out as it does not create an identif
					issue in $y=\lambda Wy+X\beta-\lambda WX\beta+\sigma\varepsilon.$
					$\operatorname{col}(W)=d$ means for any $y\in R^{n}$ there exists a
					$\bar{\beta}$ such that $Wy=WX\bar{\beta}$, and hence the SAR model becomes
					$y=\lambda WX(\bar{\beta}-\beta)+X\beta+\varepsilon$ ($\beta$ and $\lambda$
					can still be separately identified provided that...). only thing is that this
					is not an autoregression so it's a bit of a trivial case....................do
					we want to rule it out?
				\end{mycomment}

				\section{\label{sec suppl proof aux lemma} Proof of Lemma
					\ref{lemma inv sub aML}}
				
				If $\operatorname{col}(X)$ is an invariant subspace of $W$, $M_{X}%
				=M_{S(\lambda)X}$ for any $\lambda$ such that $S(\lambda)$ is invertible, and
				hence the profile quasi log-likelihood function for $(\sigma^{2},\lambda)$
				implied by the SAR model, given in equation (\ref{lik sig lam SAR}), is the
				same as that implied by the spatial error model, given by equation
				(\ref{lik sig theta}) with $\theta=\lambda$ and $A(\theta)=S(\lambda)$. It
				follows that, if $\operatorname{col}(X)$ is an invariant subspace of $W$, the
				score $s(\sigma^{2},\lambda)$ for a SAR model is given by equation
				(\ref{s(sig,theta) SEM}). The adjusted log-likelihood function $l_{\mathrm{a}%
				}(\sigma^{2},\lambda)$ is obtained by centering $s(\sigma^{2},\lambda)$ under
				the assumptions that $\mathrm{E}(\varepsilon)=0$ and $\mathrm{var}%
				(\varepsilon)=I_{n}$. If $\operatorname{col}(X)$ is an invariant subspace $W$,
				there exists a unique $k\times k$ matrix $A$ such that $WX=XA$, and hence
				$S^{-1}(\lambda)X=X(I_{k}-\lambda A)^{-1},$ for any $\lambda$ such that
				$S(\lambda)$ is invertible. Thus, when $\operatorname{col}(X)$ is an invariant
				subspace $W$, the SAR model $y=S^{-1}(\lambda)X\beta+\sigma S^{-1}%
				(\lambda)\varepsilon$ can be written as $y=X(I_{k}-\lambda A)^{-1}\beta+\sigma
				S^{-1}(\lambda)\varepsilon.$ Using this representation, it is clear that
				$\mathrm{E}(s(\sigma^{2},\lambda))$ is the same as given in equation
				(\ref{Es SEM}) for the spatial error model. Hence, $l_{\mathrm{a}}(\sigma
				^{2},\lambda)$ is given by equation (\ref{la sig lam}), i.e.,%
				\begin{equation}
				l_{\mathrm{a}}(\sigma^{2},\lambda)=-\frac{n-k}{2}\log(\sigma^{2})-\frac
				{1}{2\sigma^{2}}y^{\prime}U(\lambda)y-\frac{1}{2}\log\left(  \det\left(
				\Sigma(\lambda)\right)  \right)  -\frac{1}{2}\log\left(  \det\left(
				X^{\prime}\Sigma^{-1}(\lambda)X\right)  \right)  , \label{la sig lam 2}%
				\end{equation}
				where $U(\lambda)\coloneqq S^{\prime}(\lambda)M_{S(\lambda)X}S(\lambda)$.
				
				Consider now the quasi log-likelihood function $l(\sigma^{2},\lambda;Dy)$
				for $(\sigma^{2},\lambda)$ based on $Dy$, for any full rank $\left(
				n-k\right)  \times n$ matrix $D$ such that $DX=0$, when $y=X(I_{k}-\lambda
				A)^{-1}\beta+\sigma S^{-1}(\lambda)\varepsilon$, and $\varepsilon
				\sim\mathrm{N}(0,I_{n})$. Since $Dy\sim\mathrm{N}(0,\sigma^{2}D\Sigma
				(\lambda)D^{\prime})$, we have
				\begin{equation}
				l(\sigma^{2},\lambda;Dy)=-\frac{n-k}{2}\log(\sigma^{2})-\frac{1}{2}\log\left(
				\det\left(  D\Sigma(\lambda)D^{\prime}\right)  \right)  -\frac{1}{2\sigma^{2}%
				}y^{\prime}D^{\prime}\left(  D\Sigma(\lambda)D^{\prime}\right)  ^{-1}Dy.
				\label{lDy}%
				\end{equation}
				But the log-likelihood functions (\ref{lDy}) and (\ref{la sig lam 2}) are
				equivalent, because
				\[
				\det\left(  D\Sigma(\lambda)D^{\prime}\right)  =\det(DD^{\prime})\left(
				\det(X^{\prime}X)\right)  ^{-1}\det(X^{\prime}\Sigma^{-1}(\lambda
				)X)\det(\Sigma(\lambda))
				\]
				and%
				\[
				D^{\prime}\left(  D\Sigma(\lambda)D^{\prime}\right)  ^{-1}D=\Sigma
				^{-1}(\lambda)(I-X(X^{\prime}\Sigma^{-1}(\lambda)X)^{-1}X^{\prime}\Sigma
				^{-1}(\lambda))=U(\lambda).
				\]

				\begingroup\setlength{\bibsep}{6pt} \setstretch{1.25}
				\bibliographystyle{elsart-harv}
				\bibliography{supplement}
				\endgroup

			\end{document}